\documentclass[preprint,12pt]{elsarticle}

\usepackage[utf8]{inputenc}

\usepackage{dsfont,amssymb,amsmath,graphics,graphicx,fancyhdr,float,verbatim,
ifthen,ifpdf,algorithmic,multirow,array,color, enumerate,theorem}
\usepackage[english]{babel}
\usepackage[perpage]{footmisc}
\usepackage[ruled]{algorithm}
\usepackage[subnum]{cases}

\renewcommand{\parallel}{|\!|}
 
\usepackage[margin=1.15in]{geometry}

\usepackage{natbib} 
\biboptions{authoryear, round}
 \usepackage[
                  breaklinks = true,
                 colorlinks = true,
                 linkcolor = red,
                 urlcolor  = black, 
                 citecolor = blue,
                 anchorcolor = green,
                 ]{hyperref}

\newcommand{\bw}{\mathbf{w}}

\newcommand{\bse}{\boldsymbol{e}}

\newcommand{\bsr}{\boldsymbol{r}}

\newcommand{\bsx}{\boldsymbol{x}}
\newcommand{\bsX}{\boldsymbol{X}}
\newcommand{\bsy}{\boldsymbol{y}}

\newcommand{\bsZ}{\boldsymbol{Z}}

\newcommand{\bsalpha}{\boldsymbol{\alpha}}
\newcommand{\bsbeta}{\boldsymbol{\beta}}

\newcommand{\bsOmega}{\boldsymbol{\Omega}}

\newcommand{\bstheta}{\boldsymbol{\theta}}

\newcommand{\bsPsi}{\boldsymbol{\Psi}}
\newcommand{\bsvPsi}{\boldsymbol{\varPsi}}

\newcommand{\E}{\mathbb{E}}
\newcommand{\V}{\mathbb{V}}
\newcommand{\Pro}{\mathbb{P}}
\newcommand{\R}{\mathbb{R}}

\newcommand{\ICL}{\text{ICL}}

\journal{Neurocomputing - Elsevier}
\begin{document}

\begin{frontmatter}

\title{Robust mixture of experts modeling \\ using the skew $t$ distribution}

\author[LMNO-Unicaen]{F. Chamroukhi \corref{cor1}} \ead{faicel.chamroukhi@unicaen.fr}
\cortext[cor1]{Corresponding author: Faicel Chamroukhi\\ Universit\'e de Caen-Normandie, LMNO, UMR CNRS 6139 \\  
Campus 2, Bvd Mar\'echal Juin, 14032 Caen Cedex, France\\
Tel: +33(0) 2 31 56 73 67\\Fax:  +33(0) 2 31 56 73 20 }

\address[LMNO-Unicaen]{Normandie Universit\'e, UNICAEN, CNRS, Laboratoire de Math\'ematiques Nicolas Oresme - LMNO, 14000 Caen, France} 

\begin{abstract} 
Mixture of Experts (MoE) is a popular framework in the fields of statistics and machine learning for modeling heterogeneity in data for regression, classification and clustering. MoE for continuous data are usually based on the normal distribution. However, it is known that for data with asymmetric behavior, heavy tails and atypical observations, the use of the normal distribution is unsuitable. We introduce a new robust non-normal mixture of experts modeling using the skew $t$ distribution. The proposed  skew $t$ mixture of experts, named STMoE, handles these issues of the normal mixtures experts regarding possibly skewed, heavy-tailed and noisy data. We develop a dedicated expectation conditional maximization (ECM) algorithm  to estimate the model parameters  by monotonically maximizing the observed data log-likelihood. We describe how the presented model can be used in prediction and in model-based clustering of regression data. Numerical experiments carried out on simulated data show the effectiveness and the robustness of the proposed model in fitting non-linear regression functions as well as in model-based clustering. Then, the proposed model is applied to the real-world data of tone perception for musical data analysis, and the one of temperature anomalies for  the analysis of climate change data. The obtained results confirm the usefulness of the model for practical data analysis applications.
\end{abstract}

\begin{keyword}
mixture of experts,
skew $t$ distribution; 
EM algorithm;
ECM algorithm;
non-linear regression;
model-based clustering.
\end{keyword}

\end{frontmatter}


\section{Introduction}

Mixture of Experts (MoE) \citep{jacobsME} is a popular framework in the statistics and machine learning fields for modeling heterogeneity in data for regression, classification and clustering. They consist in a fully conditional mixture model where both the mixing proportions, known as the gating functions, and the component densities, known as the experts, are conditional on some input covariates. MoE have been investigated, in their simple form, as well as in their hierarchical form \citet{jordanHME} (e.g Section 5.12 of \citet{McLachlanFMM}) for regression and model-based cluster and discriminant analyses and in different application domains. 
MoE Have also been investigated for rank data \citet{Gormley2008MoE-Rank} and network data \citet{Gormley2010MoE-Net} with social science applications. A survey on the topic can be found in \citet{Gormley2011MoE-Survey}. A complete review of the MoE models can be found in \citet{YukselWG12}. MoE for continuous data are usually based on the normal distribution. Along this paper, we will call the MoE using the normal distribution the  normal mixture of experts, abbreviated as NMoE. However, it is well-known that the normal distribution is sensitive to outliers. Moreover, for a set of data containing a group or groups of observations with heavy tails or asymmetric behavior, the use of normal experts may be unsuitable and can unduly affect the fit of the MoE model. 
In this paper, we attempt to overcome these limitations in MoE by proposing a more adapted and robust mixture of experts model which can deal with possibly skewed, heavy-tailed data and with outliers.

Recently, the problem of sensitivity of NMoE to outliers have been considered by \citet{Nguyen2016-MoLE} where the authors proposed  a Laplace mixture of linear experts (LMoLE) for a robust modeling of non-linear regression data. The model parameters are estimated by maximizing the observed-data likelihood via a minorization-maximization (MM) algorithm. Here, we propose an alternative MoE model, by relaying on other non-normal distribution that generalizes the normal distribution, that is, the skew-$t$ distribution introduced quite recently by \citet{AzzaliniAndCapitanio2003}.  We call the proposed MoE model the skew-$t$ mixture of experts (STMoE). One may use the $t$ distribution, as in the $t$ mixture of experts (TMoE) proposed by \cite{Chamroukhi-NNMoE-2015, Chamroukhi-TMoE} which provides a natural robust extension of the normal distribution to model data with more heavy tails and to deal with possible outliers. 
The robustness of the $t$ distribution may however be not sufficient in the presence of asymmetric observations. In mixture modeling, to deal with this issue regarding skewed data, \citet{Lin07univSkewtMixture} proposed the univariate skew-$t$ mixture model which allows for accommodation of both skewness and thick tails in the data, by relying on the skew-$t$ distribution \citet{AzzaliniAndCapitanio2003}. 
For the general multivariate case using skew-$t$ mixtures, one can refer to 
\citet{Pyne2009},
\citep{Lin2010SkewtMvMixture},
\citet{LeeAndMchLachlan13non-normal-mix},
\citet{LeeAndMcLachlan13skew},
\citet{LeeAndMcLachlan14-skewtmix}, and recently, the unifying framework for previous restricted and unrestricted skew-$t$ mixtures, using the CFUST distribution \citet{LeeAndMcLachlan15-CFUST}.

The inference in the previously described approaches is performed by maximum likelihood estimation via the expectation-maximization (EM) algorithm or its extensions \citep{dlr,McLachlanEM2008}, in particular the  expectation conditional maximization (ECM) algorithm \citep{meng_and_rubin_ECM_93}. \citet{Fruhwirth10BayesSkewMixtures} have also considered the Bayesian inference framework for namely the skew-$t$ mixtures.

For the regression context, the robust modeling of regression data has been studied namely by \citet{Bai2012,Wei2012,Ingrassia2012}  who considered a mixture of linear regressions using the $t$ distribution. In the same context of regression, \citet{Song2014} proposed the mixture of Laplace regressions, which has been then extended by \citet{Nguyen2016-MoLE} to the case of mixture of experts, by introducing the Laplace mixture of linear experts (LMoLE). Recently, \citet{Zeller15SkewNMixReg}  introduced the  scale mixtures of skew-normal distributions for robust mixture regressions. However, unlike our proposed STMoE model, the regression mixture models of \citet{Wei2012}, \citet{Bai2012}, \cite{Ingrassia2012}, \citet{Song2014}, \citet{Zeller15SkewNMixReg} do not consider conditional mixing proportions, that is, mixing proportions depending on some input variables, as in the case of mixture of experts, which we investigate here. In addition, the approaches of \citet{Wei2012}, \citet{Bai2012}, \cite{Ingrassia2012} and \citet{Song2014} do not consider both the problem of robustness to outliers together with the one of dealing with possibly asymmetric data. 

Here we consider the mixture of experts framework for non-linear regression problems and model-based clustering of regression data, and we attempt to overcome the limitations of the NMoE model for dealing with asymmetric, heavy-tailed data and which may contain outliers. We investigate the use of the  skew $t$ distribution for the experts, rather than the commonly used normal distribution. We propose the skew-$t$ mixture of experts (STMoE) model  which allows for accommodation of both skewness and heavy tails in the data and which is  obust to outliers. This model corresponds to an extension of the unconditional skew $t$ mixture model \citep{Lin07univSkewtMixture}, to the  mixture of experts (MoE) framework, where the mixture means are regression functions and the mixing proportions are also covariate-varying. 

For the model inference, we develop a dedicated expectation conditional maximization (ECM) algorithm to estimate the model parameters by monotonically maximizing the observed data log-likelihood. The expectation-maximization algorithm and its extensions \citep{dlr,McLachlanEM2008} are indeed very popular and successful estimation algorithms for mixture models in general and for mixture of experts in particular.
Moreover, the EM algorithm for MoE has been shown by \citet{NgM-EM-IRLS-04} to be monotonically maximizing the MoE likelihood. The authors have showed that the EM (with Iteratively Reweighted Least Squares (IRLS) in this case) algorithm has stable convergence and  the log-likelihood is monotonically increasing when a learning rate smaller than one is adopted for the IRLS procedure within the M-step of the EM algorithm.
They have further proposed an expectation conditional maximization (ECM) algorithm to train MoE, which also has desirable numerical properties. 
The MoE has also been considered in the Bayesian framework, for example one can cite the Bayesian MoE \cite{Waterhouse96bayesianMoE,Waterhouse1997} and the Bayesian hierarchical MoE \citet{Bishop_BayesianMoE}. 
Beyond the Bayesian parametric framework, the MoE models have also been investigated within the Bayesian non-parametric framework.
We cite for example the  Bayesian non-parametric MoE model  \citep{Rasmussen01infiniteMoE}
and the Bayesian non-parametric hierarchical MoE approach of \citet{ShiMT05_Hierarchical_GPR} using Gaussian Processes experts for regression. For further models on mixture of experts for regression, the reader can be referred to for example the book of \citet{ShiGPR_Book2011}. 
In this paper, we investigate semi-parametric models under the maximum likelihood estimation framework.

The remainder of this paper is organized as follows. In Section \ref{sec: MoE} we briefly recall the normal MoE framework. In Section 
Then, in Section \ref{sec: STMoE}, we present the STMoE model and in  Section \ref{sec: MLE for the STMoE} the parameter estimation technique using the ECM algorithm.
We then investigate in Section \ref{sec: Prediction using the STMoE} the use of the proposed model for non-linear regression and for prediction. We also show in Section \ref{sec: MBC using the STMoE} how the model can be used in a model-based clustering prospective.
In Section \ref{sec: Model selection for the STMoE}, we discuss the model selection. Section \ref{sec: Experimental study} is dedicated to the experimental study to assess the proposed model. Finally, in Section \ref{sec: Conclusion}, conclusions are drawn and we open a future work.

\section{Mixture of experts for continuous data}
\label{sec: MoE}

Mixtures of experts \citep{jacobsME,jordanHME} are used in a variety of contexts including regression, classification and clustering. 
Here we consider the MoE framework for fitting (non-linear) regression functions and clustering of univariate continuous data . 
The aim of regression is to explore the relationship of an observed random variable $Y$ given a covariate vector $\bsX \in \R^p$ via conditional density functions for $Y|\bsX = \bsx$ of the form $f (y|\bsx)$, rather than only exploring the unconditional distribution of $Y$. 
Thanks to their great flexibility, mixture models \citep{McLachlanFMM} has took much attention for non-linear regression problems and we  distinguish in particular the classical mixture of regressions model  \cite{Quandt1972, QuandtANDRamsey1978, DeVeaux1989, JonesANDMcLachlan1992, Gaffney99trajectoryclustering, VieleANDTong2002, FariaANDSoromenho2010,HunterANDYoung} and mixture of experts for regression analysis \citep{jacobsME,jordanHME,YoungANDHunter}. 
The univariate mixture of regressions model assumes that the observed pairs of data $(\bsx,y)$ where $y \in \R$ is the response for some covariate $\bsx \in \R^p$, are generated from $K$ regression functions and are governed by a hidden categorical random variable $Z$ indicating from which component each observation is generated.
Thus, the mixture of regressions model decomposes the nonlinear regression model density $f (y|\bsx)$ into a convex weighted sum of $K$ regression component models $f_k(y|\bsx)$ and can be defined as follows: %
\begin{eqnarray}
f(y|\bsx;\bsvPsi) &=& \sum_{k=1}^K \pi_k f_k(y|\bsx; \bsvPsi_k)
\label{eq: mixture of regressions}
\end{eqnarray}where the $\pi_k$'s defined by $\pi_k = \Pro(Z = k)$ and represent the non-negative mixing proportions that sum to 1. The model parameter vector is given by $\bsvPsi = (\pi_1,\ldots,\pi_{K-1},\bsvPsi^T_1,\ldots,\bsvPsi^T_K)^T$,  $\bsvPsi_k$ being the parameter vector of the $k$th component density.

\subsection{The mixture of experts (MoE) model}

Although similar, the mixture of experts \citep{jacobsME} differ from regression mixture models in many aspects. One of the main differences is that the MoE model consists in a fully conditional mixture while in the regression mixture, only the component densities are conditional. Indeed, the mixing proportions are constant for the regression mixture, while in the MoE, they are modeled as a function of the inputs, generally modeled by logistic or a softmax function. 
Mixture of experts (MoE) for regression analysis \citep{jacobsME,jordanHME} extend the model (\ref{eq: mixture of regressions}) by modeling the mixing proportions as function of some covariates $\bsr \in \R^q$. 
The mixing proportions, known as the gating functions in the context of MoE, are modeled by the multinomial logistic model and are defined by:
\begin{eqnarray}
\pi_{k}(\bsr;\bsalpha) =\Pro(Z=k|\bsr;\bsalpha) =\frac{\exp{(\bsalpha_k^T\bsr)}}{\sum_{\ell=1}^K\exp{(\bsalpha_{\ell}^T \bsr)}}
\label{eq: multinomial logistic}
\end{eqnarray}where $\bsr \in \R^q$ is a covariate vector, $\bsalpha_{k}$ is the $q$-dimensional coefficients vector associated with $\bsr$ and $\bsalpha = (\bsalpha^T_1,\ldots,\bsalpha^T_{K-1})^T$ is the parameter vector of the logistic model, with $\bsalpha_K$ being the null vector. 
Thus, the MoE model consists in a fully conditional mixture model where both the mixing proportions (the gating functions) and the component densities (the experts) are conditional on predictors (respectively $\bsr$ and $\bsx$). 

\subsection{The normal mixture of experts (NMoE) model and maximum likelihood estimation}
\label{sec: MLE for the NMoE}
In the case of mixture of experts for regression, it is usually assumed that the experts are normal, that is, follow a normal distribution. A $K$-component normal mixture of experts (NMoE)  ($K>1$) has the following formulation:
\begin{eqnarray}
f(y|\bsr,\bsx;\bsvPsi) &=& \sum_{k=1}^K \pi_k(\bsr;\bsalpha) \text{N}\!\left(y; \mu(\bsx;\bsbeta_k), \sigma_k^2\right)
\label{eq: normal MoE}
\end{eqnarray}which involves, in the semi-parametric case, component means defined as parametric (non-)linear regression functions $\mu(\bsx;\bsbeta_k)$.  
 
The NMoE model parameters are estimated by maximizing the observed data log-likelihood by using the EM algorithm \citep{dlr, jacobsME, jordanHME, jordan_and_xu_1995, NgM-EM-IRLS-04, McLachlanEM2008}.  Suppose we observe an i.i.d sample of $n$ individuals $(y_1,\ldots,y_n)$ with their respective associated covariates $(\bsx_1,\ldots,\bsx_n)$ and $(\bsr_1,\ldots,\bsx_r)$.  Then, under the NMoE model, the observed data log-likelihood for the parameter vector $\bsvPsi$ is given by:
\begin{equation}
\log L(\bsvPsi)  = \sum_{i=1}^n  \log  \sum_{k=1}^K \pi_k(\bsr_i;\bsalpha) \text{N}\!\left(y_i; \mu(\bsx;\bsbeta_k), \sigma_k^2\right).
\label{eq: log-lik normal MoE}
\end{equation}
The E-Step at the $m$th iteration of the EM algorithm for the NMoE model requires the calculation of the following posterior probability that the individual $(y_i, \bsx_i,\bsr_i)$ belongs to expert $k$, given a parameter estimation $\bsvPsi^{(m)}$:
\begin{eqnarray}
\tau_{ik}^{(m)} = \Pro(Z_i=k|y_{i},\bsx_i,\bsr_i;\bsvPsi^{(m)})
 =  \frac{\pi_k(\bsr;\bsalpha^{(m)}) \text{N}\!\left(y_i;\mu_k(\bsx_i; \beta^{(m)}_k),{\sigma^2_k}^{(m)}\right)}{f(y_i|\bsr_i,\bsx_i;\bsvPsi^{(m)})}.
\label{eq: posterior prob NMoE}
\end{eqnarray}
Then, the M-step calculates the parameter update $\bsvPsi^{(m+1)}$ by maximizing the well-known $Q$-function (the expected complete-data log-likelihood), that is:
\begin{equation}
\bsvPsi^{(m+1)} =  \arg \max_{\bsvPsi \in \bsOmega} Q(\bsvPsi;\bsvPsi^{(m)})
\label{eq: arg max Q}
\end{equation}where $\bsOmega$ is the parameter space.  
For example, in the case of  normal mixture of linear experts (NMoLE) where each expert's mean has the flowing linear form: 
\begin{equation}
\mu(\bsx;\bsbeta_k) = \bsbeta_k^T \bsx, 
\label{eq: linear regression mean}
\end{equation}where $\bsbeta_k \in \R^p$ is the  vector of regression coefficients of component $k$, the updates for each of the expert component parameters consist in analytically solving a weighted Gaussian linear regression problem and are given by:
\begin{eqnarray}
\bsbeta_k^{(m+1)}  &=& \Big[\sum_{i=1}^{n}\tau^{(m)}_{ik}  \bsx_i\bsx^T_i \Big]^{-1} \sum_{i=1}^{n} 
 \tau^{(q)}_{ik}  y_i \bsx_i,
\label{eq: beta_k update for NMoE}\\ 
{\sigma^2_{k}}^{(m+1)} &= &
\frac{\sum_{i=1}^n\tau_{ik}^{(m)}\left(y_i - {\bsbeta^T_{k}}^{(m+1)}\bsx_i\right)^2}{\sum_{i=1}^n\tau_{ik}^{(m)}}\cdot
\label{eq: sigma2k update NMoE}
\end{eqnarray}
For the mixing proportions, the parameter update $\bsalpha^{(m+1)}$  cannot however be obtained in a closed form. It is calculated by IRLS \citep{jacobsME,jordanHME,Chen1999,Green1984,Chamroukhi-IJCNN-2009}. 

However, the normal distribution is not adapted to deal with asymmetric and heavy tailed data. It is also known that the normal distribution is sensitive to outliers.  
In the proposal, we address these issues regarding the skewness, heavy tails and atypical observations in the data, by proposing a robust MoE modeling  by using the skew-$t$ distribution, recently introduced by \citet{AzzaliniAndCapitanio2003}, for the expert components rather than the usually used normal one.
The proposed skew $t$ mixture of experts (STMoE) allows for simultaneously accommodating asymmetry and heavy tails in the data and is also robust to outliers.

\section{The skew $t$ mixture of experts (STMoE) model}
\label{sec: STMoE}

The proposed  skew $t$ mixture of experts (STMoE) model is a MoE model in which the expert components have a skew-$t$ density, rather than the standard normal one as in the NMoE model. The skew-$t$ distribution \citet{AzzaliniAndCapitanio2003}, which is a robust generalization the skew-normal distribution \citep{Azzalini1985,Azzalini1986}, as well as its stochastic and hierarchical representations, which will be used to define the proposed STMoE model, are recalled in the following section.  

\subsection{The skew $t$ distribution}
Let us denote  by $t_{\nu}(.)$ and $T_{\nu}(.)$  respectively  the  pdf  and  cdf of  the  standard $t$ distribution  with  degrees  of  freedom $\nu$.
The skew $t$ distribution, introduced by \citet{AzzaliniAndCapitanio2003}, can be characterized as follows. Let $U$ be an univariate random variable with a standard skew-normal distribution $U \sim \text{SN}(0,1,\lambda)$ (which can be shortened as $U \sim \text{SN}(\lambda)$) with pdf given by (\ref{eq: Skew-normal density})). The skew-normal distribution is recalled in \ref{ssec: Skew-normal distribution}.
Then, let $W$ be an univariate random variable independent of $U$ and following the Gamma distribution, that is, $W \sim \text{Gamma}(\frac{\nu}{2},\frac{\nu}{2})$. 
A random variable $Y$ having the following representation:
\begin{equation}
Y = \mu + \sigma \frac{U}{\sqrt{W}}
\label{eq: stochastic representation skew-t}
\end{equation}follows the skew $t$ distribution $\text{ST}(\mu,\sigma^2,\lambda,\nu)$ with location parameter $\mu$, scale parameter $\sigma$, skewness parameter $\lambda$ and degrees of freedom $\nu$, whose density is defined by: 
\begin{equation}
f(y;\mu,\sigma^2,\lambda,\nu) = \frac{2}{\sigma} ~ t_{\nu}(d_y) ~  T_{\nu + 1} \left(\lambda ~ d_y \sqrt{\frac{\nu+1}{\nu+d^2_y}}\right)
\label{eq: skew-t density}
\end{equation} where $d_y = \frac{y-\mu}{\sigma}$.
From the hierarchical distribution of the skew-normal (\ref{eq: hierarchical representation skew-normal}), a further hierarchical representation of the stochastic representation (\ref{eq: stochastic representation skew-t}) of the skew $t$ distribution is given by:
\begin{eqnarray} 
Y_i|u_i, w_i&\sim & \text{N}\left(\mu + \delta |u_i|, \frac{1-\delta^2}{w_i}\sigma^2\right), \nonumber \\
U_i|w_i&\sim & \text{N}(0,\frac{\sigma^2}{w_i}),\label{eq: hierarchical representation skew-t}\\
W_i&\sim& \text{Gamma}\left(\frac{\nu}{2},\frac{\nu}{2}\right). 
\nonumber
\end{eqnarray}

\subsection{The skew $t$ mixture of experts (STMoE) model}
The skew proposed $t$ mixture of experts (STMoE) model extends the skew $t$ mixture model, which was first introduced by \citet{Lin07univSkewtMixture},  to the MoE framework. 
In the skew-$t$ mixture model of \citet{Lin07univSkewtMixture}, 
the mixing proportions and the components means are constant, that is, they are not predictor-depending. In the proposed STMoE, however, we consider skew-$t$ expert components in which both the mixing proportions and the mixture component means are predictor-depending. More specifically, we use polynomial regressors for the components, as well as multinomial logistic regressors for the mixing proportions.
A $K$-component mixture of skew $t$ experts (STMoE) is therefore defined by:
\begin{eqnarray}
f(y|\bsr,\bsx;\bsvPsi) &=& \sum_{k=1}^K \pi_k(\bsr;\bsalpha)  \,\text{ST}(y; \mu(\bsx;\bsbeta_k), \sigma_k^2, \lambda_k,\nu_k)\cdot
\label{eq: skew-t MoE}
\end{eqnarray}The parameter vector of the STMoE model is $\bsvPsi = (\bsalpha^T_1,\ldots,\bsalpha^T_{K-1},\bsvPsi^T_1,\ldots,\bsvPsi^T_K)^T$ where $\bsvPsi_k = (\bsbeta^T_k,\sigma^2_k,\lambda_k,\nu_k)^T$ is the parameter vector for the $k$th skew $t$ expert component whose density is defined by 
\begin{equation}
f\big(y|\bsx;\mu(\bsx;\bsbeta_k),\sigma^2,\lambda,\nu\big) = \frac{2}{\sigma} ~ t_{\nu}(d_y(\bsx)) ~  T_{\nu + 1} \left(\lambda ~ d_y(\bsx) \sqrt{\frac{\nu+1}{\nu+d^2_y(\bsx)}}\right)
\label{eq: skew-t density}
\end{equation}where $d_y(\bsx) = \frac{y-\mu(\bsx;\bsbeta_k)}{\sigma}\cdot$ 

It can be seen that, when the robustness parameter $\nu_k \rightarrow \infty$ for each $k$, the STMoE model (\ref{eq: skew-t MoE}) reduces to a skew-normal mixture of experts model (SNMoE)  (see \cite{Chamroukhi-NNMoE-2015}). 
 On the other hand, if the skewness parameter $\lambda_k = 0$ for each $k$, the STMoE model    reduces to the $t$ mixture of experts model (TMoE) (e.g., see \cite{Chamroukhi-NNMoE-2015,Chamroukhi-TMoE}).  Moreover, when  $\nu_k \rightarrow \infty$ and  $\lambda_k = 0$ for each $k$, the STMoE approaches the standard NMoE model (\ref{eq: normal MoE}). This therefore makes the STMoE very flexible as it generalizes the previously described MoE models to accommodate situations with asymmetry, heavy tails, and outliers. 
 
\subsection{Hierarchical representation of the STMoE model}
 By introducing the binary latent component-indicators $Z_{ik}$ such that $Z_{ik}=1$ iff $Z_i =k$, $Z_i$ being the hidden class label of the $i$th observation,  a hierarchical model for the STMoE model can be derived as follows.
From the hierarchical representation (\ref{eq: hierarchical representation skew-t}) of the skew $t$ distribution, a hierarchical model for the proposed STMoE model (\ref{eq: skew-t MoE}) can be derived from its stochastic representation (\ref{eq: stochastic representation skew-t MoE}) given in \ref{apx: stochastic representation of the STMoE}, and is as follows:
\begin{eqnarray} 
Y_i|u_i, w_i, Z_{ik}=1, \bsx_i &\sim& \text{N}\left(\mu(\bsx_i;\bsbeta_k) + \delta_k |u_i|, \frac{1-\delta^2_k}{w_i}\sigma^2_k\right), \nonumber \\
U_i|w_i, Z_{ik}=1&\sim & \text{N}\left(0,\frac{\sigma^2_k}{w_i}\right),\label{eq: hierarchical representation skew-t MoE}\\
W_i|Z_{ik}=1&\sim& \text{Gamma}\left(\frac{\nu_k}{2},\frac{\nu_k}{2}\right)\nonumber\\
\bsZ_i|\bsr_i &\sim & \text{Mult}\big(1;\pi_1(\bsr_i;\bsalpha),\ldots,\pi_K(\bsr_i;\bsalpha) \big).\nonumber
\end{eqnarray}The variables $U_i$ and $W_i$ are treated as hidden in this hierarchical representation, which facilitates the inference scheme and will be used to derive the maximum likelihood estimation of the STMoE model parameters $\bsvPsi$ by using the ECM algorithm.

\subsection{Identifiability of the STMoE model} 

\cite{Jiang_and_tanner_NN_99} have established that  ordered, initialized, and irreducible MoEs are identifiable. Ordered implies that there exist a certain ordering relationship on the experts parameters $\bsvPsi_k$ such that $(\bsalpha^T_1, \bsvPsi^T_1)^T \prec  \ldots \prec (\bsalpha^T_K, \bsvPsi^T_K)^T $; initialized implies that $\bw_K$, the parameter vector of the $K$th logistic proportion, is the null vector, and irreducible implies that $\bsvPsi_{k} \neq \bsvPsi_{k\prime}$ for any $k \neq k\prime$.
For the proposed STMoE, 
ordered implies that there exist a certain ordering relationship  such that  $(\bsbeta^T_1,\sigma^2_1,\lambda_1,\nu_1)^T \prec  \ldots \prec (\bsbeta^T_K,\sigma^2_K,\lambda_K,\nu_K)^T $; initialized implies that $\bw_K$ is the null vector, as assumed in the model, and finally,  irreducible implies that if $k \neq k\prime$, then one of the following conditions holds: 
 $\bsbeta_k\neq \bsbeta_{k\prime}$,
 $\sigma_k\neq \sigma_{k\prime}$,
 $\lambda_k\neq \lambda_{k\prime}$ or
 $\nu_k\neq \nu_{k\prime}$. 
Then, we can establish the identifiability of  ordered and initialized irreducible STMoE models 
by applying Lemma 2 of \cite{Jiang_and_tanner_NN_99}, which requires the validation of the following nondegeneracy condition. The set $\{\text{ST}(y; \mu(\bsx;\bsbeta_1), \sigma_1^2, \lambda_1,\nu_1),\ldots,\text{ST}(y; \mu(\bsx;\bsbeta_{4K}), \sigma_{4K}^2, \lambda_{4K},\nu_{4K})\}$ contains $4K$ linearly independent functions of $y$, for any $4K$ distinct quadruplet $(\mu(\bsx;\bsbeta_k), \sigma_k^2, \lambda_k,\nu_k)$ for $k=1,\ldots,4K$. 
Thus,  via Lemma 2 of \cite{Jiang_and_tanner_NN_99} we have any ordered and initialized irreducible STMoE is identifiable.
 
\section{Maximum likelihood estimation of the STMoE model}
\label{sec: MLE for the STMoE}
The unknown parameter vector $\bsvPsi$ of the STMoE model is estimated by maximizing the following observed-data log-likelihood given an observed i.i.d sample of $n$ observations, that is,
 the responses $(y_1,\ldots,y_n)$ and the corresponding predictors $(\bsx_1,\ldots,\bsx_n)$ and $(\bsr_1,\ldots,\bsr_n)$:
\begin{equation}
\log L(\bsvPsi) = \sum_{i=1}^n  \log  \sum_{k=1}^K \pi_k(\bsr_i;\bsalpha) \text{ST}(y; \mu(\bsx_i;\bsbeta_k), \sigma_k^2, \lambda_k,\nu_k)\cdot
\label{eq: log-lik skew-t MoE}
\end{equation}
We perform this iteratively by a dedicated ECM algorithm. 
The  complete data consist of the observations  as well as the latent variables $(u_1,\ldots,u_n)$ and $(w_1,\ldots,w_n)$, and the latent component labels $(z_1,\ldots,z_n)$. Then, from the hierarchical representation of the STMoE (\ref{eq: hierarchical representation skew-t MoE}), the complete-data log-likelihood of $\bsvPsi$ is given by:
{\small \begin{eqnarray}
\log L_c(\bsvPsi)  & = &  \sum_{i=1}^n \sum_{k=1}^K Z_{ik} \Big[ \log\left(\Pro\left(Z_i=k|\bsr_i\right)\right) +   \log\left(f\left(w_i|Z_{ik}=1\right)\right) + \nonumber \\
& & \log\left(f\left(u_i|w_i, Z_{ik}=1\right)\right)+ \log\left(f\left(y_i|u_i,Z_{ik}=1,\bsx_i\right)\right)\Big] \nonumber \\
& = & \log L_{1c}(\bsalpha)  + \sum_{k=1}^K \big[\log L_{2c}(\bstheta_k) + \log L_{3c}(\nu_k) \big]
\label{eq: complete log-likelihood STMoE}
\end{eqnarray}}
where $\bstheta_k = (\bsbeta^T_k,\sigma_k^2,\lambda_k)^T$ and
{\small \begin{eqnarray*}
\log L_{1c}(\bsalpha) &= & \sum_{i=1}^{n} \sum_{k=1}^K Z_{ik} \log \pi_k(\bsr_i;\bsalpha), \\
\log L_{2c}(\bstheta_k)
&=&\sum_{i=1}^{n}  Z_{ik} \Big[ -  \log (2 \pi) - \log (\sigma^2_k) - \frac{1}{2} \log (1 - \delta^2_k) 
 - \frac{w_i ~ d^2_{ik}}{2(1 - \delta^2_k)} + \frac{w_i~ u_i ~ \delta_k ~ d_{ik}}{(1 - \delta^2_k)\sigma_k}  - \frac{ w_i ~ u_i^2}{2(1 - \delta^2_k)\sigma^2_k}  
\Big],\\
\log L_{3c}(\nu_k) &=& \sum_{i=1}^{n}  Z_{ik}  \Big[ - \log \Gamma \left(\frac{\nu_k}{2}\right) + \left(\frac{\nu_k}{2}\right) \log \left(\frac{\nu_k}{2}\right)    +  \left(\frac{\nu_k}{2}\right)  \log (w_i)
 - \left(\frac{\nu_k}{2}\right) w_i\Big].
\label{eq: complete log-lik decomposition skew-t MoE}
\end{eqnarray*}}

\subsection{The ECM algorithm for the STMoE model}
\label{ssec:ECM STMoE}
The ECM algorithm for the STMoE model starts with an initial parameter vector $\bsvPsi^{(0)}$ and alternates between the E- and CM- steps until convergence.
\subsection{E-Step}
The E-Step of the CEM algorithm for the STMoE calculates the $Q$-function, that is the conditional expectation of the complete-data log-likelihood (\ref{eq: complete log-likelihood STMoE}),  given the observed data $\{y_i,\bsx_i,\bsr_i\}_{i=1}^n$ and a current parameter estimation $\bsvPsi^{(m)}$, $m$ being the current iteration.
From (\ref{eq: complete log-likelihood STMoE}), the $Q$-function is given by:
{\small \begin{equation}
Q(\bsvPsi;\bsvPsi^{(m)})=Q_{1}(\bsalpha;\bsvPsi^{(m)})+\sum_{k=1}^K \left[Q_{2}(\bstheta_k,\bsvPsi^{(m)})+ Q_{3}(\nu_k,\bsvPsi^{(m)})\right],
\label{eq: Q-function decomposition STMoE}
\end{equation}}
where
{\small \begin{eqnarray*}
Q_{1}(\bsalpha;\bsvPsi^{(m)}) &\!\! =\!\! & \sum_{i=1}^{n} \sum_{k=1}^K \tau^{(m)}_{ik} \log \pi_k(\bsr_i;\bsalpha), \label{eq: Q_alpha STMoE}\\ 
Q_{2}(\bstheta_k;\bsvPsi^{(m)}) &\!\!=\!\! & \sum_{i=1}^{n} \tau^{(m)}_{ik}\Bigg[ -  \log (2 \pi) - \log (\sigma^2_k) - \frac{1}{2} \log (1 - \delta^2_k)
- \frac{ w^{(m)}_{ik} ~ d^2_{ik}}{2(1 - \delta^2_k)} + \frac{\delta_k ~ d_{ik} ~ e_{1,ik}^{(m)}}{(1 - \delta^2_k)\sigma_k}  - \frac{e_{2,ik}^{(m)}}{2(1 - \delta^2_k)\sigma^2_k} \Bigg],\label{eq: Q_Psik STMoE}\\
Q_{3}(\nu_k;\bsvPsi^{(m)}) &\!\! =\!\! & \sum_{i=1}^{n} \tau^{(m)}_{ik}  \left[- \log \Gamma \left(\frac{\nu_k}{2}\right) + \left(\frac{\nu_k}{2}\right) \log \left(\frac{\nu_k}{2}\right)    
 - \left(\frac{\nu_k}{2}\right) ~ w^{(m)}_{ik} 
  +  \left(\frac{\nu_k}{2}\right) e^{(m)}_{3,ik} \right]\cdot \label{eq: Q_nuk STMoE}
\end{eqnarray*}}It can be seen that computing the $Q$-function requires the following conditional expectations:
\begin{eqnarray*} 
\tau_{ik}^{(m)} &=& \E_{{\bsvPsi^{(m)}}}\left[Z_{ik}|y_i,\bsx_i,\bsr_i\right],\label{eq: E[Zi|yi] definition STMoE}\\
w_{ik}^{(m)} &=&\E_{{\bsvPsi^{(m)}}}\left[W_{i}|y_i, Z_{ik}=1,\bsx_i,\bsr_i\right],\label{eq: E[Wi|yi,Zik] definition STMoE}\\
e_{1,ik}^{(m)} &=& \E_{{\bsvPsi^{(m)}}}\left[W_{i}U_{i}|y_i, Z_{ik}=1,\bsx_i,\bsr_i\right],\label{eq: E[WiUi|yi,Zik] definition STMoE}\\
e_{2,ik}^{(m)} &=& \E_{{\bsvPsi^{(m)}}}\left[W_{i}U^2_{i}|y_i, Z_{ik}=1,\bsx_i,\bsr_i\right],\label{eq: E[WiUi2|yi,Zik] definition STMoE}\\
e_{3,ik}^{(m)} &=& \E_{{\bsvPsi^{(m)}}}\left[\log(W_{i})|y_i, Z_{ik}=1,\bsx_i,\bsr_i\right]\label{eq: E[logWi|yi,Zik] definition STMoE}\cdot
\end{eqnarray*}
Following the expressions of these conditional expectations given namely in the case of the standard skew $t$ mixture  model  \citep{Lin07univSkewtMixture},  the conditional expectations for the case of the proposed STMoE model can be expressed similarly as:
{\small \begin{eqnarray}
\tau_{ik}^{(m)} &=& \frac{\pi_k(\bsr;\bsalpha^{(m)}) ~\text{ST}\Big(y_i;\mu(\bsx_i;\bsbeta_k^{(m)}),\sigma^{2(m)}_k,\lambda^{(m)}_k, \nu^{(m)}_k\Big)}{f(y_i|\bsr_i,\bsx_i;\bsvPsi^{(m)})}, \label{eq: posterior prob STMoE}\\ 
w_{ik}^{(m)} &=& \left(\frac{\nu^{(m)}_k+1}{\nu^{(m)}_k+{d^2_{ik}}^{(m)}}\right)
\times 
\frac{T_{\nu^{(m)}_k+3} \left(M^{(m)}_{ik}\sqrt{\frac{\nu^{(m)}_k+3}{\nu^{(m)}_k + 1}}\right)}
{T_{\nu^{(m)}_k+1} \left(M^{(m)}_{ik}\right)},
\label{eq: E[Wi|yi,Zik] expression STMoE}
\end{eqnarray}}where
$M^{(m)}_{ik} = \lambda^{(m)}_k ~ d^{(m)}_{ik} \sqrt{\frac{\nu^{(m)}_k +1}{\nu^{(m)}_k+{d^2_{ik}}^{(m)}}}$,
{\small \begin{eqnarray}
\!\!\!\! e_{1,ik}^{(m)} &\!\!\!\!  = \!\!\!\!  & \delta_{k}^{(m)} \left(y_i -\mu_k(\bsx_i;\bsbeta^{(m)})\right) w_{ik}^{(m)} 
+ \Bigg[\frac{\sqrt{1 -{\delta^2_{k}}^{(m)}}}{\pi f(y_i|\bsr_i,\bsx_i;\bsvPsi^{(m)})} \left( \frac{{d^2_{ik}}^{(m)}}{\nu^{(m)}_{k} (1 - {\delta^2_k}^{(m)})} +1 \right)^{-(\frac{\nu^{(m)}_{k}}{2}+1)}
\Bigg],
\label{eq: E[WiUi|yi,Zik] expression STMoE}
\end{eqnarray}}
{\small \begin{eqnarray}
\!\!\!\! e_{2,ik}^{(m)} &\!\!\!\!  =\!\!\!\!  & {\delta^2_{k}}^{(m)} \left(y_i -\mu_k(\bsx_i;\bsbeta^{(m)})\right)^2 w_{ik}^{(m)} 
+  \Bigg[\left(1-{\delta^2_{k}}^{(m)}\right) {\sigma^2_k}^{(m)} \nonumber \\
&\!\!\!\! \!\!\!\! &
+ \frac{\delta_{k}^{(m)} \left(y_i -\mu_k(\bsx_i;\bsbeta^{(m)})\right) \sqrt{1 -{\delta^2_{k}}^{(m)}}}{\pi f(y_i|\bsr_i,\bsx_i;\bsvPsi^{(m)})}
\times \left( \frac{{d^2_{ik}}^{(m)}}{\nu^{(m)}_{k} (1 - {\delta^2_k}^{(m)})} +1 \right)^{-(\frac{\nu^{(m)}_{k}}{2}+1)}
\Bigg],
\label{eq: E[WiUi2|yi,Zik] expression STMoE}
\end{eqnarray}}
{\small \begin{eqnarray}
\!\!\!\! e_{3,ik}^{(m)} &\!\!\!\! =\!\!\!\! & w^{(m)}_{ik} - \log\left(\frac{\nu^{(m)}_k + {d^2_{ik}}^{(m)}}{2}\right) - \left(\frac{\nu^{(m)}_k +1}{\nu^{(m)}_k + {d^2_{ik}}^{(m)}}\right) 
+ \psi\left(\frac{\nu^{(m)}_k +1}{2}\right)\nonumber \\
&\!\!\!\! \!\!\!\! & + \frac{\lambda^{(m)}_{k} d_{ik}^{(m)}\left({d^2_{ik}}^{(m)}  - 1\right)}
{\sqrt{\left(\nu^{(m)}_k+ 1\right)\left(\nu^{(m)}_k + {d^2_{ik}}^{(m)}\right)^3}}
\times 
\frac{t_{\nu^{(m)}_k+1} \left(M^{(m)}_{ik}\right)}
{T_{\nu^{(m)}_k+1} \left(M^{(m)}_{ik}\right)}\cdot
\label{eq: E[logWi|yi,Zik] expression STMoE}
\end{eqnarray}}
We note that, for (\ref{eq: E[logWi|yi,Zik] expression STMoE}), we adopted a one-step-late (OSL) approach to compute the conditional expectation $e_{3,ik}^{(m)} $ as described in \citet{LeeAndMcLachlan14-skewtmix}, by setting the integral part in the expression of the corresponding conditional expectation given in \citep{Lin07univSkewtMixture} to zero, rather than using a Monte Carlo approximation.
We also mention that, for the multivariate skew $t$ mixture models, recently \citet{LeeAndMcLachlan15-CFUST} presented a series-based truncation approach, which exploits an exact representation of this conditional expectation and which can also be used in place of (\ref{eq: E[logWi|yi,Zik] expression STMoE}).

\subsection{M-Step}
\label{ssec: M-Step ECM STMoE}
The M-step maximizes the $Q$-function (\ref{eq: Q-function decomposition STMoE})  with respect to $\bsvPsi$ and provides the parameter vector update $\bsvPsi^{(m+1)}$. 
From (\ref{eq: Q-function decomposition STMoE}), it can be seen that the maximization of $Q$ can be performed by separately maximizing $Q_{1}$ with respect to the parameters $\bsalpha$ of the mixing proportions, and for each expert $k$ $(k=1,\ldots,K)$, $Q_{2}$ with respect to $(\bsbeta^T_k,\sigma^2_k)^T$ and $\lambda_k$, and $Q_{3}$ with respect to $\nu_k$. The maximization of $Q_2$ and $Q_3$ is carried out by conditional maximization (CM) steps by updating $(\bsbeta_k,\sigma_k^2)$ and then updating $(\lambda,\nu_k)$ with the given updated parameters.
This leads to the following CM steps.
On the $(m+1)$th iteration of the M-step, the STMoE model parameters are updated as follows.
\paragraph{CM-Step 1}Calculate $\bsalpha^{(m+1)}$ maximizing the function $Q_{1}(\bsalpha;\bsvPsi^{(m)})$:
\begin{equation}
\bsalpha^{(m+1)} =  \arg \max_{\bsalpha} Q_{1}(\bsalpha;\bsvPsi^{(m)}).
\label{eq: arg max_alpha Q-function SNMoE}
\end{equation}
Contrarily to the case of the standard  mixture model and mixture of regression models, this maximization in the case of the proposed STMoE does not exist in closed form. It is performed iteratively by Iteratively Reweighted Least Squares (IRLS).

\paragraph{The Iteratively Reweighted Least Squares (IRLS) algorithm:}
\label{par: IRLS M-Step SNMoE}
The IRLS algorithm is used to maximize $Q_{1}(\bsalpha,\bsvPsi^{(m)})$ with respect to the parameter vector $\bsalpha$ of the softmax function in the M step at each iteration $m$ of the ECM algorithm.  
The IRLS is a Newton-Raphson algorithm, which consists in starting with a vector $\bsalpha^{(0)}$, and, at the $l+1$ iteration, updating the estimation of $\bsalpha$ as follows:
\begin{equation}
\bsalpha^{(l+1)}=\bsalpha^{(l)}-\Big[\frac{\partial^2 Q_{1}(\bsalpha,\bsvPsi^{(m)})}{\partial \bsalpha \partial \bsalpha^T}\Big]^{-1}_{\bsalpha=\bsalpha^{(l)}} \frac{\partial Q_{1}(\bsalpha,\bsvPsi^{(m)})}{\partial \bsalpha}\Big|_{\bsalpha=\bsalpha^{(l)}}
\label{eq: IRLS update}
\end{equation}
where $\frac{\partial^2 Q_{1}(\bsalpha,\bsvPsi^{(m)})}{\partial \bsalpha \partial \bsalpha^T}$ and $\frac{\partial Q_{1}(\bsalpha,\bsvPsi^{(m)})}{\partial \bsalpha}$ are respectively the Hessian matrix and the gradient vector of $Q_{1}(\bsalpha,\bsPsi^{(m)})$. At each IRLS iteration the Hessian and the gradient are evaluated at $\bsalpha = \bsalpha^{(l)}$ and are  computed similarly as in \citet{chamroukhi_et_al_NN2009}. 
The parameter update $\bsalpha^{(m+1)}$  is taken at convergence of the IRLS algorithm (\ref{eq: IRLS update}).  
Then,  for  $k=1\ldots,K$,
\paragraph{CM-Step 2} Calculate $(\bsbeta_k^{T(m+1)},{\sigma^2_k}^{(m+1)})^T$ by maximizing $Q_{2}(\bstheta_k;\bsvPsi^{(m)})$ w.r.t $(\bsbeta^T_k,\sigma^2_k)^T$. 
For the  skew $t$ mixture of linear experts (STMoLE) case, where the expert means are linear regressors, that is, of the form  (\ref{eq: linear regression mean}), this maximization can be performed in a closed form and provides the following updates:
{\small \begin{eqnarray} 
\!\!\!\! \bsbeta_k^{(m+1)}  &\!\! =\!\! & \Big[\sum_{i=1}^{n}\tau^{(q)}_{ik} w_{ik}^{(m)} \bsx_i\bsx^T_i \Big]^{-1} \sum_{i=1}^{n} 
\tau^{(q)}_{ik}\left(w_{ik}^{(m)}y_i -  \bse^{(m)}_{1,ik} \delta_{k}^{(m+1)} \right)\bsx_i,
\label{eq: beta_k update for STMoE} \\
\!\!\!\!  {\sigma^2_{k}}^{(m+1)} &\!\! =\!\! &
\frac{\sum_{i=1}^n\tau_{ik}^{(m)} \Big[w_{ik}^{(m)}\left(\bsy_i - {\bsbeta^T_{k}}^{(m+1)}\bsx_i\right)^2
 - 2 \delta_{k}^{(m+1)} \bse^{(m)}_{1,ik} (y_i - {\bsbeta^T_{k}}^{(m+1)}\bsx_i) 
+\bse^{(m)}_{2,ik}\Big]}
{2 \left( 1 - {\delta^2_{k}}^{(m)} \right)\sum_{i=1}^n\tau_{ik}^{(m)}}\cdot
\label{eq: sigma2k update STMoE}
\end{eqnarray}}
\paragraph{CM-Step 3} The skewness parameters $\lambda_k$ are updated by maximizing $Q_{2}(\bstheta_k;\bsvPsi^{(m)})$ w.r.t $\lambda_k$, with $\bsbeta_k$ and $\sigma^2_k$ fixed at  the update $\bsbeta_k^{(m+1)}$ and ${\sigma^2_{k}}^{(m+1)}$, respectively.
It can be easily shown that the maximization  to obtain $\delta_{k}^{(m+1)}$ $(k=1,\ldots,K)$ consists in solving the following equation in $\delta_k$:
{\small \begin{eqnarray} 
\delta_{k} (1-\delta^2_{k}) \sum_{i=1}^n\tau_{ik}^{(m)} + (1+\delta^2_{k}) \sum_{i=1}^n\tau_{ik}^{(m)} \frac{d^{(m+1)}_{ik} e^{(m)}_{1,ik}}{\sigma_{k}^{(m+1)}} -  \delta_{k} \sum_{i=1}^n\tau_{ik}^{(m)} \Big[w_{ik}^{(m)}  {d^2_{ik}}^{(m+1)}+ \frac{e^{(m)}_{2,ik}}{{\sigma^2_{k}}^{(m+1)}}  \Big]=0 \cdot
\label{eq: deltak update STMoE}
\end{eqnarray}}

\paragraph{CM-Step 4} Similarly, the degrees of freedom $\nu_k$ are updated by maximizing $Q_{3}(\nu_k;\bsvPsi^{(m)})$ w.r.t $\nu_k$ with $\bsbeta_k$ and $\sigma^2_k$ fixed at  $\bsbeta_k^{(m+1)}$ and ${\sigma^2_{k}}^{(m+1)}$, respectively.
An update $\nu^{(m+1)}_k$ is calculated as solution of the following equation in $\nu_k$:
\begin{equation}
- \psi\left(\frac{\nu_k}{2}\right) + \log\left(\frac{\nu_k}{2}\right) +1 + 
\frac{\sum_{i=1}^n\tau_{ik}^{(m)} \left(\bse^{(m)}_{3,ik} - w^{(m)}_{ik}\right)}{\sum_{i=1}^n\tau_{ik}^{(m)}} = 0.
\label{eq: nu update STMoE}
\end{equation}
The two scalar non-linear equations (\ref{eq: deltak update STMoE}) and (\ref{eq: nu update STMoE})  can be solved  similarly as in the TMoE model, that is with a root finding algorithm, such as Brent's method \citep{Brent1973}.

As mentioned before, one can see that, when the robustness parameter $\nu_k \rightarrow \infty$ for all the components, the parameter updates for the  STMoE model correspond  to those of the SNMoE model (see \cite{Chamroukhi-NNMoE-2015}). 
On the other hand,  when the skewness parameters $\lambda_k = 0$, the STMoE parameter updates correspond  to those of the TMoE model (\cite{Chamroukhi-NNMoE-2015}). Finally,  when  both the degrees of freedom $\nu_k \rightarrow \infty$ and  the skewness $\lambda_k = 0$, we obtain the parameter updates of the standard NMoE model. The STMoE therefore provides a more general  framework for inferring flexible MoE models
and  attempts to simultaneously accommodate data with asymmetric distribution heavy tails and outliers.

Here the ECM algorithm is used to infer the STMoE model parameters.
We note that there is a good generalization of the EM algorithms for non-probabilistic problems which does not require to construct a complete-data likelihood nor  a probabilistic interpretation of the maximization problem, that is the Minorization-Maximization (MM) algorithm \citep{MM-tutorial}. The MM algorihm, used in the MoE framewrok namely by \citet{Gormley2008MoE-Rank,Nguyen2016-MoLE} can also be a good alternative to the ECM algorithm used here. 
 On the other hand, the ECM algorithm divides the space of model-parameters to perform sequentially the optimization in each sub-space. It may also be convenient to divide the space of the hidden variables and to alternate the optimisation, cyclically within each sub-space. This scheme is known as the Alternating ECM (AECM) algorithm \citep{AECMalgorithm}.

\section{Prediction using the STMoE}
\label{sec: Prediction using the STMoE}
The goal in  regression is to be able to make predictions for the response variable(s) given some new value of the predictor variable(s) on the basis of a model trained on a set of training data. 
In regression analysis using mixture of experts, the aim is therefore to predict the response $y$ given new values of the predictors $(\bsx,\bsr)$, on the basis of a MoE model characterized by a parameter vector $\hat \bsvPsi$ inferred from a set of training data, here, by maximum likelihood via EM. 
These predictions can be expressed in terms of the predictive distribution of $y$, which is obtained by substituting the maximum likelihood parameter $\hat\bsvPsi$ into (\ref{eq: mixture of regressions})-(\ref{eq: multinomial logistic}) to give:
\begin{equation*}
f(y|\bsx,\bsr;\hat \bsvPsi) = \sum_{k=1}^K \pi_{k}(\bsr;\hat \bsalpha)  f_k(y|\bsx; \hat \bsvPsi_k).
\label{eq: predictive MoE}
\end{equation*}Using $f$, we might then predict $y$ for a given set of $\bsx$'s and $\bsr$'s as the expected value under $f$, 
that is by calculating the prediction $\hat y = \E_{{\it \hat\bsvPsi}}(Y|\bsr,\bsx)$. We thus need to compute the expectation of the mixture of experts model.
It is easy to show (see for example Section 1.2.4 in \citet{sylvia_fruhwirth_book_2006}) that the mean and the variance of a mixture of experts distribution of the form (\ref{eq: predictive MoE}) are respectively given by
{\small
\begin{eqnarray}
\!\!\!\!\! \E_{{\it \hat\bsvPsi}}(Y|\bsr,\bsx) &\! =\! & \sum_{k=1}^K \pi_k(\bsr;\hat \bsalpha_{n}) \E_{{\it \hat\bsvPsi}}(Y|Z=k,\bsx),
\label{eq: mean of MoE}\\ 
\!\!\!\!\! \V_{{\it \hat\bsvPsi}}(Y|\bsr,\bsx) &\! =\! & \sum_{k=1}^K \pi_k(\bsr;\hat \bsalpha_{n})
\big[\left(\E_{{\it \hat\bsvPsi}}(Y|Z=k,\bsx)\right)^2  + \V_{{\it \hat\bsvPsi}}(Y|Z=k,\bsx) \big] - \big[\E_{{\it \hat\bsvPsi}}(Y|\bsr,\bsx)\big]^2,
\label{eq: variance of MoE}
\end{eqnarray}}where $\E_{{\it \hat\bsvPsi}}(Y|Z=k,\bsx)$ and $\V_{{\it \hat\bsvPsi}}(Y|Z=k,\bsx)$  are respectively the component-specific (expert) means and variances. 
 The mean and the variance for the two MoE models described here are given as follows.

\paragraph{NMoE}For the NMoE model, the normal expert means and variances are respectively given by
$\E_{{\it \hat\bsvPsi}}(Y|Z=k,\bsx) = \hat \bsbeta^T_{k} \bsx$
and 
$\V_{{\it \hat\bsvPsi}}(Y|Z=k,\bsx) = \hat \sigma^2_{k}$. Then, from (\ref{eq: mean of MoE}) it follows that the mean of the NMoE is given by
\begin{eqnarray}\E_{{\it \hat\bsvPsi}}(Y|\bsr,\bsx) &=& \sum_{k=1}^K \pi_k(\bsr;\hat \bsalpha_{n}) \hat \bsbeta^T_{k} \bsx.
\label{eq: mean of NMoE}
\end{eqnarray}

\paragraph{STMoE}The mean and the variance for a skew $t$ random variable, for this scalar case, can be easily computed as in Section 4.2 in \citet{AzzaliniAndCapitanio2003} for a non-zero location parameter. Thus, for the STMoE model, the expert means for $\hat \nu_{k}>1$, are given by 
$$\E_{{\it \hat\bsvPsi}}(Y|Z=k,\bsx) = \hat \bsbeta^T_{k} \bsx + \hat \sigma_{k} ~ \hat \delta_{k} ~ \xi(\hat \nu_{k})$$
and the expert variances for $\hat \nu_{k}>2$ are given by 
$$\V_{{\it \hat\bsvPsi}}(Y|Z=k,\bsx) = \left(\frac{\hat \nu_{k}}{\hat \nu_{k} - 2}  - \hat \delta^2_{k} ~ \xi^2(\hat \nu_{k})\right) \hat \sigma^2_{k},$$
where $\xi(\hat \nu_{k}) = \sqrt{\frac{\hat \nu_{k}}{\pi}} \frac{\Gamma\left(\frac{\hat \nu_{k}}{2} - \frac{1}{2}\right)}{\Gamma \left(\frac{\hat \nu_{k}}{2}\right)}$. Then, following (\ref{eq: mean of MoE}), the mean of the proposed STMoE is thus given by:
\begin{eqnarray}
\E_{{\it \hat\bsvPsi}}(Y|\bsr,\bsx) &=& \sum_{k=1}^K \pi_k(\bsr;\hat \bsalpha)\Big(\hat \bsbeta^T_{k} \bsx + \hat \sigma_{k} ~ \hat \delta_{k} ~ \xi(\hat \nu_{k})\Big).
\label{eq: mean of STMoE}
\end{eqnarray}

Finally, the variance for each MoE model is obtained by using (\ref{eq: variance of MoE}) with the specified corresponding  means and variances calculated in the above.

\section{Model-based clustering using the STMoE} 
\label{sec: MBC using the STMoE}

The MoE models can also be used for a model-based clustering perspective  to provide a partition of the regression data into $K$ clusters. 
Model-based clustering using the proposed STMoE consists in assuming that the observed data $\{\bsx_i,\bsr_i,y_i\}_{i=1}^n$ are  generated from a $K$ component mixture of skew $t$ experts, with parameter vector $\bsvPsi$ where the STMoE components are  interpreted as clusters and hence  associated to clusters. 
The problem of clustering therefore becomes the one of estimating the MoE parameters $\bsvPsi$, which is performed here by using the dedicated ECM algorithm presented in section \ref{ssec:ECM STMoE}. 
Once the parameters are estimated, the provided posterior component memberships $\tau_{ik}$ given by (\ref{eq: posterior prob STMoE}) represent a fuzzy partition of the data. 
A hard partition of the data can then be obtained from the posterior memberships by applying the MAP rule, that is: 
\begin{eqnarray}
\hat{z}_i = \arg \max_{k=1}^K \hat \tau_{ik}
\label{eq: MAP rule for clustering}
\end{eqnarray}where $\hat{z}_i$ represents the estimated cluster label for the $i$th individual.

\section{Model selection for the STMoE}
\label{sec: Model selection for the STMoE}
One of the issues in mixture model-based clustering is model selection.  
The problem of model selection for the STMoE models presented  here in its general form is equivalent to the one of choosing the optimal number of experts $K$, the degree $p$ of the polynomial regression and the degree $q$ for the logistic regression. 
The optimal value  of the triplet $(K,p,q)$ can be computed by using some model
selection criteria such as the Akaike Information Criterion (AIC) \citep{AIC}, the Bayesian Information Criterion (BIC) \citep{BIC} or the Integrated Classification Likelihood criterion (ICL)  \citep{ICL}, etc. 
The AIC and BIC are are penalized observed log-likelihood criteria which can be defined as functions to be maximized and are respectively given by:
\begin{eqnarray*}
\mbox{AIC}(K,p,q) &=& \log L(\hat{\bsPsi}) - \eta_{\bsvPsi},\\
\mbox{BIC}(K,p,q) &=&  \log L(\hat{\bsvPsi}) - \frac{\eta_{\bsvPsi} \log(n)}{2}.
\end{eqnarray*}The ICL criterion consists in a penalized complete-data log-likelihood and can be expressed as follows:
\begin{equation*}
\ICL(K,p,q) = \log L_c(\hat{\bsvPsi}) - \frac{\eta_{\bsvPsi} \log(n)}{2}.
\end{equation*}In the above, $\log L(\hat{\bsvPsi})$ and  $\log L_c(\hat{\bsvPsi})$ are respectively the incomplete (observed) data log-likelihood and the complete data log-likelihood, obtained at convergence of the ECM algorithm for the corresponding MoE model and $\eta_{\bsvPsi}$ is the number of free model parameters. The number of free parameters $\eta_{\bsvPsi} $ is given by $\eta_{\bsvPsi} = K(p+q+3)-q-1$ for the NMoE model and $\eta_{\bsvPsi} = K(p+q+5)-q-1$ for the proposed STMoE model. Indeed, for each component, the STMoE have two additional parameters to be estimated, which are the robustness and the skewness parameters. 
 
However, note that in MoE it is common to use mixing proportions modeled as logistic transformation of linear functions of the covariates, that is the covariate vector  in (\ref{eq: multinomial logistic}) is given by $\bsr_i = (1, r_i)^T$ (corresponding to $q=2$), $r_i$ being an univariate covariate variable. This is also adopted in this work.
Moreover, for the case of linear experts, that is when the experts are linear regressors with parameter vector $\bsbeta_k$ for which the corresponding covariate vector $\bsx_i$ in  (\ref{eq: linear regression mean})  is given by $\bsx_i = (1, x_i)^T$ (corresponding to $p=2$), $r_i$ being an univariate covariate variable, the model selection reduces to choosing the number of experts $K$.
Here we mainly consider this linear case. However, for a general use of the proposed STMoE model, even though the model selection criteria such as AIC, BIC, ICL can be easily computed, the direct model selection is difficult due to the large model space dimension $\nu_{\bf \Psi}$. A searching strategy is then required to optimise the way of exploring the model space.

\section{Experimental study}  
\label{sec: Experimental study}

This section is dedicated to the evaluation of the proposed approach on simulated data and real-world data . 
We evaluated the performance of proposed ECM algorithm\footnote{The codes have been implemented in Matlab and are available upon request from the author.} for the STMoE model in terms of modeling, robustness to outliers and clustering.

\subsection{Initialization and stopping rules}
\label{ssec: initialization and stopping}
The parameters $\bsalpha_k$ ($k=1, \ldots, K-1$) of the mixing proportions are initialized randomly, 
including an initialization at the null vector for one run (corresponding to equal mixing proportions).
Then, the common parameters $(\bsbeta_k,\sigma^2_k)$ ($k=1,\ldots, K$) are initialized from a random partition of the data into $K$ clusters. This corresponds to fitting a normal mixture of experts where the initial values of the parameters are respectively  given by  
(\ref{eq: beta_k update for NMoE}) and (\ref{eq: sigma2k update NMoE}) with the posterior memberships $\tau_{ik}$ replaced by the hard assignments $Z_{ik}$ issued from the random partition.
For the STMoE model, the robustness parameters $\nu_k$ ($k=1,\ldots, K$) is initialized randomly in the range [1, 200] and the skewness parameters $\lambda_k$ ($k=1,\ldots, K$) is initialized by randomly  initializing the parameter $\delta_k$ in  $(-1,1)$ from the relation $\lambda_k = \frac{\delta_k}{\sqrt{1 -\delta^2_k}}$.
Then, the proposed ECM algorithm for each model is stopped when the relative variation of the observed-data log-likelihood 
$\frac{\log L(\bsvPsi^{(m+1)})- \log L(\bsvPsi^{(m)})}{|\log L(\bsvPsi^{(m)})|}$ reaches a prefixed threshold (for example $\epsilon=10^{-6}$).  
For each model, this process is repeated 10 times and the solution corresponding the highest log-likelihood is finally selected.

\subsection{Experiments on simulation data sets}
In this section we perform an experimental study on simulated data sets to apply and assess the proposed model. 
Two sets of experiments have been performed.  
The first experiment aims at observing the effect of the sample size on the estimation quality and the second one aims at observing the impact of the presence of outliers in the data on the estimation quality, that is the robustness of the models.

\subsubsection{Experiment 1}
For this first experiment on simulated data, each simulated sample consisted of $n$ observations with increasing values of the sample size $n: 50, 100, 200, 500, 1000$.  
The simulated data are generated from a two component mixture of linear experts, that is $K=2, p=q=1$.
The covariate variables $(\bsx_i, \bsr_i)$ are simulated such that $\bsx_i = \bsr_i = (1,x_i)^T$ where $x_i$ is
simulated uniformly over the interval $(-1, 1)$. 
We consider each of the two models for data generation (NMoE and STMoE), that is, given the covariates, the response $y_i|\{\bsx_i,\bsr_i;\bsvPsi\}$ is simulated according to the generative process of the models (\ref{eq: normal MoE}) and  (\ref{eq: hierarchical representation skew-t MoE}). 
For each generated sample, we fit each of the two models.   
Thus, the results are reported for the two models with data generated from each of them. We consider the mean square error (MSE) between each component of the true parameter vector and the estimated one, which is given by $\parallel \bsvPsi_j - \hat{\bsvPsi}_j\parallel^2$.
The squared errors are averaged on 100 trials. 
The used simulation parameters $\bsvPsi$ for each model are given in Table \ref{tab: simulation parameters situation 1}.
\begin{table}[H]
\centering
\small
\begin{tabular}{l | l l l l l}
\hline
& \multicolumn{5}{c}{parameters} \\
\hline
\hline
component 1 & $\bsalpha_1=(0, 10)^T$&$\bsbeta_1=(0,1)^T$ & $\sigma_1=0.1$ &$\lambda_1 = 3$   & $\nu_1 = 5$\\
component 2 & $\bsalpha_2=(0, 0)^T$&$\bsbeta_2=(0,-1)^T$  &$\sigma_2=0.1$ &$\lambda_2 = -10$ & $\nu_2 = 7$\\
 \hline
\end{tabular}
\caption{Parameter values used in simulation.}
\label{tab: simulation parameters situation 1}
\end{table}

\subsubsection{Obtained results}
 
 Table \ref{tab. MSE for the parameters: STMoE->STMoE} shows the obtained results in terms of the MSE for  the STMoE. 
 One can observe that, for the  proposed model, the parameter estimation error is decreasing as $n$ increases, which is related the convergence property of the maximum likelihood estimator.   One can also observe that the error decreases significantly  for $n\geq 500$, especially for the regression coefficients and the scale parameters.
{\setlength{\tabcolsep}{4pt
\begin{table}[H]
\centering
{\footnotesize
\begin{tabular}{c c c  c c c c c c c c c c}
\hline
param. & $\alpha_{10}$ & $\alpha_{11}$ & $\beta_{10}$ & $\beta_{11}$ & $\beta_{20}$ & $\beta_{21}$ & $\sigma_{1}$& $\sigma_{2}$ & $\lambda_{1}$ & $\lambda_{2}$ & $\nu_{1}$ & $\nu_{2}$ \\ 
$n$		& & & & & & & & & & & & \\
 \hline
 \hline
$50$    & 525 & 5737 & 0.965 & 2.440 & 4.388 & 0.667 & 0.954 & 0.608 & 3115 & 16095 & 15096 & 4643\\
$100$  & 457 & 1815 & 0.847 & 0.852 & 0.742 & 0.660 & 0.844 & 0.303 & 2013 & 7844 & 5360 & 263\\
$200$  & 247 & 785 & 0.816 & 0.348 & 0.473 & 0.556 & 0.362 & 0.297 & 700 & 3847 & 3135 & 167\\ 
$500$  & 31  & 565 & 0.363 & 0.091 & 0.314 & 0.398 & 0.091 & 0.061 & 7.8 & 1078 & 223 & 8.6\\ 
$1000$& 8.5    & 68 & 0.261 & 0.076 & 0.233 & 0.116 & 0.026 & 0.002 & 2.8 & 554 & 49.4 & 0.79\\ 
\hline
\end{tabular}
}
\caption{\label{tab. MSE for the parameters: STMoE->STMoE} MSE $\times 10^3$ between each component of the estimated parameter vector of the STMoE model and the actual one for a varying sample size $n$.}
\end{table}
} 
In addition to the previously showed results, we plotted in Figures 
\ref{fig. TwoClust-NMoE_NMoE} and 
\ref{fig. TwoClust-NMoE_STMoE} the estimated quantities provided by applying respectively the NMoE model and the proposed STMoE model, and their true counterparts for $n=500$ for the same the data set which was generated according the NMoE model. 
The upper-left plots show the estimated mean function, the estimated expert component mean functions, and the corresponding true ones.
The upper-right plots show the estimated mean function with the estimated confidence region computed as plus and minus twice the estimated (pointwise) standard deviation of the model as presented in Section \ref{sec: Prediction using the STMoE}, and their true counterparts. 
The bottom-left plots show the true expert component mean functions and the true partition, and the bottom-right plots show their estimated counterparts.

One can clearly see that the estimations provided by  the proposed model are very close to the true ones which  correspond to those of the NMoE model in this case. 
This shows  that the proposed algorithm performs well and 
provides an additional support to the fact that the corresponding proposed STMoE model is good generalization of the normal mixture of experts (NMoE), as it clearly approaches the NMoE as shown in this simulated examples.
\begin{figure}[!h]
   \centering 
   \begin{tabular}{cc}
   \includegraphics[width=6.2cm]{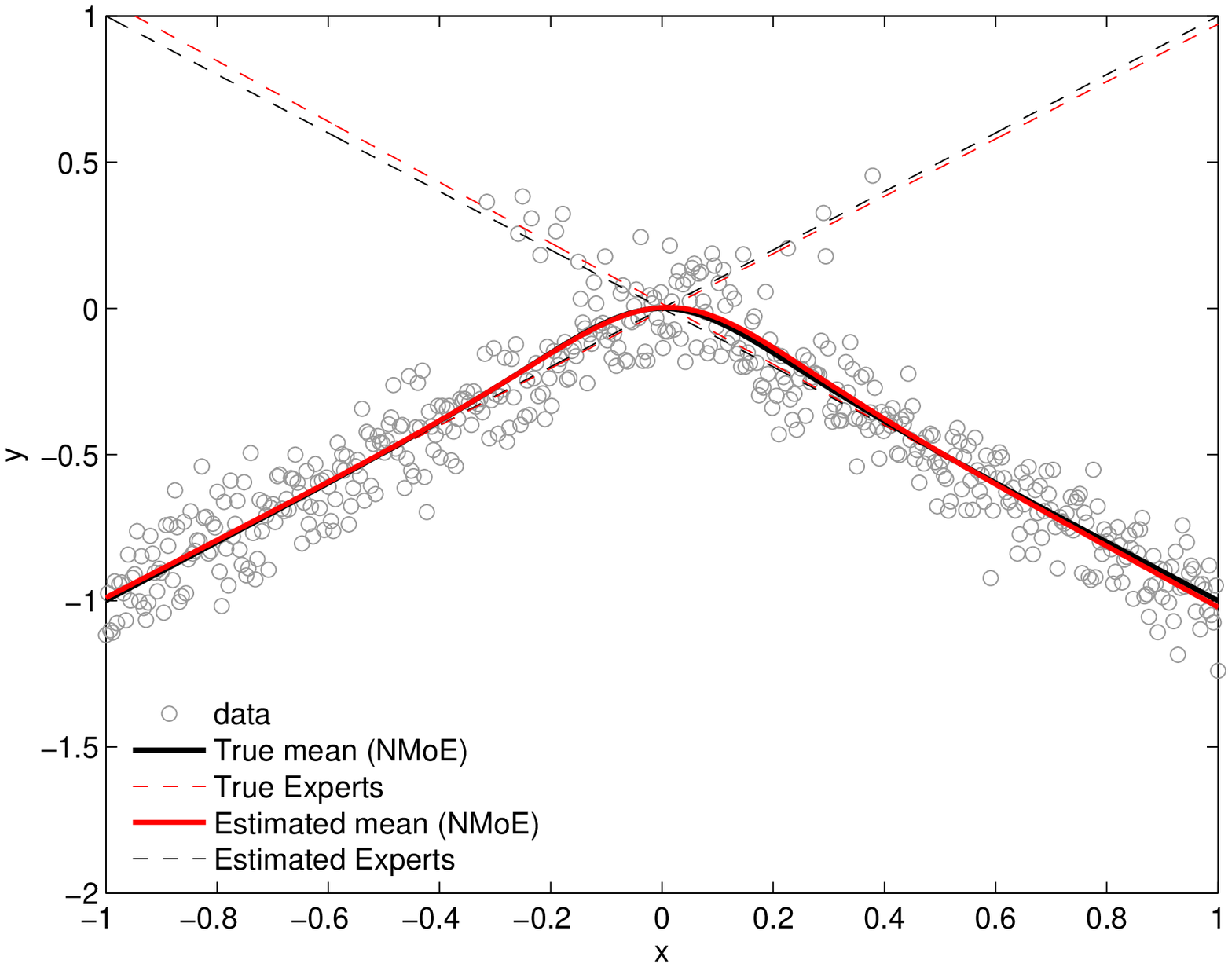}&  
   \includegraphics[width=6.2cm]{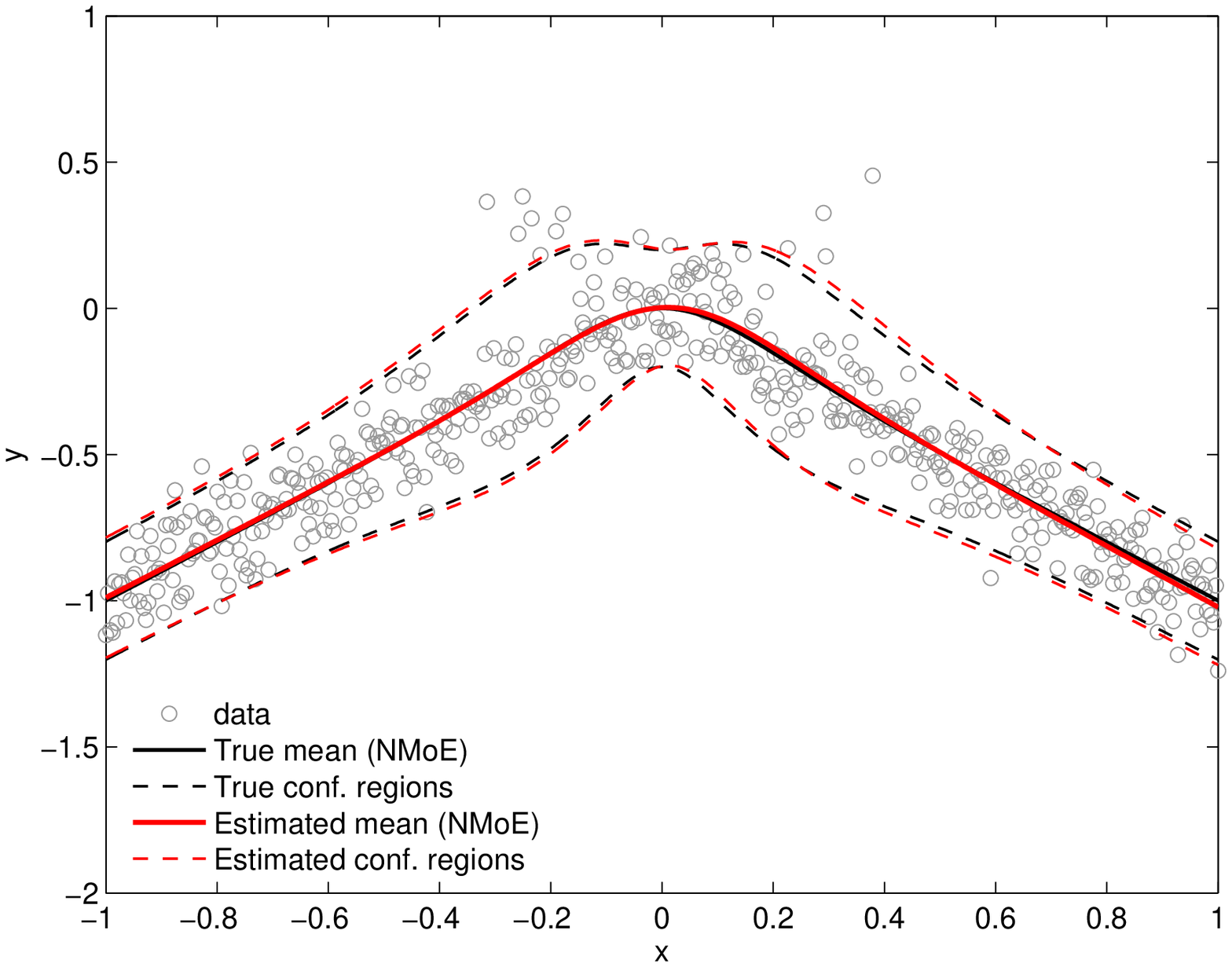}\\
\hspace{0.2cm}\includegraphics[width=6cm]{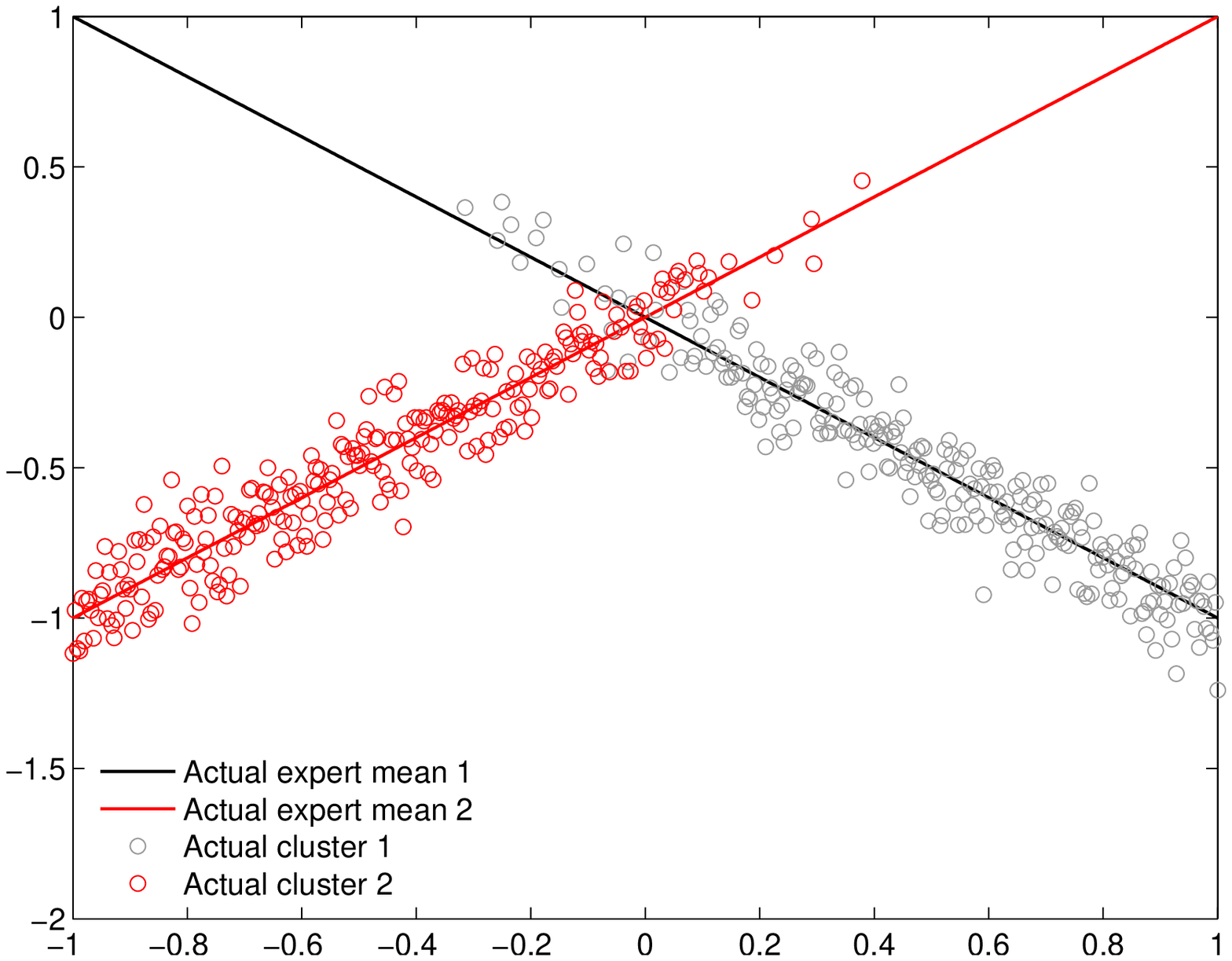}&
\hspace{0.2cm}\includegraphics[width=6cm]{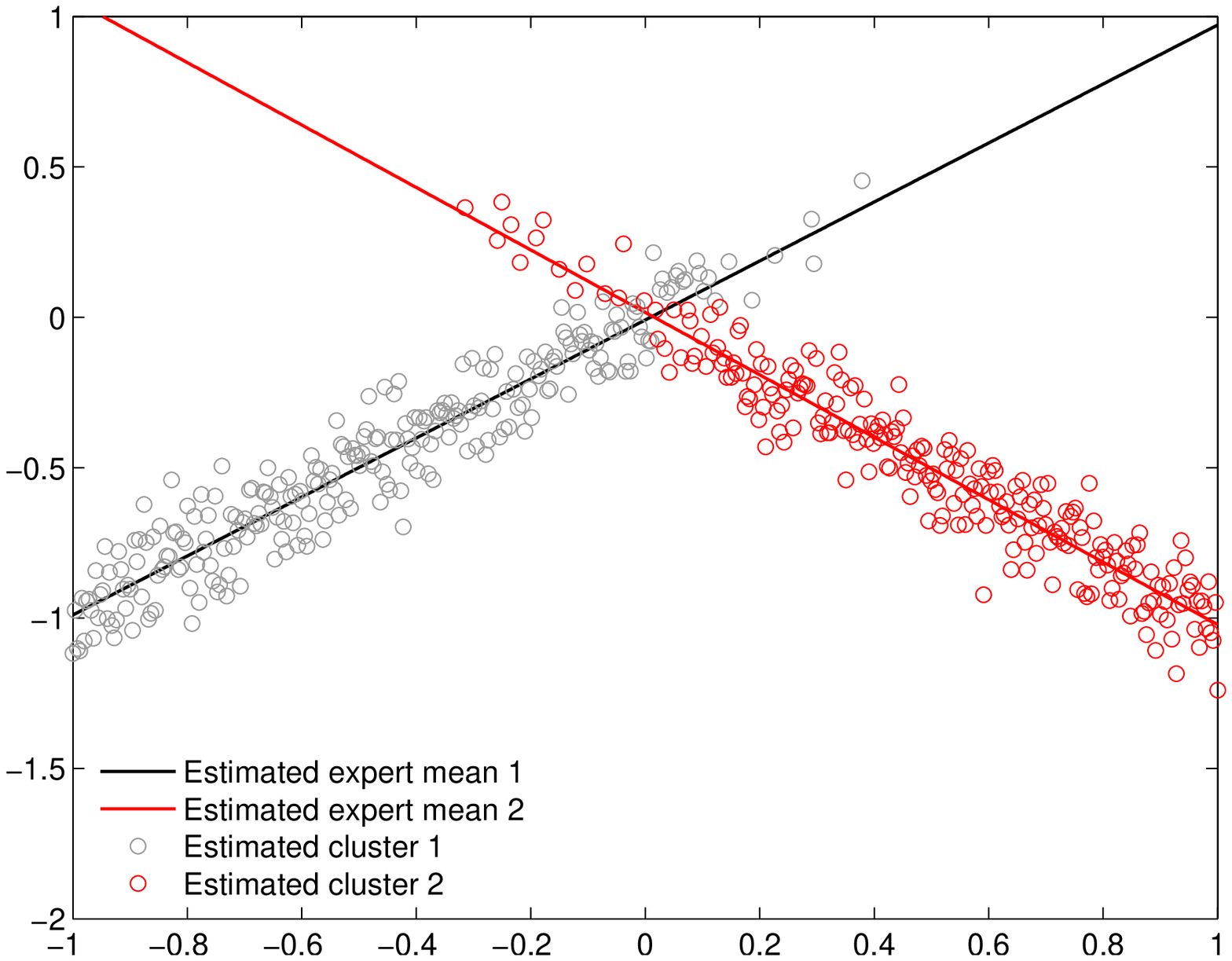}
   \end{tabular}
      \caption{\label{fig. TwoClust-NMoE_NMoE}Fitted NMoE model to a data set generated according to the NMoE model.}
\end{figure}
\begin{figure}[!h]
   \centering  
   \begin{tabular}{cc}
   \includegraphics[width=6.2cm]{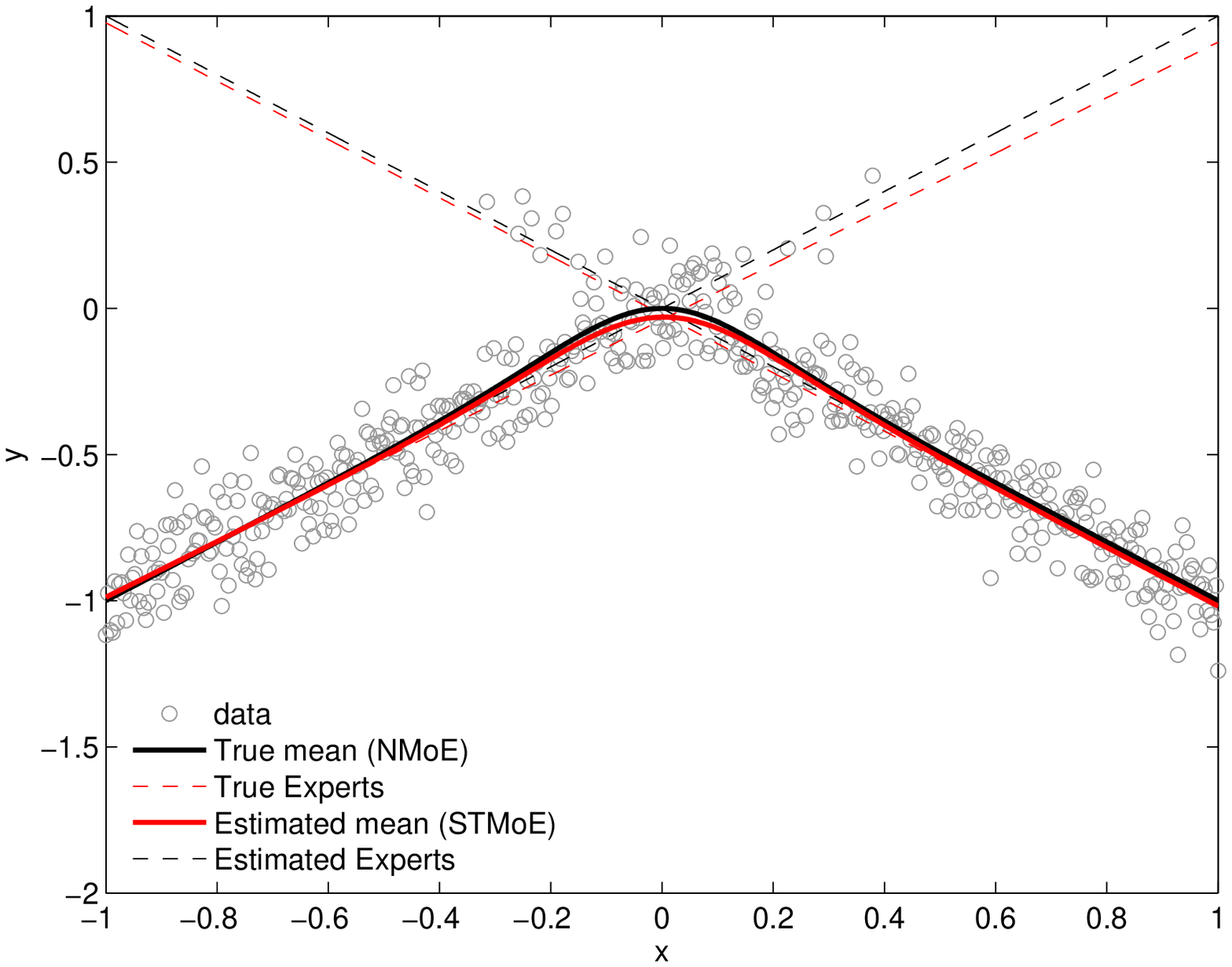}&  
   \includegraphics[width=6.2cm]{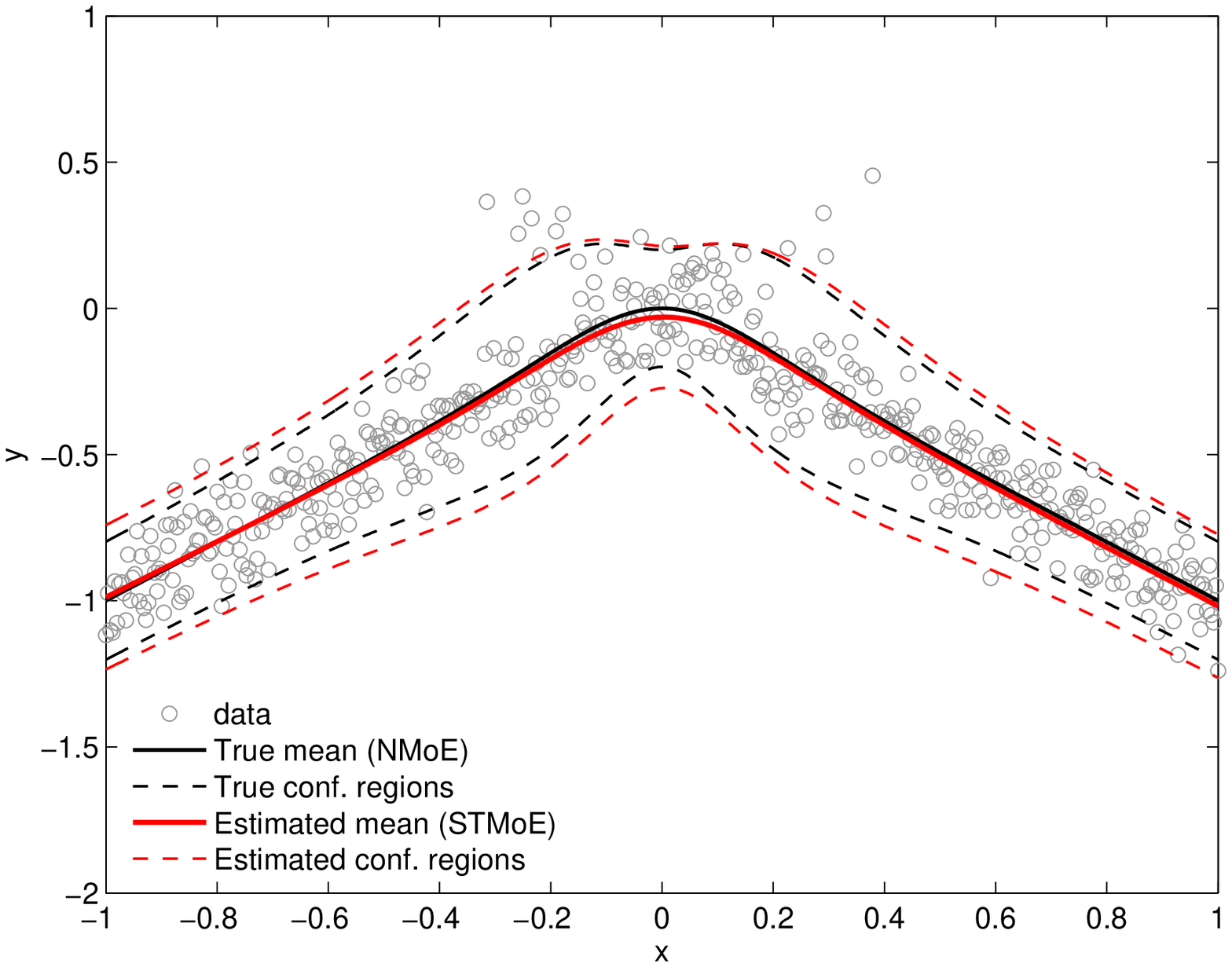}\\
  \hspace{0.2cm}\includegraphics[width=6cm]{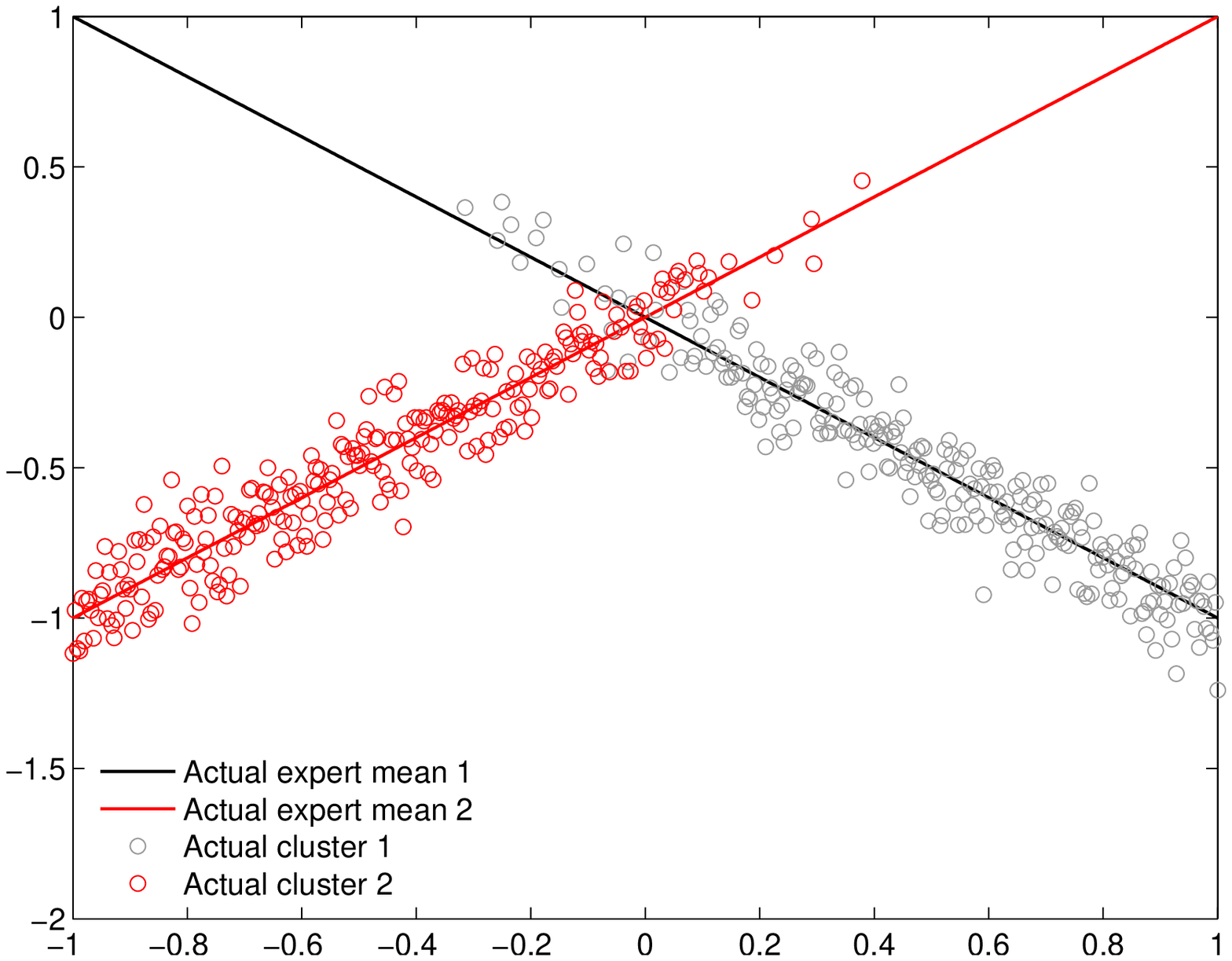}&
   \hspace{0.2cm}\includegraphics[width=6cm]{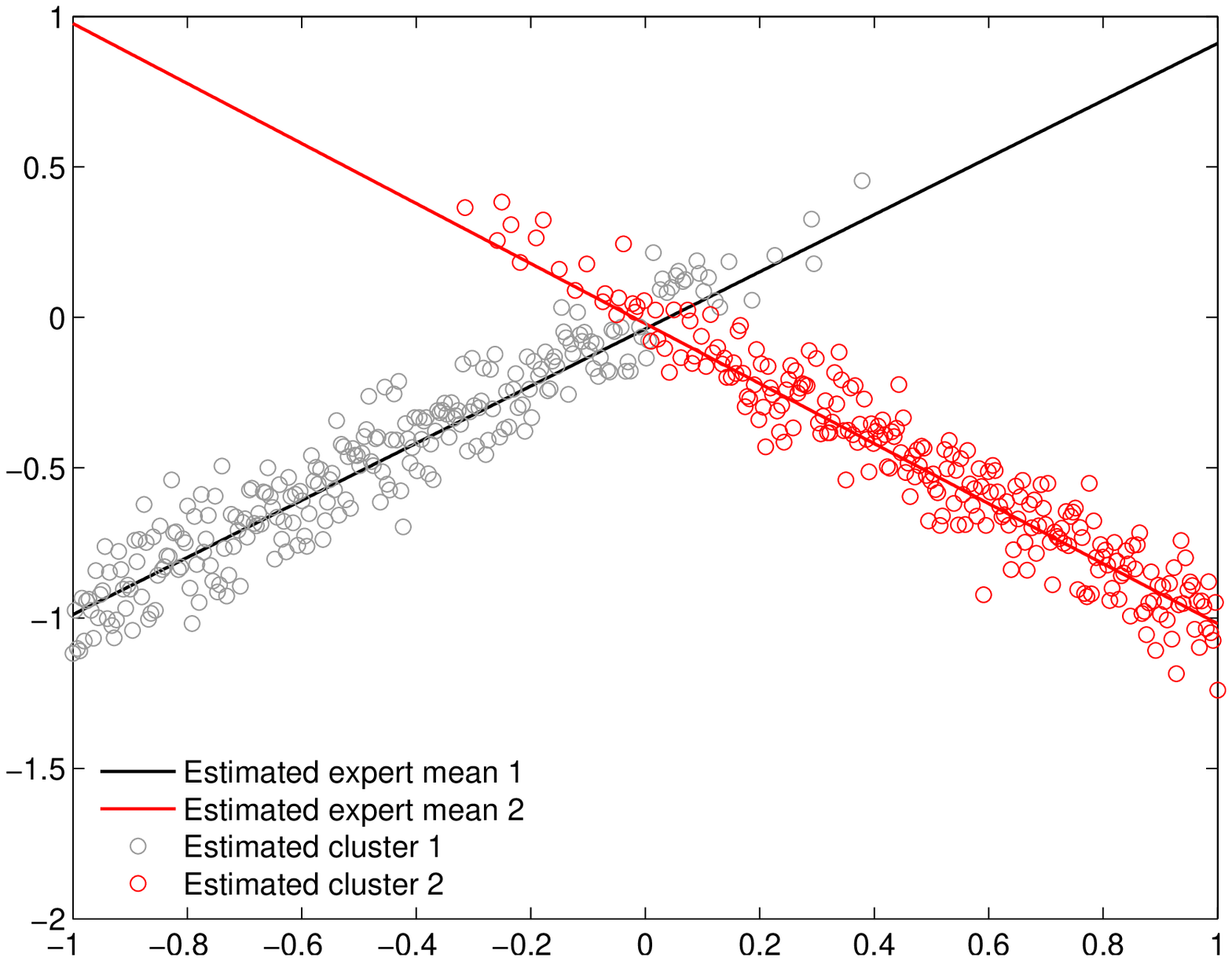}
      \end{tabular}
      \caption{\label{fig. TwoClust-NMoE_STMoE}Fitted STMoE model to a data set generated according to the NMoE model.}
\end{figure}

Figure \ref{fig. TwoClust-All->All-estimated models} shows the true and estimated MoE mean functions and component mean functions by fitting the proposed STMoE model to a simulated data set of $n=500$ observations. Each model was considered for data generation. The upper plot corresponds to the NMoE model and the  bottom plot corresponds to the STMoE model.
Finally, Figure \ref{fig. TwoClust-All->All-estimated partitions} shows the corresponding true and estimated partitions.
Again, one can clearly see that both the estimated models are precise. The fitted functions are close to the true ones.
In addition, one can also see that the partitions estimated by the STMoE model are close the actual partitions.
The proposed STMoE model can therefore be used as alternative to the NMoE model for both regression and model-based clustering. 
\begin{figure}[!h]
   \centering  
   \begin{tabular}{cc}
   \includegraphics[width=6.2cm]{TwoClust-NMoE_NMoE_Means.eps}&  
  \includegraphics[width=6.2cm]{TwoClust-NMoE_NMoE_Means_ConfidRegions.eps}\\
   \includegraphics[width=6.2cm]{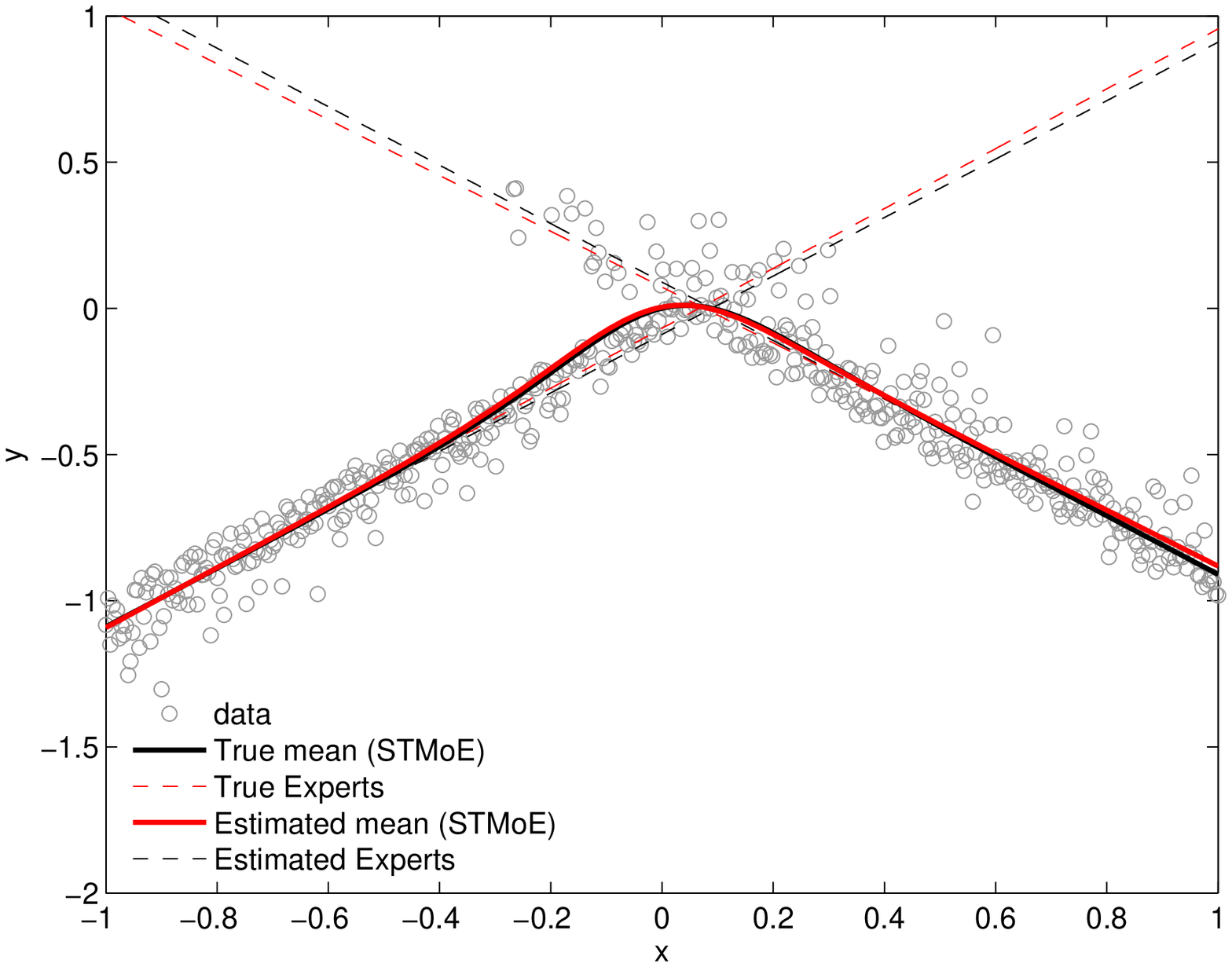}&  
   \includegraphics[width=6.2cm]{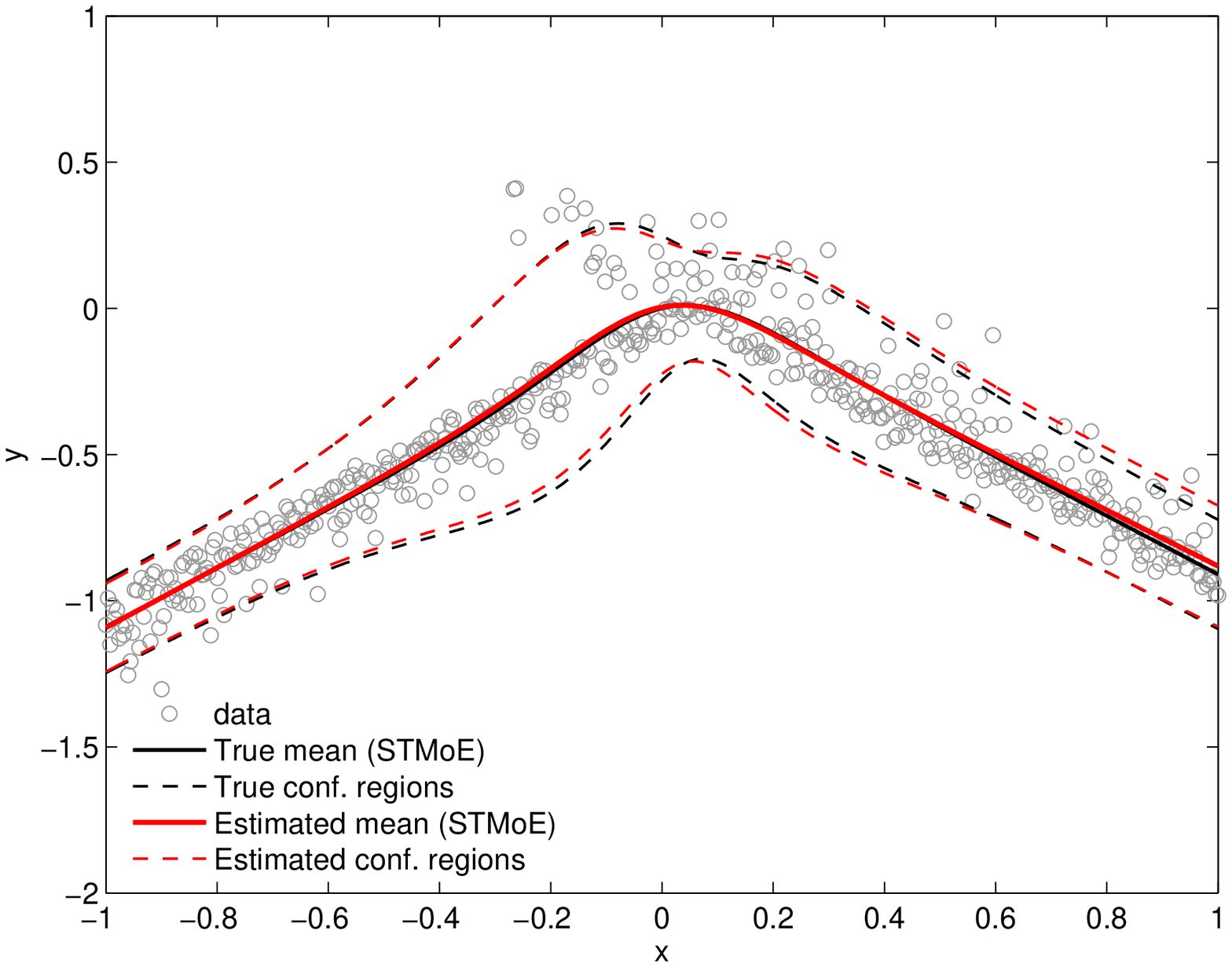}
\end{tabular}
   \caption{\label{fig. TwoClust-All->All-estimated models}The true and estimated mean function and expert mean functions by fitting the standard NMoE model (up) and the proposed STMoE model (bottom)  to a simulated data set of $n=500$ observations generating according to the corresponding model.}
\end{figure}
\begin{figure}[!h]
   \centering 
\begin{tabular}{cc}
  \hspace{0.2cm}\includegraphics[width=6cm]{TwoClust-NMoE_NMoE_TruePartition.eps}&
  \hspace{0.2cm}\includegraphics[width=6cm]{TwoClust-NMoE_NMoE_EstimatedPartition.eps}\\
  \hspace{0.2cm}\includegraphics[width=6cm]{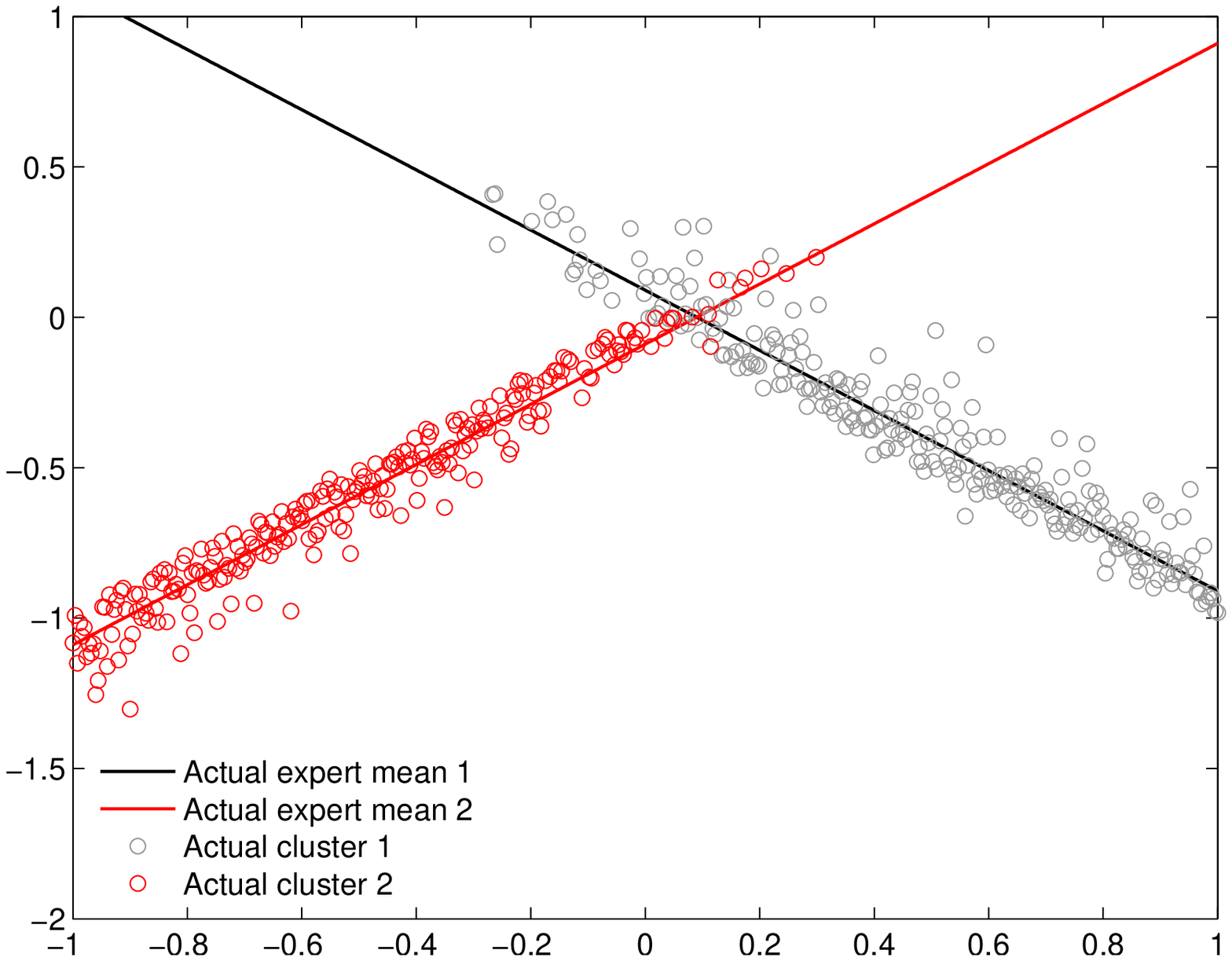}&
  \hspace{0.2cm}\includegraphics[width=6cm]{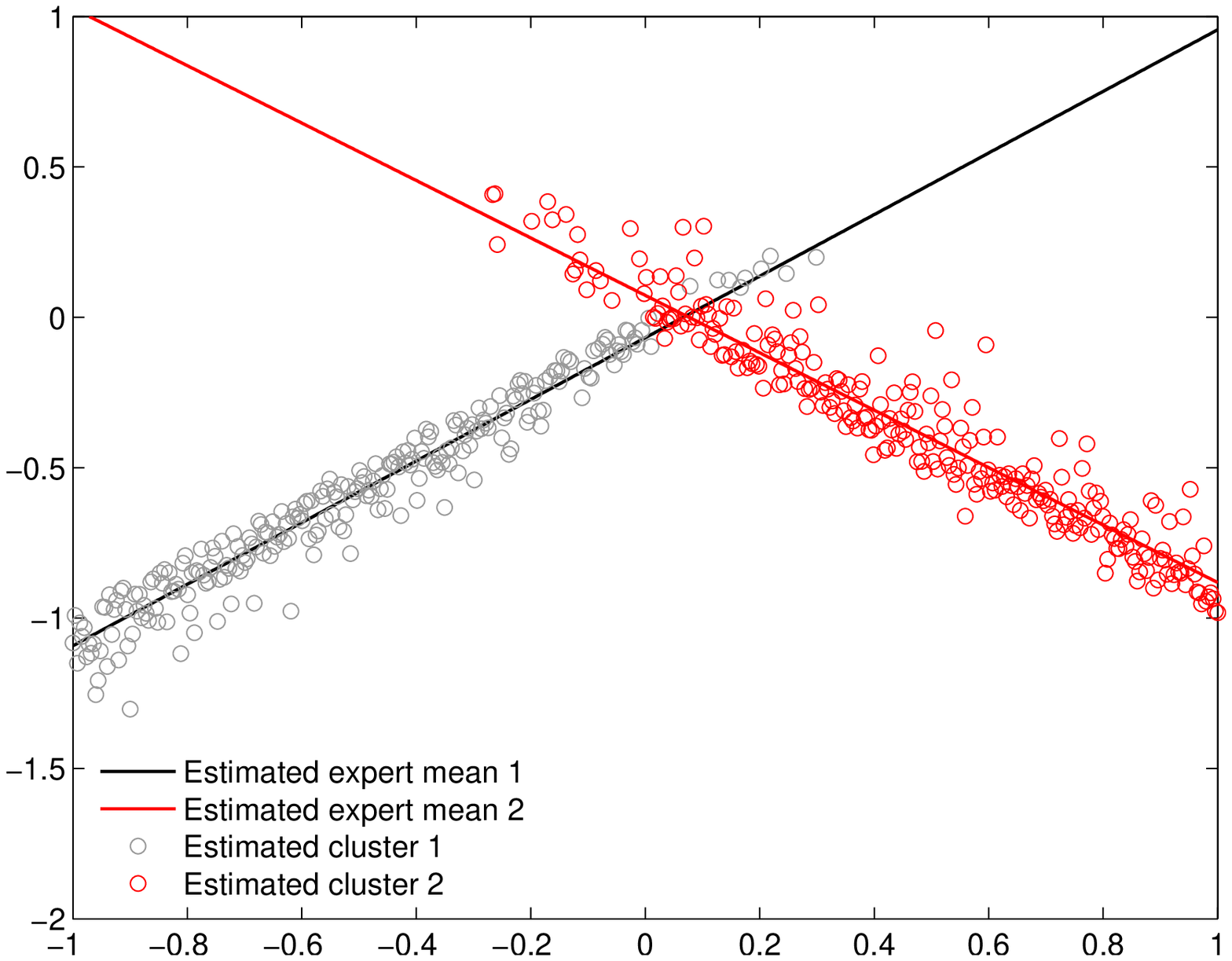} 
   \end{tabular}
   \caption{\label{fig. TwoClust-All->All-estimated partitions}The true and estimated partitions by fitting 
    the standard NMoE model (up) and the proposed STMoE model (bottom) to the simulated data sets  shown in Figure \ref{fig. TwoClust-All->All-estimated models}.}
\end{figure}

\subsubsection{Experiment 2}

In this experiment we examine the robustness of the proposed STMoE model to outliers  versus the standard NMoE one. 
For that, we considered each of the two models (NMoE and STMoE) for data generation. For each generated sample, each of the two  models in considered for the inference. The data were generated  exactly in the same way as in Experiment 1, except for some observations which were generated with a probability $c$ from a class of outliers. We considered the same class of outliers as in \citet{Nguyen2016-MoLE}, that is the predictor $x$  is generated uniformly over the interval $(-1, 1)$ and the response $y$ is set the value $-2$. 
We apply the MoE models by setting the covariate vectors as before, that is,  $\bsx = \bsr  = (1, x)^T$. 
We considered varying probability of outliers $c = 0\%, 1\%, 2\%, 3\%, 4\%, 5\%$ and the sample size of the generated data is $n=500$. An example of simulated sample containing $5\%$ outliers is shown in Figure \ref{fig. TwoClust-Outliers-NMoE_NMoE}. 
As a criterion of evaluation of the impact of the outliers on the quality of the results, we considered the MSE between the true regression mean function and the estimated one. This MSE is calculated as
$\frac{1}{n}\sum_{i=1}^n\!\parallel \! \E_{{\it \bsvPsi}}(Y_i|\bsr_i,\bsx_i) - \E_{{\it \hat\bsvPsi}}(Y_i|\bsr_i,\bsx_i)\!\parallel^2$ where the expectations are computed as in Section \ref{sec: Prediction using the STMoE}.

\subsubsection{Obtained results}

Table \ref{tab. MSE for the mean function - Noisy simulated data : All->all} shows,  for each of the two models, the results in terms of mean squared error (MSE) between the true mean function and the estimated one, for an increasing number of outliers in the data.
First, one can see that, when there is no outliers ($c=0\%$) and when the data follow a NMoE distribution, the error of fitting a NMoE is very slightly less than the one of fitting the proposed STMoE model. The STMoE then is still competitive.
However, when the data do not contain outliers and follow a STMoE distribution, fitting a NMoE is restrictive since the error of this one is high compared to the one obtained by fitting a STMoE model.
More importantly, it can be seen that, as expected, when there is outliers in the data, including the situations with only few atypical data points (1\% and 2 \%), the NMoE does not provide an adapted fit and is clearly outperformed by the proposed STMoE model. This includes the two situations, that is, including when the data are not generated according to the STMoE model.
The errors of the NMoE model are high compared to those of the STMoE.
This confirms that te STMoE is much more robust to outliers compared to the normal mixture of experts because the expert components in the STMoE model follow the robust skew $t$ distribution. The NMoE is  sensitive to outliers.
On the other hand it can be seen that, when the number of outliers is increasing, increase in the error of the NMoE is more pronounced compared to the one of STMoE model. The error for the STMoE may indeed slightly increase, remain stable or even decrease in some situations. This provides an additional support to the expected robustness of the STMoE compared to the NMoE.
{\setlength{\tabcolsep}{6pt
\begin{table}[H]
\centering
{\small
\begin{tabular}{l l c c c c c c }
\hline
& \hspace{1cm}$c$& $0\%$ & $1\%$ & $2\%$ & $3\%$ & $4\%$ & $5\%$\\
\multicolumn{2}{c}{Model} & & & & & &\\
 \hline
 \hline 
\multirow{2}{*}{\rotatebox[origin=c]{0}{NMoE}}
& NMoE    & 	\textbf{0.178}	& 1.057 &1.241 & 3.631 &	 13.25 &	28.96 \\ 
& STMoE  & 	0.258	& \textbf{0.741} & \textbf{0.794} & \textbf{0.696} &	\textbf{0.697} &	\textbf{0.626} \\
 \hline 
 \hline 
 \multirow{2}{*}{\rotatebox[origin=c]{0}{STMoE}} 
& NMoE    & 0.710 & 0.7238 & 1.048 & 6.066 & 12.45 & 31.64 \\ 
& STMoE  & \textbf{0.280} & \textbf{0.186} & \textbf{0.447} & \textbf{0.600} & \textbf{0.509} & \textbf{0.602} \\
 \hline
\end{tabular}
}
\caption{\label{tab. MSE for the mean function - Noisy simulated data : All->all}MSE $\times 10^3$ between the estimated mean function and the true one for each of the two models for a varying probability $c$ of outliers for each simulation. The first column indicates the model used for generating the data and the second one indicates the model used for inference.}
\end{table} 
}

Then, in order to highlight the robustness to noise of the TMoE and STMoE models, in addition to the previously shown numerical results, 
figures 
\ref{fig. TwoClust-Outliers-NMoE_NMoE} 
and
\ref{fig. TwoClust-Outliers-NMoE_STMoE}
show an example of results obtained on the same data set by, respectively, the NMoE and the STMoE. The data are generated by the NMoE model and contain $c=5\%$ of outliers. 

In this example, we clearly see that  the NMoE model is severely affected by the outliers. It provides a rough fit especially for the second component whose estimation is corresponds to a rough approximation due to the atypical data.
However, one can  see that the STMoE model clearly provides a precise fit; the estimated mean function  and expert components are very close to the true ones. The  STMoE is robust to outliers, in terms of estimating the true model as well as in terms of estimating the true partition of the data (as shown in the middle plots of the data).  
Notice that for the STMoE, the confidence regions are not shown because for this situation the estimated degrees of freedom are less than $2$ ($1.6097$  and  $1.5311$) for the STMoE); Hence the variance for this model in that case is not defined (see Section \ref{sec: Prediction using the STMoE}). 
The  STMoE model  provides indeed components with small degrees of freedom corresponding to highly heavy tails, which allow to handle outliers in this noisy case. While the variance is not estimable here, re-sampling techniques can be used to evaluate it, such as the techniques of \cite{OHagan2015} for producing standard errors and confidence intervals for mixture parameters in model-based clustering.
\begin{figure}[!h]
   \centering 
\begin{tabular}{cc}
   \includegraphics[width=6.2cm]{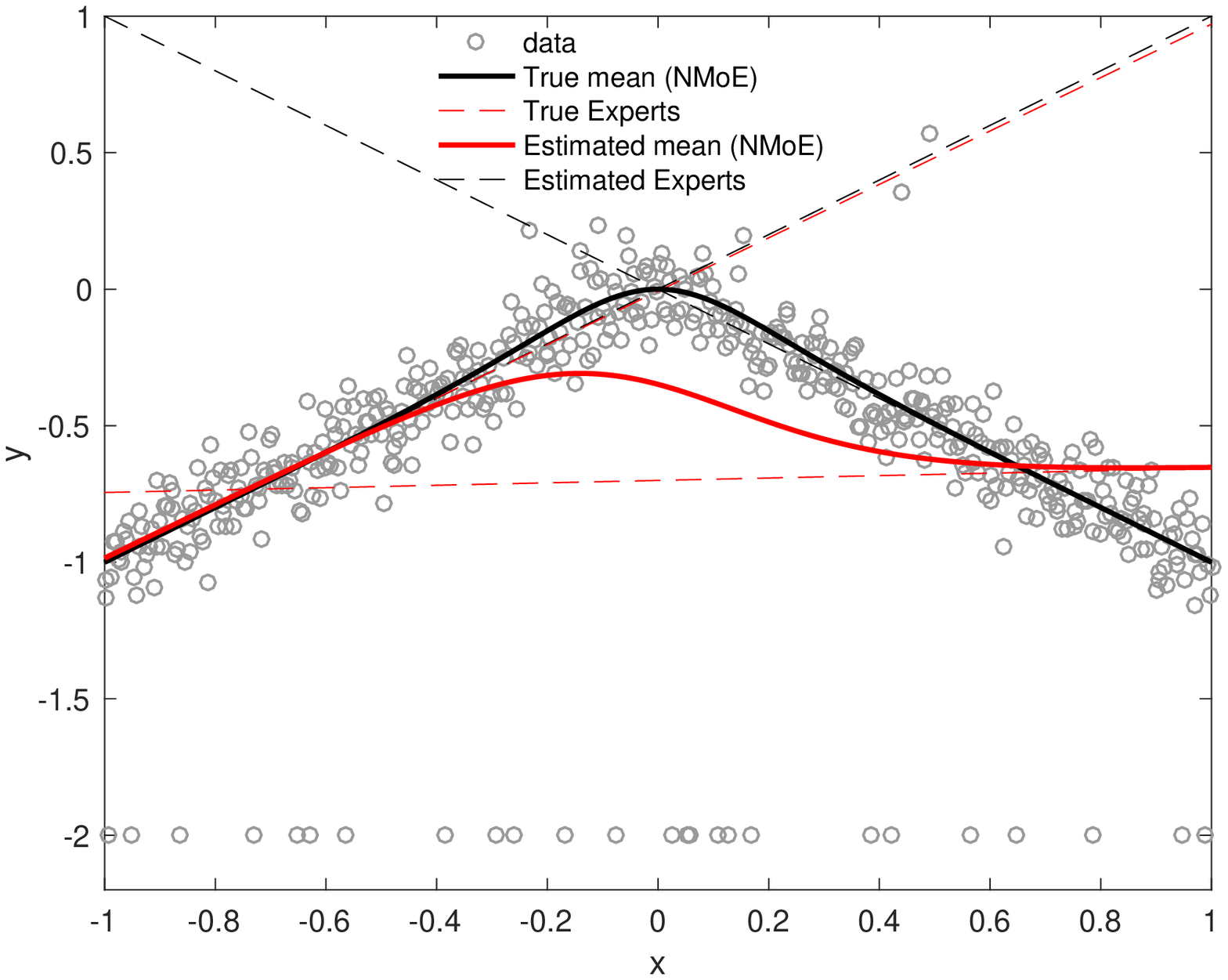}&  
   \includegraphics[width=6.2cm]{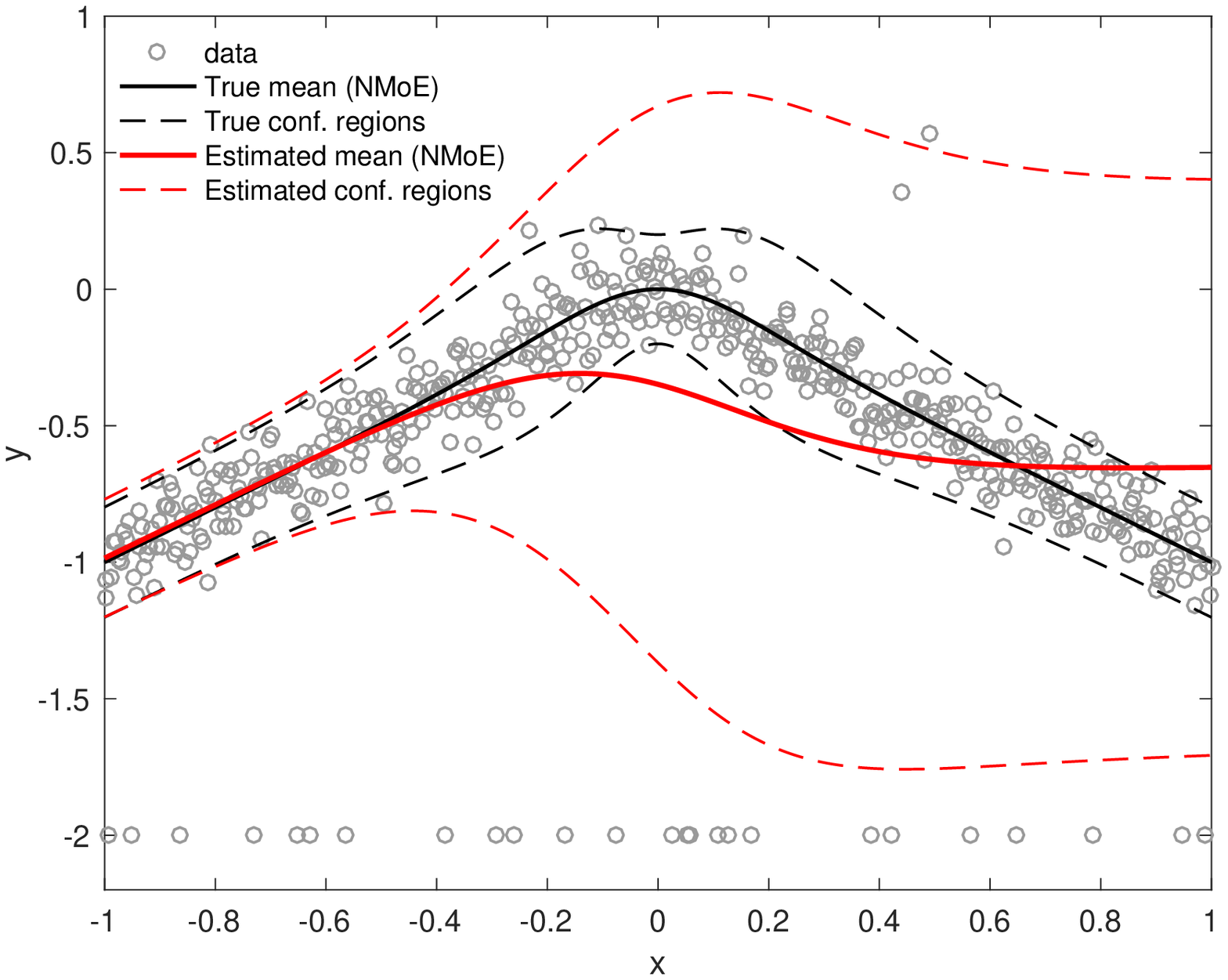}\\
  \includegraphics[width=6.2cm]{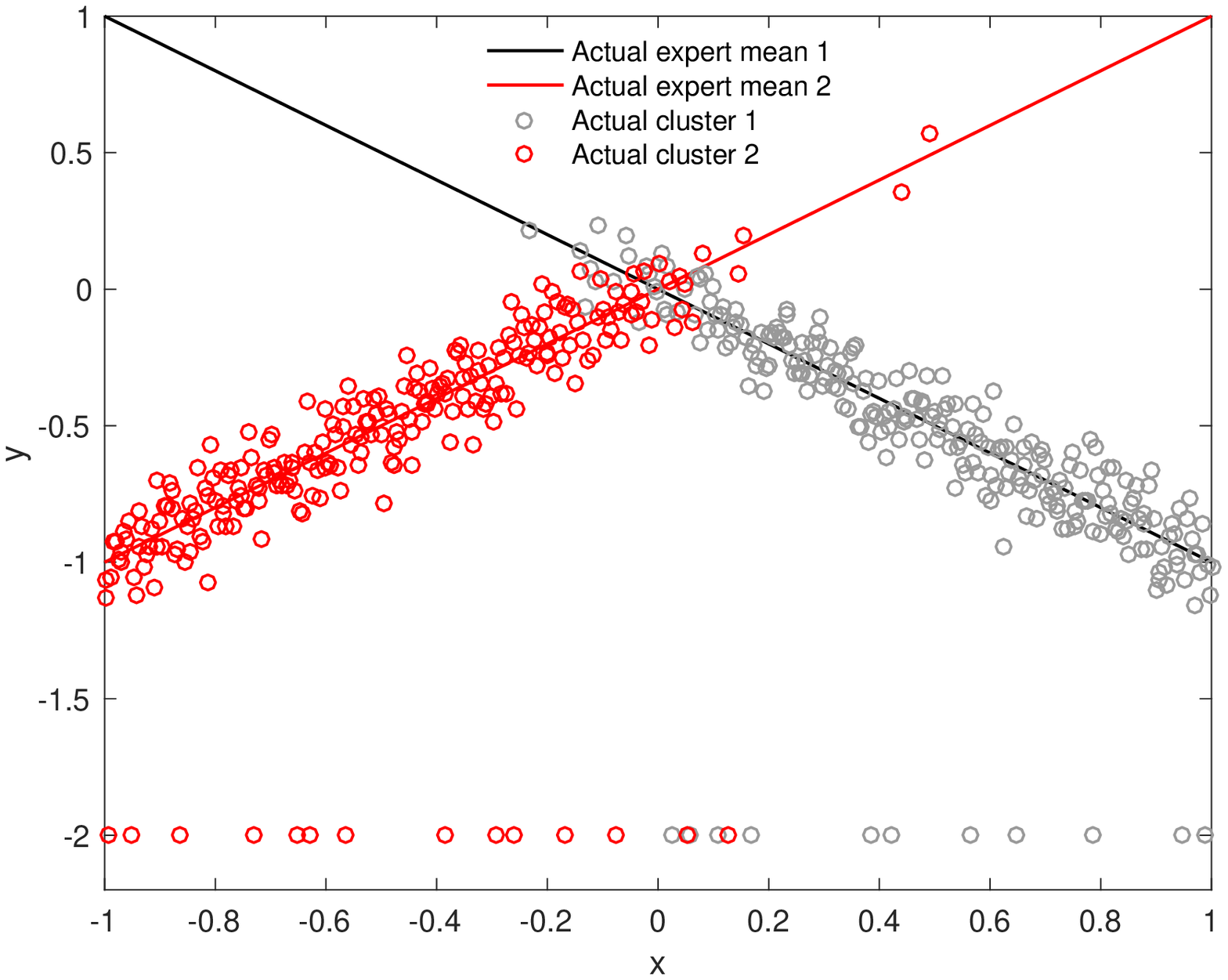}&
 \includegraphics[width=6.2cm]{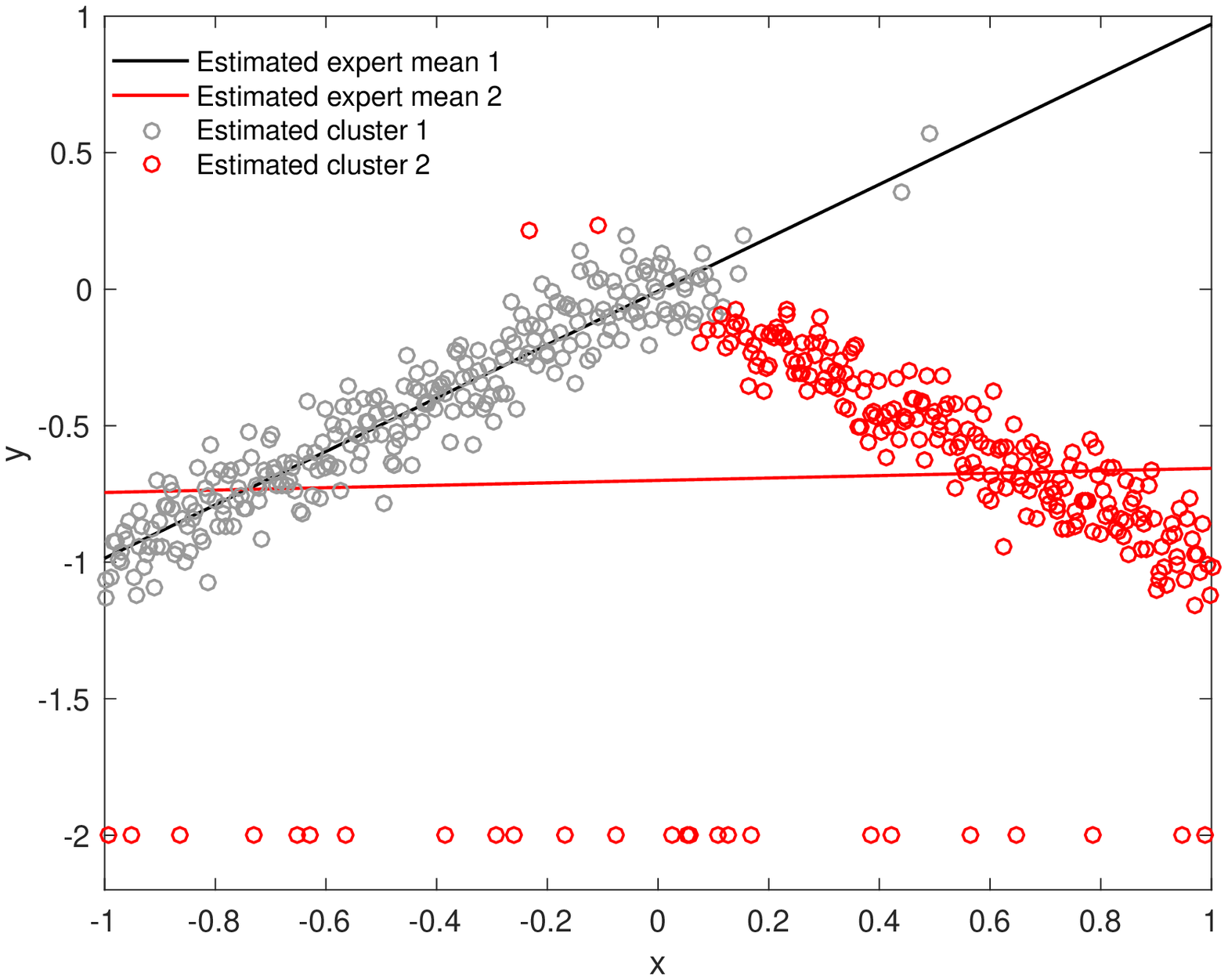} \\
   \includegraphics[width=6.2cm]{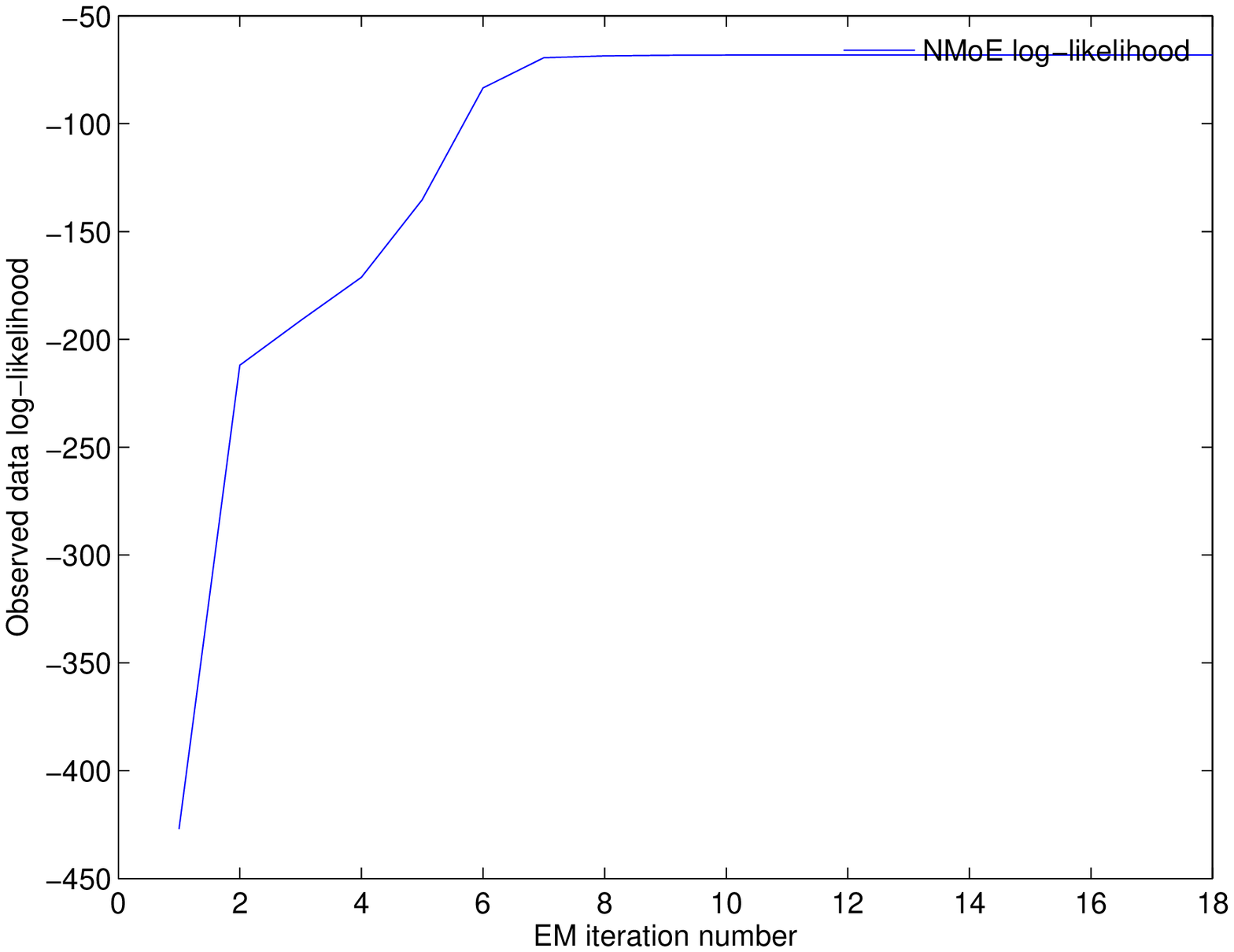} &
   \includegraphics[width=6.2cm]{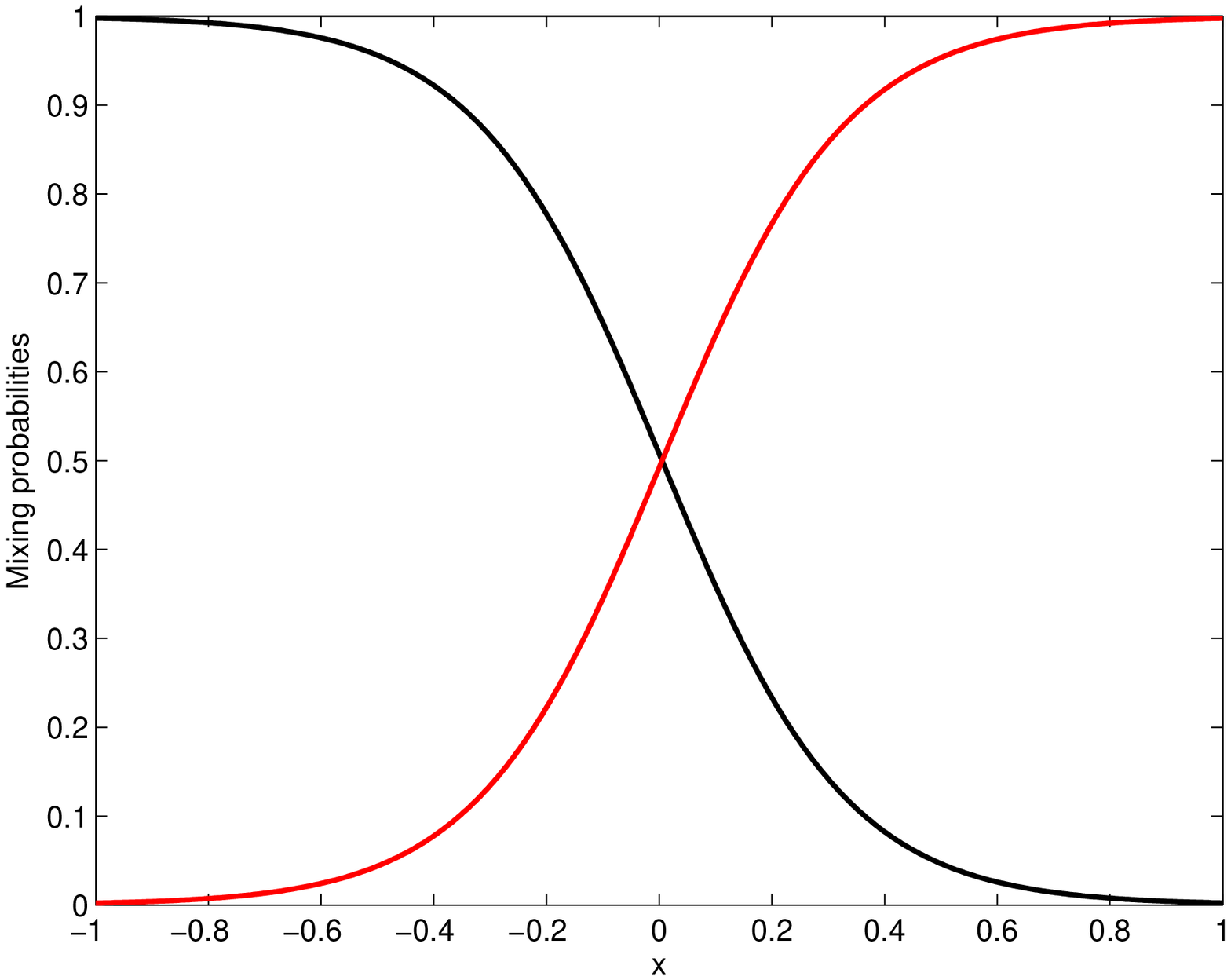}\\
   \end{tabular}
      \caption{\label{fig. TwoClust-Outliers-NMoE_NMoE}Fitted NMoE model to a data set of $n=500$ observations generated according to the NMoE model  and including $5\%$ of outliers.}
\end{figure}
\begin{figure}[!h]
   \centering  
   \begin{tabular}{cc}
   \includegraphics[width=6.2cm]{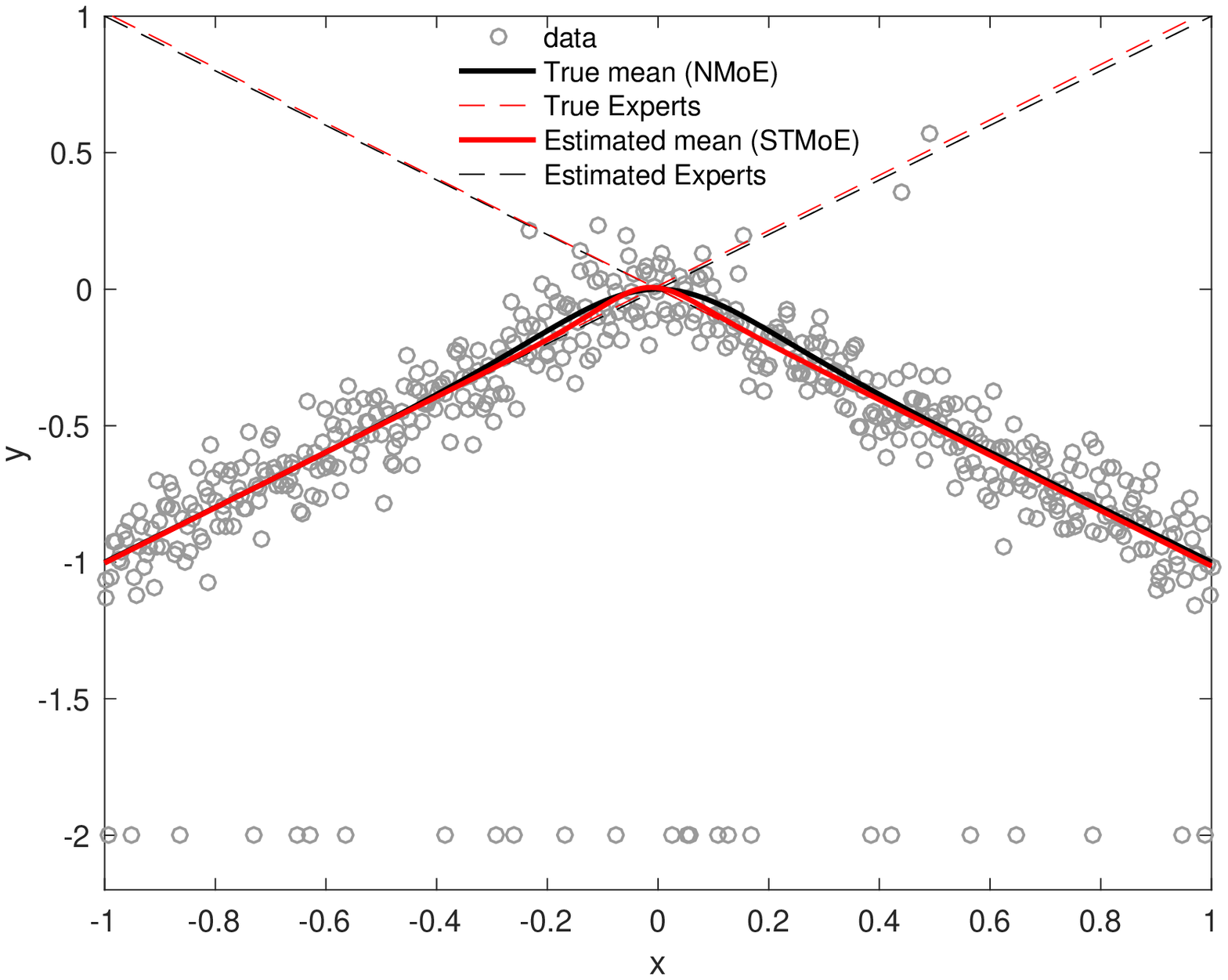} &
   \includegraphics[width=6.2cm]{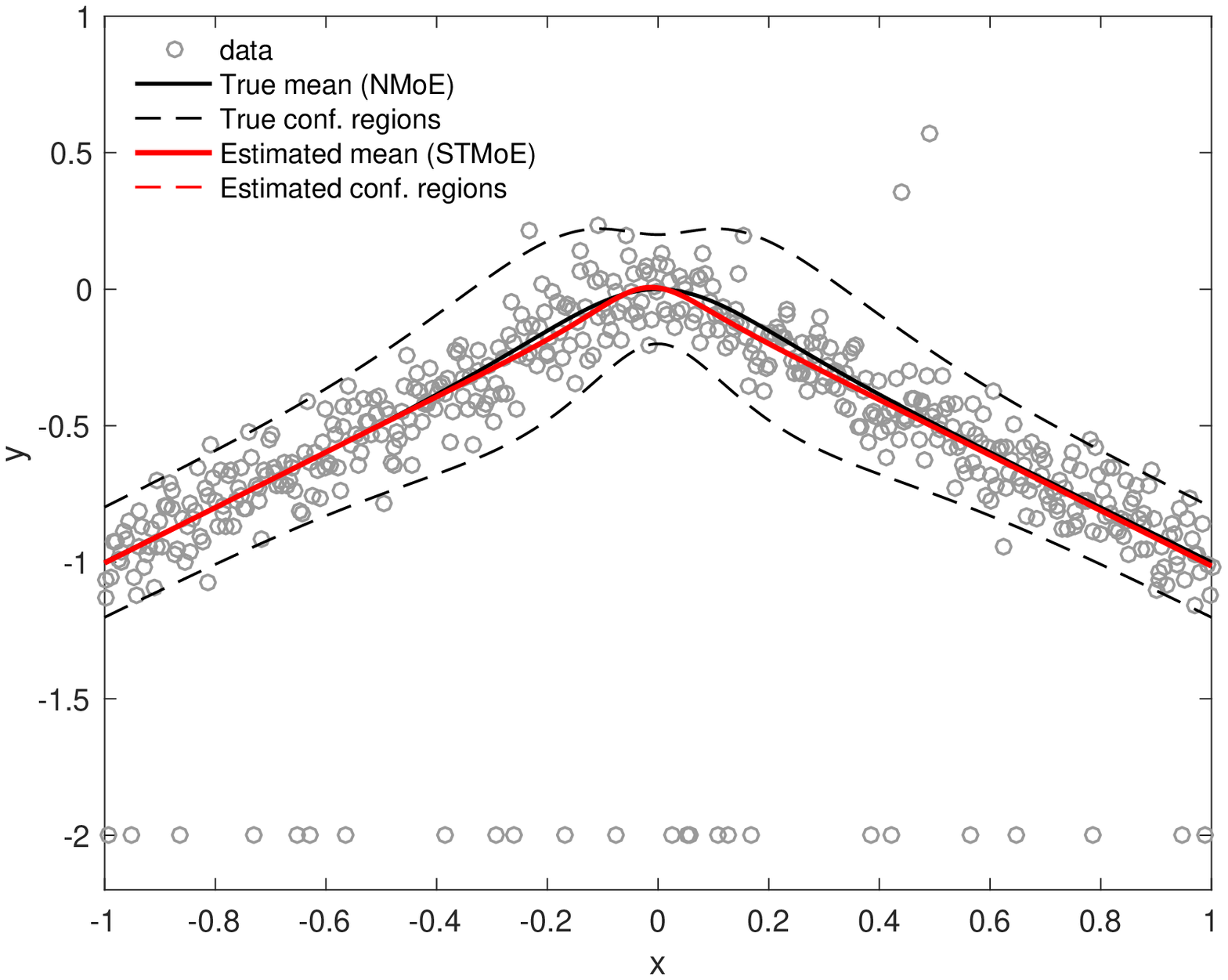}\\
   \includegraphics[width=6.2cm]{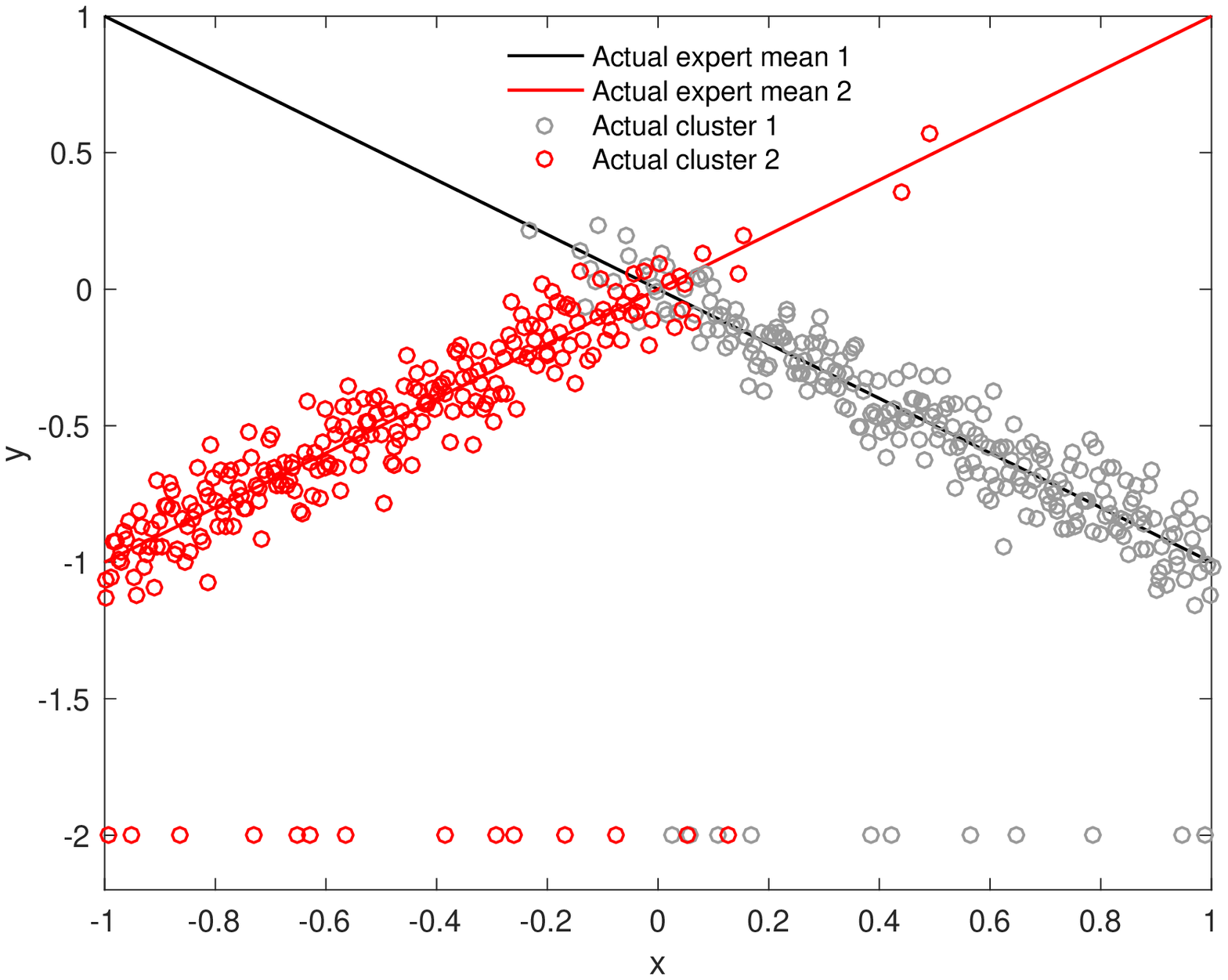}&
  \includegraphics[width=6.2cm]{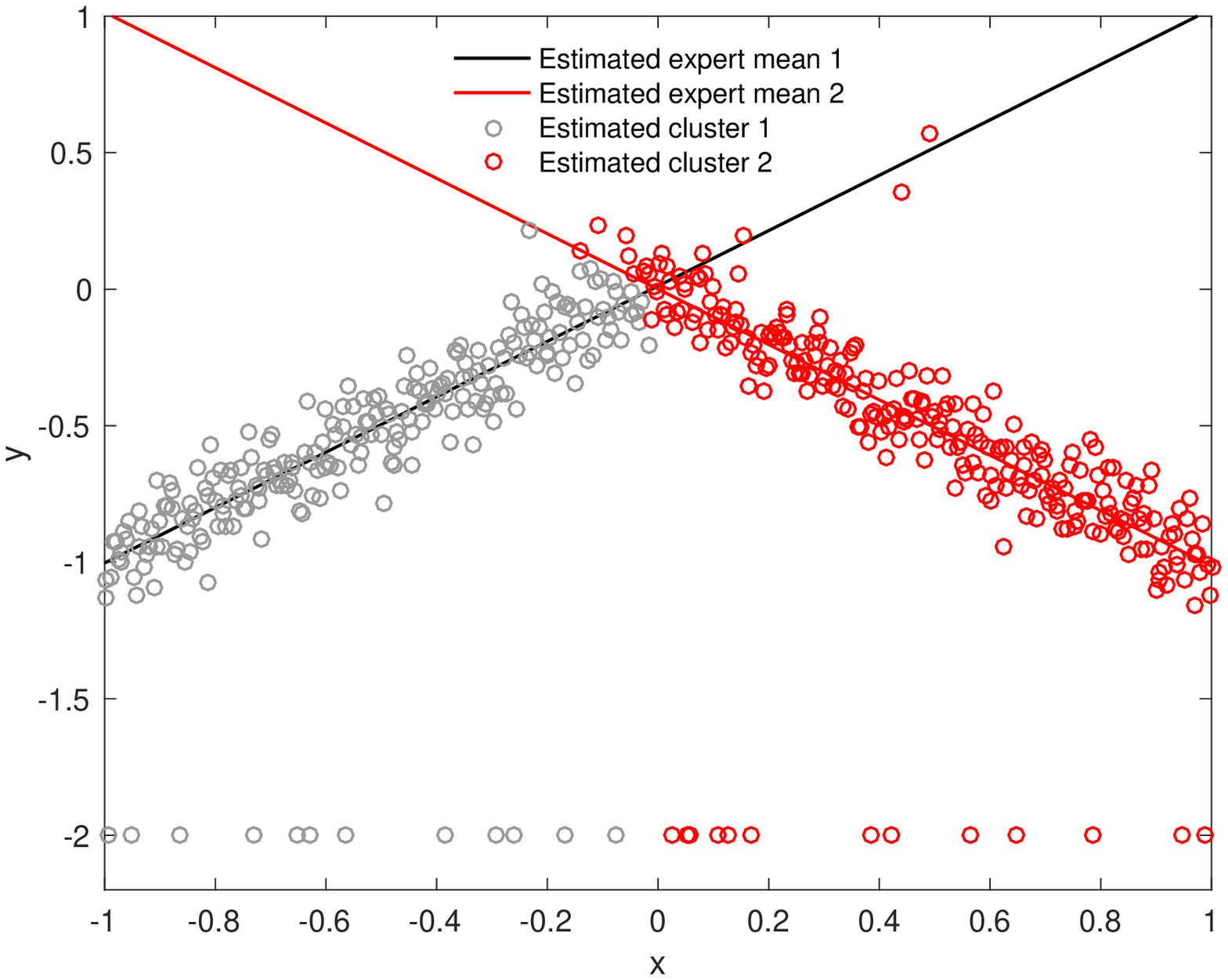}\\
   \includegraphics[width=6.2cm]{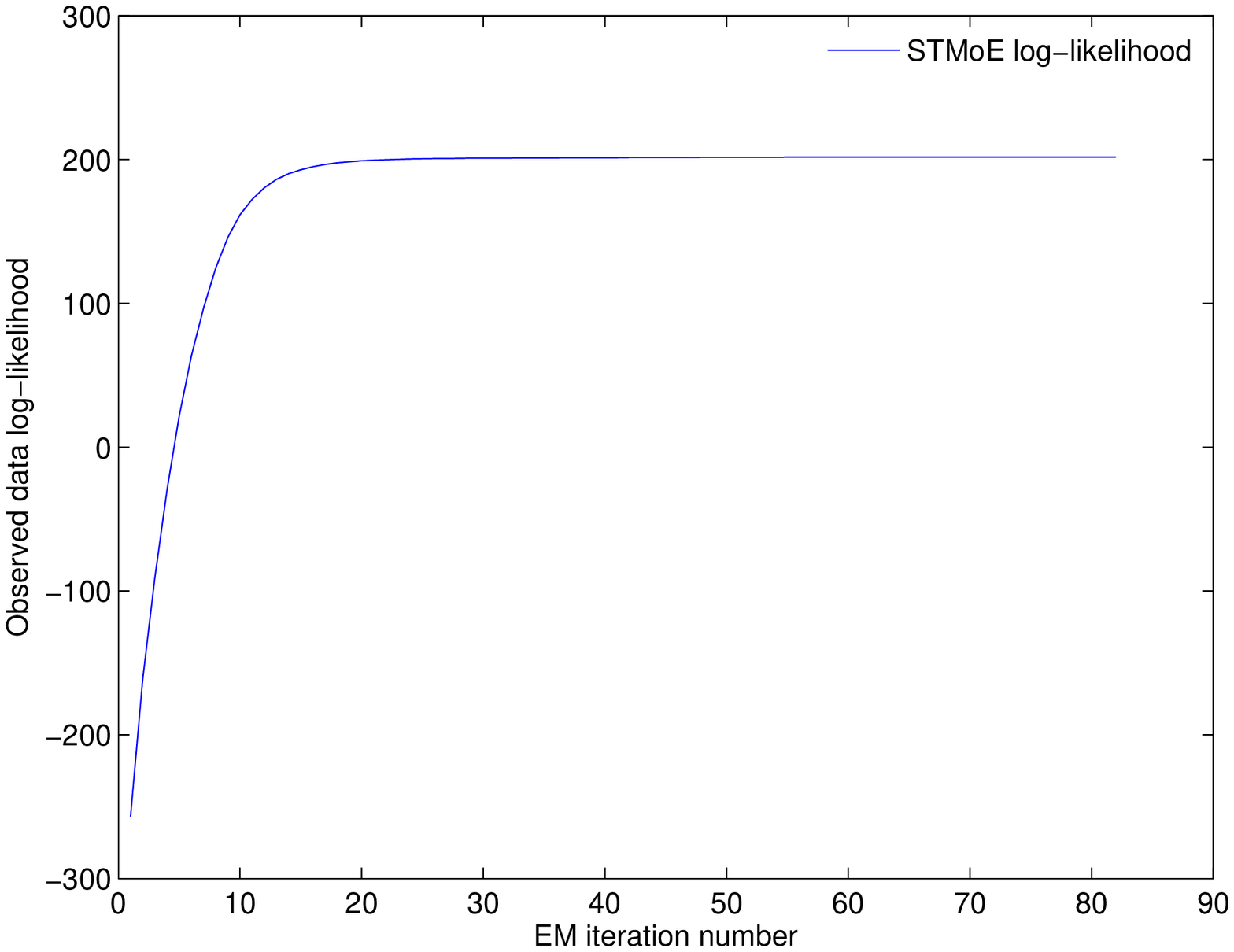} & 
   \includegraphics[width=6.2cm]{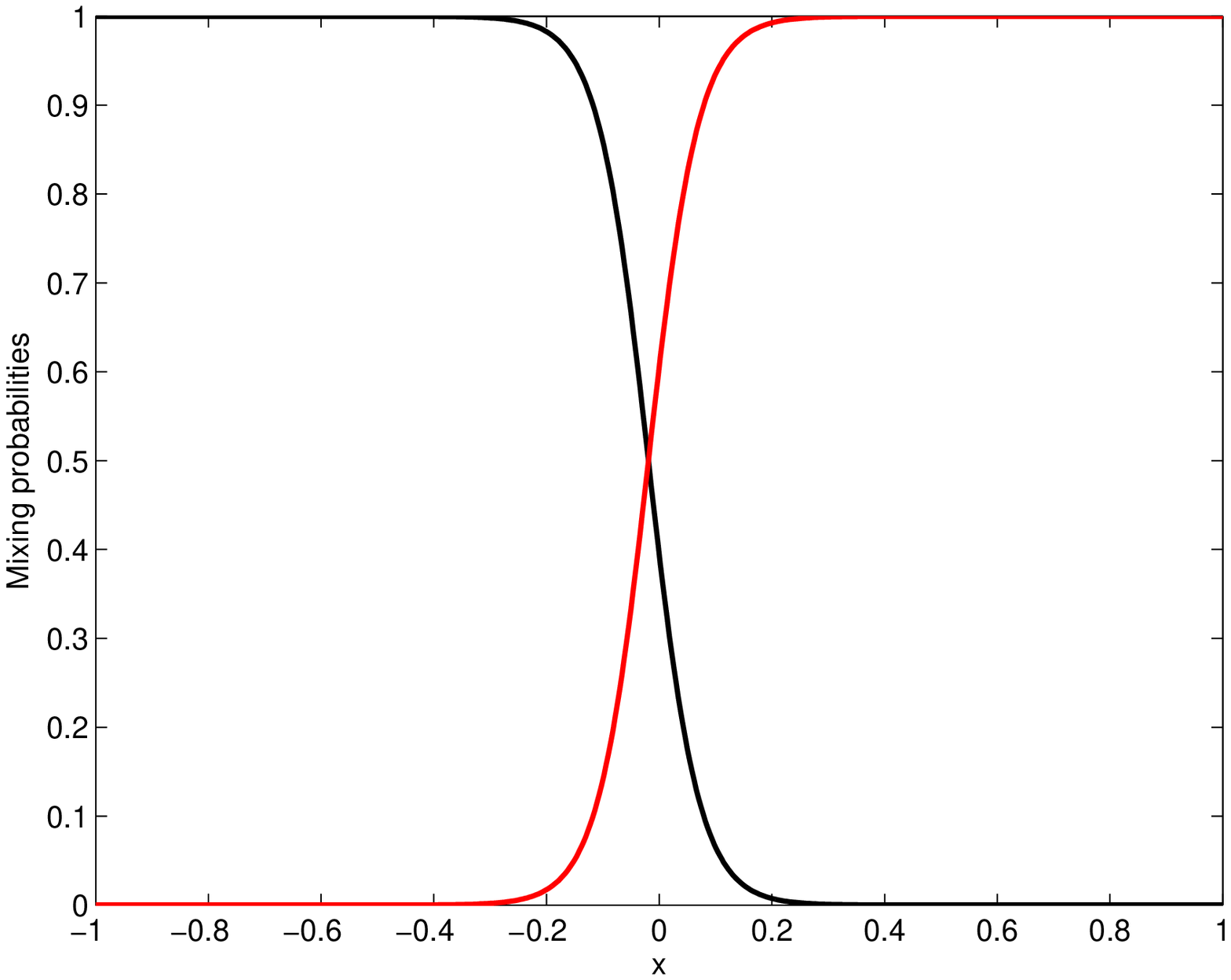}  
   \end{tabular}
      \caption{\label{fig. TwoClust-Outliers-NMoE_STMoE}Fitted STMoE model to a data set of $n=500$ observations generated according to the NMoE model  and including $5\%$ of outliers.}
\end{figure}

\subsection{Application to two real-world data sets}

In this section, we consider an application to two real-world data sets: the tone perception data set and the temperature anomalies data set shown in Figure \ref{fig. Tone and temperature anomalies data}.
\begin{figure}[H]
   \centering 
   \begin{tabular}{cc}
   \includegraphics[width=6.2cm]{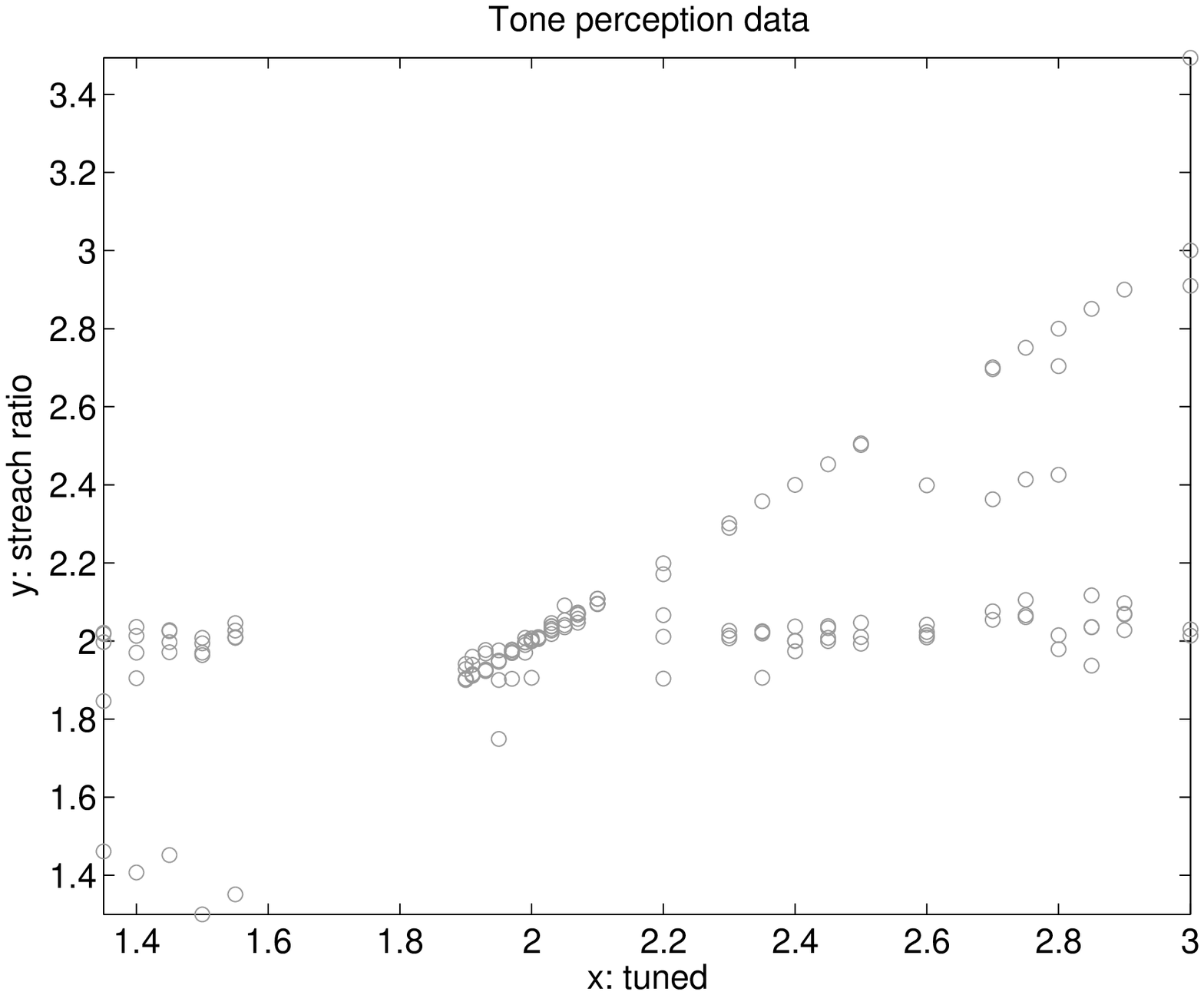} & 
   \includegraphics[width=6.2cm]{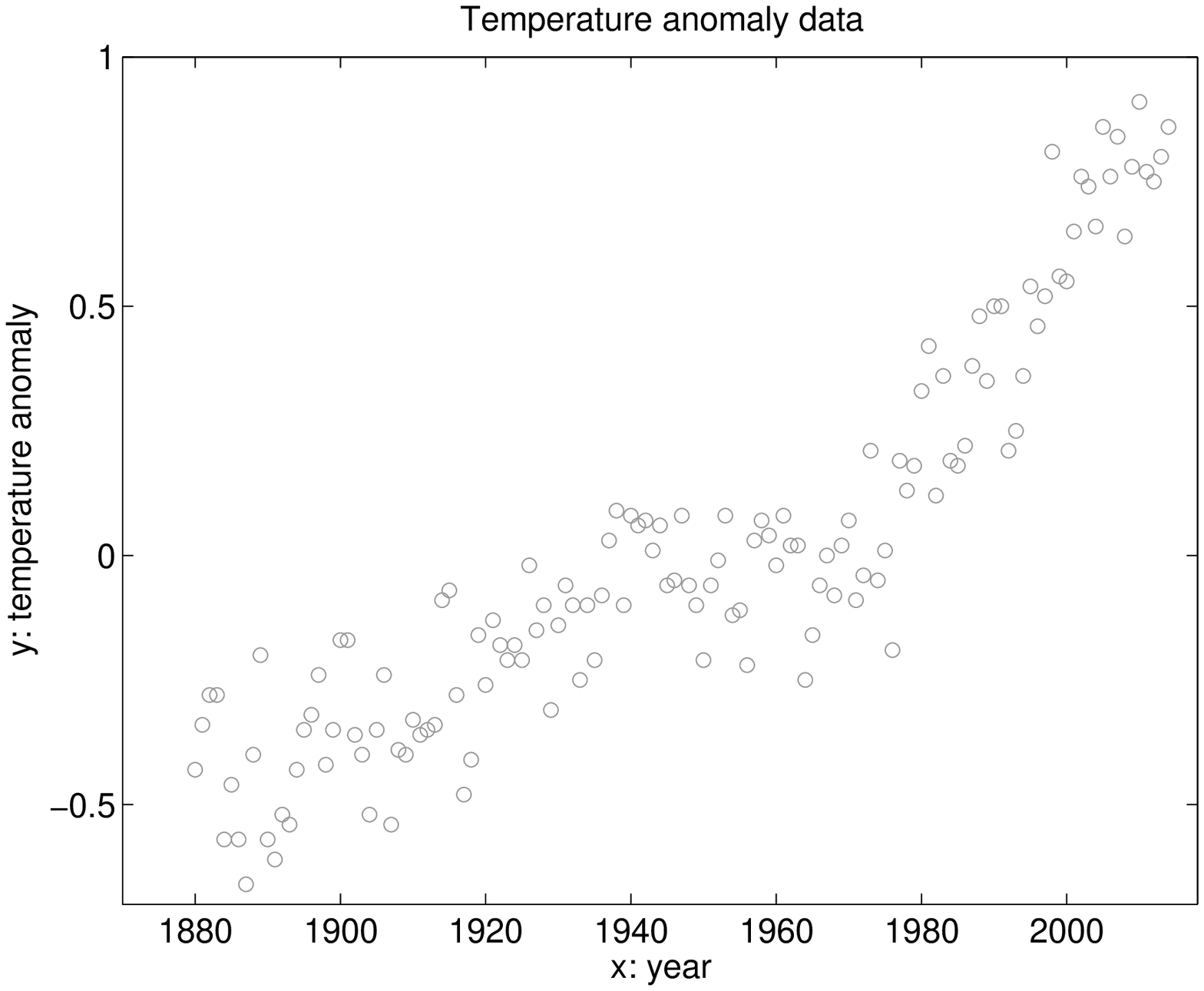} 
   \end{tabular}
      \caption{\label{fig. Tone and temperature anomalies data}Scatter plots of the tone perception data and the temperature anomalies data.}
\end{figure}

\subsubsection{Tone perception data set} The first analyzed data set is the real tone perception data set\footnote{Source: \url{http://artax.karlin.mff.cuni.cz/r-help/library/fpc/html/tonedata.html}} which goes back to \citet{Cohen1984}. It was recently studied by \citep{Bai2012,Ingrassia2012} and \citet{Song2014} by using robust regression mixture models based on, respectively, the $t$ distribution and the Laplace distribution.
In the tone perception experiment, a pure fundamental tone was played to a trained musician. Electronically generated overtones were added, determined by a stretching ratio (``stretch ratio" = 2) which corresponds to the harmonic pattern usually heard in traditional definite pitched instruments. 
The musician was asked to tune an adjustable tone to the octave above the fundamental tone and a ``tuned'' measurement gives the ratio of the adjusted tone to the fundamental. The obtained data consists of $n=150$ pairs of ``tuned'' variables, considered here as  predictors  ($x$), and their corresponding ``strech ratio'' variables considered as responses ($y$). 
To apply the proposed MoE models, we set the response $y_i (i=1,\ldots,150)$ as the ``strech ratio'' variables and the covariates $\bsx_i = \bsr_i = (1,x_i)^T$ where $x_i$ is the ``tuned'' variable of the $i$th observation. We also follow the study in \citet{Bai2012} and \citet{Song2014} by using two mixture components. Model selection results are given later in Table \ref{tab. Model selection Tone data}.

Figure  \ref{fig. Original Tone data and all models} shows the scatter plots of the tone perception data and the linear expert components of the fitted NMoE model and the proposed  STMoE model. 
One can observe that we obtain a good fit with the two models. 
The NMoE  fit differs very slightly from the one of the  STMoE.
The two regression lines may correspond to correct tuning and tuning to the first overtone, respectively, as analyzed in \citet{Bai2012}. 
\begin{figure}[H]
   \centering 
   \begin{tabular}{cc}
   \includegraphics[width=6.2cm]{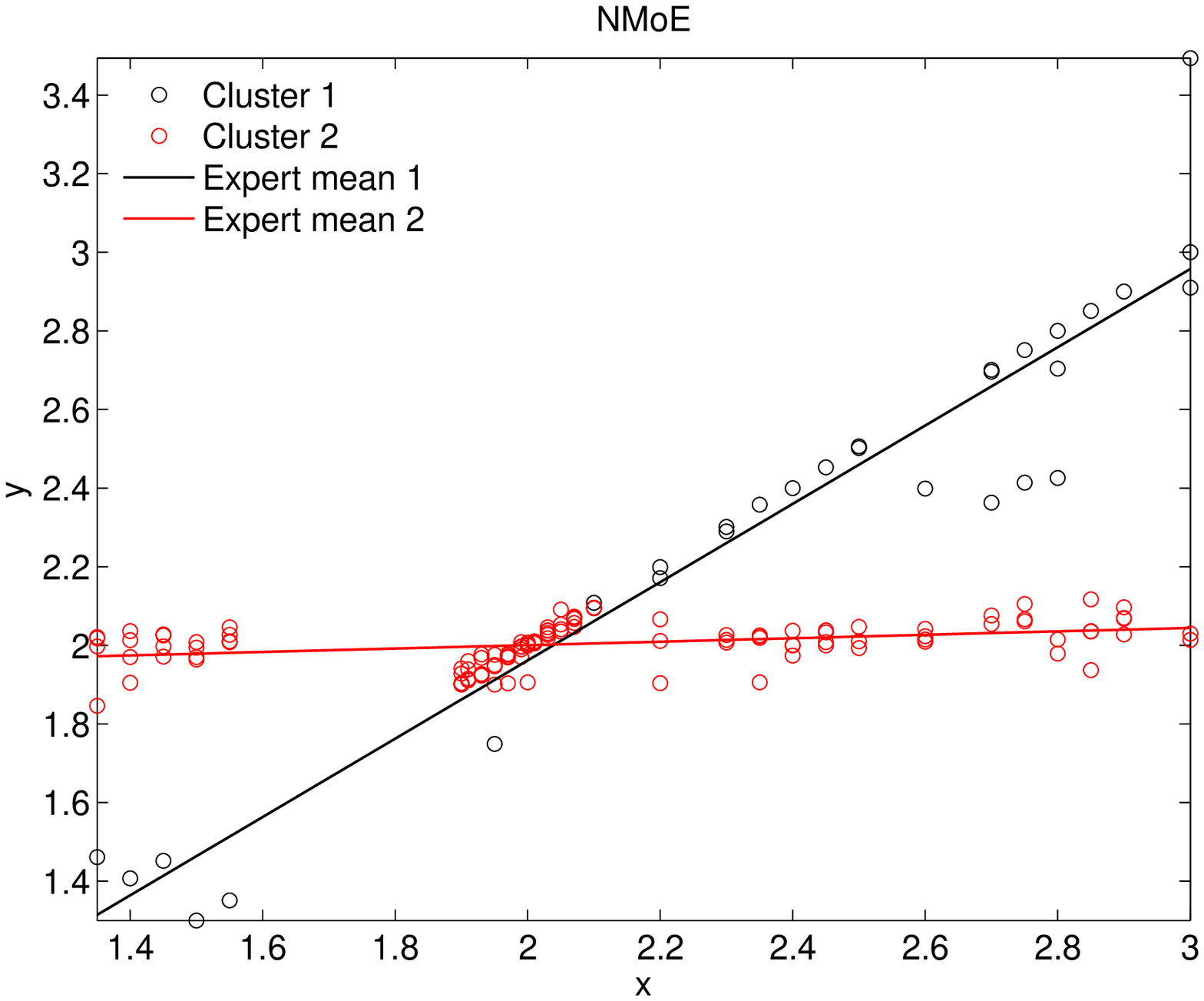} &
   \includegraphics[width=6.2cm]{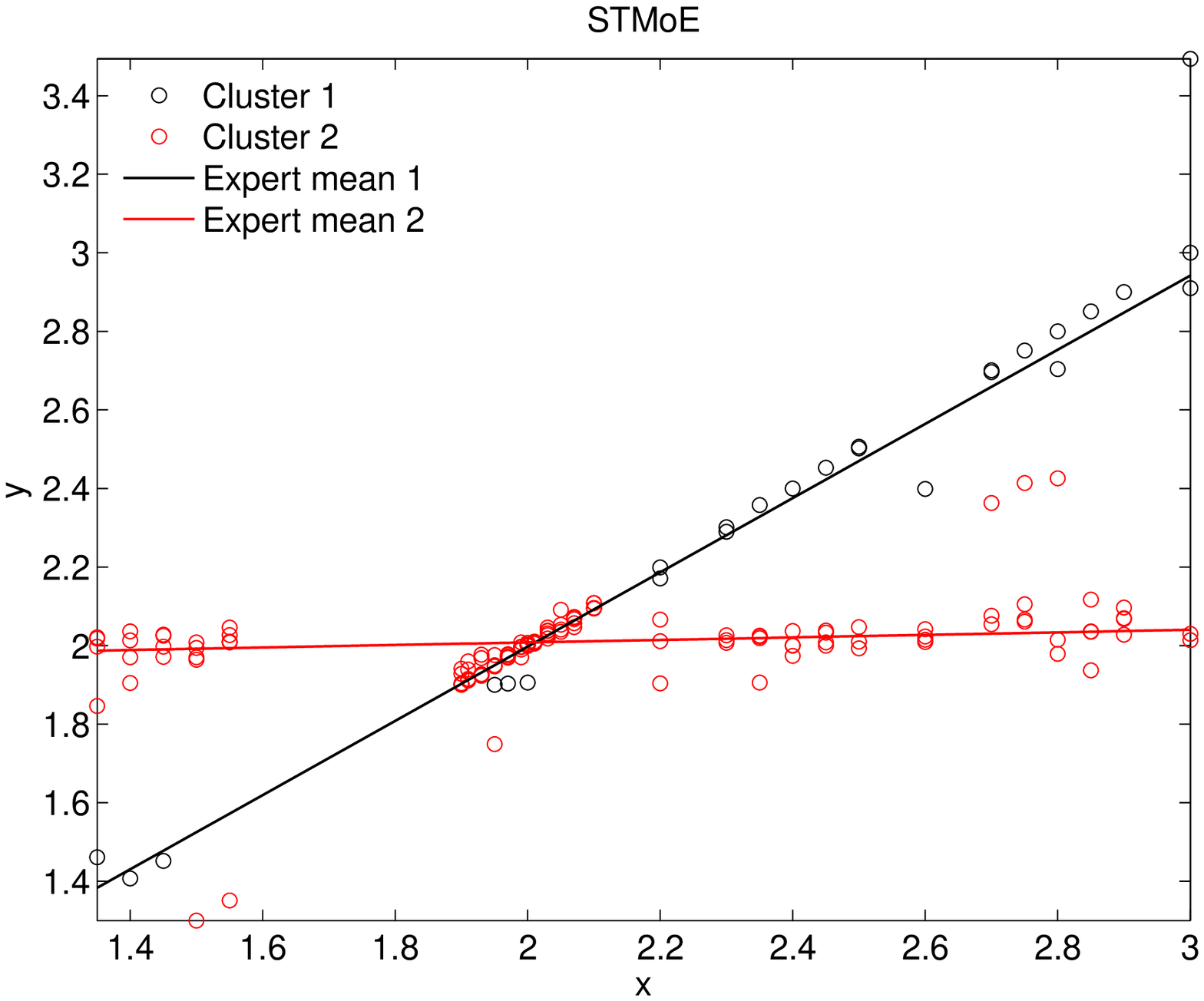}\\
      \end{tabular}
      \caption{\label{fig. Original Tone data and all models}The fitted MoLE to the original tone data set with the NMoE model (left) and the STMoE model (right). The predictor $x$ is the actual tone ratio and the response $y$ is the perceived tone ratio.}
\end{figure}
Figure \ref{fig. Tone data and all loglik models} shows the log-likelihood profiles for each of the two  models. It can namely be seen that training the skew $t$ mixture of experts may take more iterations than the normal MoE model. The STMoE has indeed more parameters to estimate (additional skewness and robustness parameters). However, in terms of computing time, the algorithm is fast and converges in only few seconds  (around 10 seconds for this example) on a personal laptop with 2,9 GHz processor and 8 GB memory.  
\begin{figure}[H]
   \centering 
   \begin{tabular}{cc}
   \includegraphics[width=6.2cm]{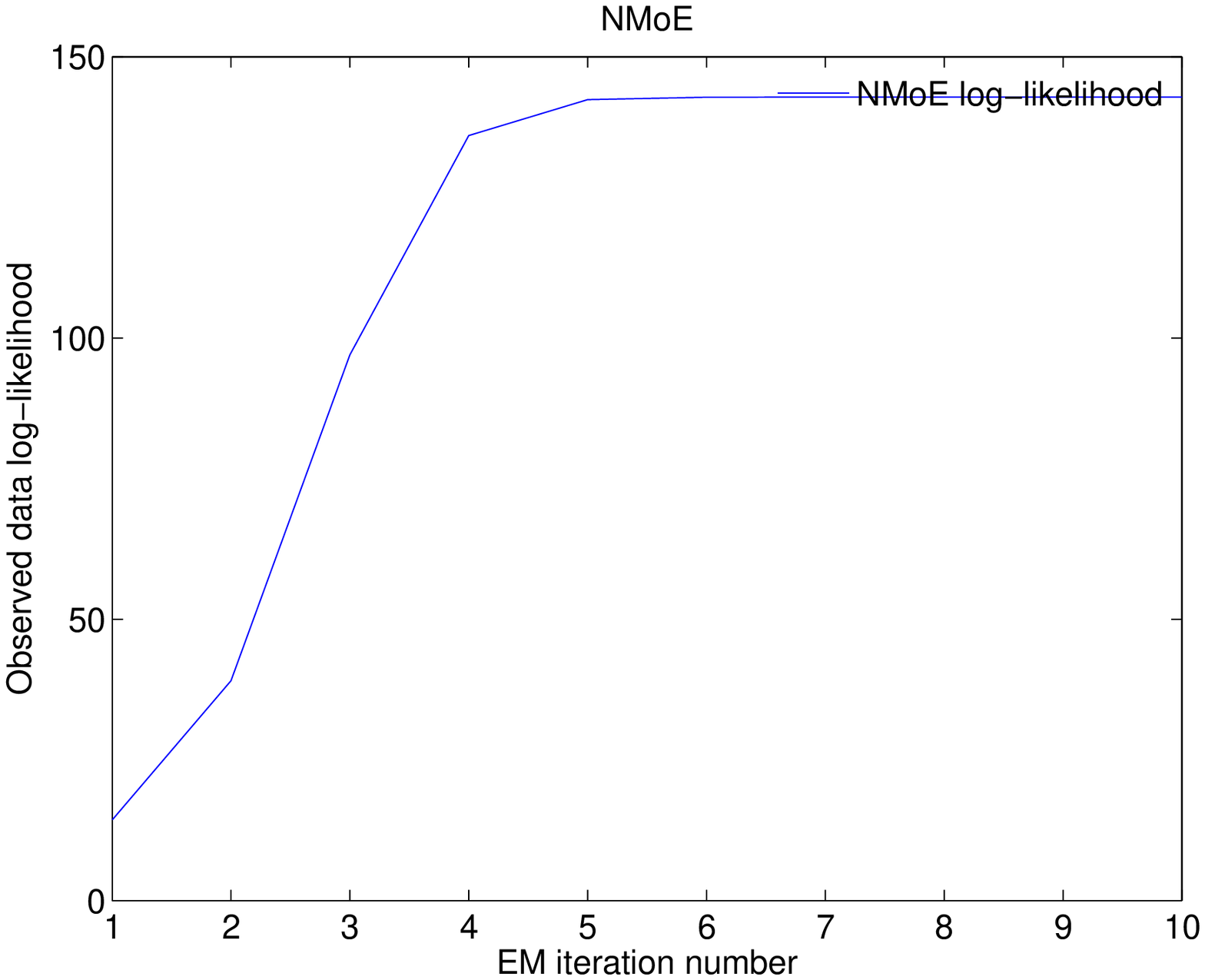} &
   \includegraphics[width=6.2cm]{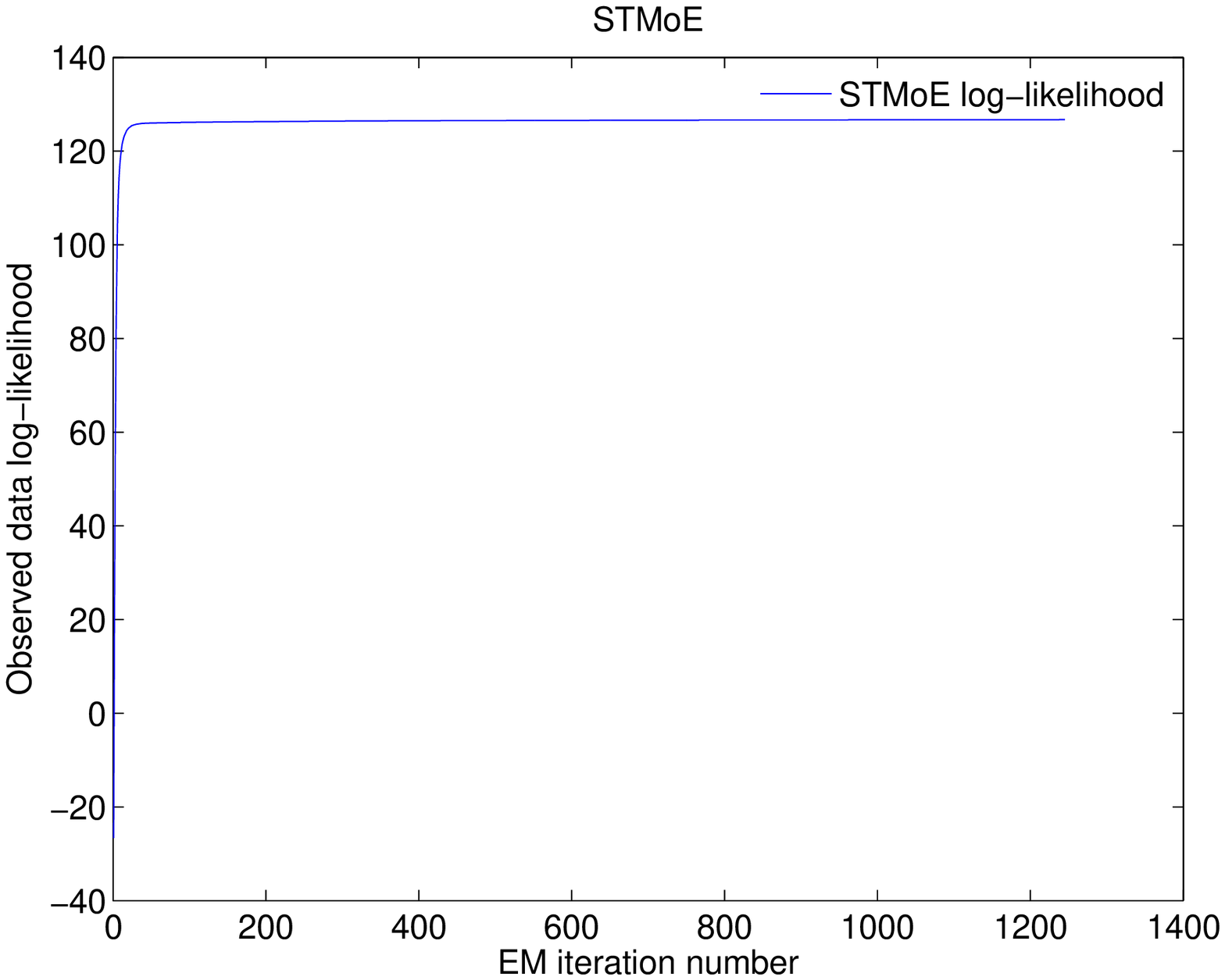}\\
   \end{tabular}
      \caption{\label{fig. Tone data  and all loglik models}The log-likelihood during the EM iterations when fitting the MoLE models to the original tone data set with  the NMoE model (left) and  the STMoE model (right).}
\end{figure}
The values of estimated parameters for the tone perception data set are given in Table \ref{tab. Estimated parameters for the tone perception data set}.
One can see that the regression coefficients are very similar for the two models. 
One can also see that the STMoE model retrieves a skewed component and with  high degrees of freedom compared to the other component. This one may be seen as approaching the one of a skew-normal MoE model, while the second one in approaching a $t$ distribution, that is the one of a $t$-MoE model.
{\setlength{\tabcolsep}{3pt
\begin{table}[H]
\centering
{\small \begin{tabular}{l c c  c c c c c c c c c c}
\hline
param. & $\alpha_{10}$ & $\alpha_{11}$ & $\beta_{10}$ & $\beta_{11}$ & $\beta_{20}$ & $\beta_{21}$ & $\sigma_{1}$& $\sigma_{2}$ & $\lambda_{1}$ & $\lambda_{2}$ & $\nu_{1}$ & $\nu_{2}$ \\ 
model		& & & & & & & & & & & & \\
 \hline
 \hline
NMoE  	&  -2.690 &	0.796 & 	-0.029 &	0.995 & 1.913 &	 0.043 &	0.137 & 0.047 &	 - & - & - & -	\\
 STMoE	&  -3.044  &	0.824& -0.058    &	  0.944 & 1.944 &	 0.032     &	0.200 & 0.032 &	93.386 & -0.011 &	19.070 & 1.461\\
 \hline
\end{tabular}}
\caption{\label{tab. Estimated parameters for the tone perception data set}Values of the estimated MoE parameters for the original Tone perception data set.}
\end{table}
} 

We also performed a model selection procedure on this data set to choose the best number of MoE components for a number of components between 1 and 5. We used BIC, AIC, and ICL. Table  \ref{tab. Model selection Tone data} gives the obtained values of the  model selection criteria. 
One can see that for the NMoE model, the three criteria overestimate the number of components, but for both BIC and ICL, the solution with two components is also likely and is the most competitive to the selected one with 5 components. In deed, it can be seen that, if the number mixture of components is fixed at 4 rather than 5, both BIC and ICL would select the right number of components in that case.
AIC performs poorly for the two models and overestimates the number of components. 
On the other hand, for the proposed STMoE model, both BIC and ICL retrieve the correct number of components. Then, one can conclude that the BIC and the ICL are the criteria that one would suggest for the analysis of this data with the proposed model. 
{\setlength{\tabcolsep}{3pt
\begin{table}[H]
\centering
{\small 
\begin{tabular}{l |ccc | ccc}
\hline
	& \multicolumn{3}{c|}{NMoE}	& \multicolumn{3}{c}{STMoE}\\
\cline{2-7}
K			 &   BIC 	 & 	AIC 	&    ICL		 &   BIC 	 & 	AIC 	&    ICL \\
\hline
\hline
	1		& 1.866  &  	6.382    &	1.866	& 69.532  & 77.059 &  69.532\\
	2		&122.805&  134.847&  107.384& \textbf{92.435} & 110.499 &  \textbf{82.455}\\
	3		&118.193&  137.763&   76.524& 77.9753&  106.576&   52.564\\
	4		&121.703&  148.798&   94.460 &  77.7092 & 116.847&   56.365\\
	5		& \textbf{141.696}&\textbf{176.318} & \textbf{123.655}& 79.043 & \textbf{128.719} &  67.748\\
\hline 
\end{tabular}
\caption{\label{tab. Model selection Tone data}Choosing the number of expert components $K$ for the original tone perception data by using the information criteria BIC, AIC, and ICL. Underlined value indicates the highest value for each criterion.}
}
\end{table}}
 
Now we examine the sensitivity of the MoE models to outliers based on this real data set.  For this, we adopt the same scenario used in \citet{Bai2012} and \citet{Song2014} (the last and more difficult scenario) by adding 10 identical pairs $(0,4)$ to the original data set as outliers in the $y$-direction, considered as high leverage outliers. We apply the MoE models in the same way as before.

The upper plots in Figure \ref{fig. Tone data with outliers and all models} clearly show that the normal  mixture of experts  fit is  sensitive to outliers.
However, note that for this situation, compared to the normal regression mixture result in \citet{Bai2012}, and the Laplace regression mixture and the $t$ regression mixture results in \citet{Song2014}, the fitted NMoE model is affected less severely by the outliers. 
This may be attributed to the fact that the mixing proportions here are depending on the predictors, which is not the case in these regression mixture models, namely thoses of \citet{Bai2012} and  \citet{Song2014}.  One can also see that, even the regression mean functions are  affected severely by the outliers, the provided partitions are still reasonable and similar to those provided in the previous non-noisy case.
Then, the bottom plots in Figure \ref{fig. Tone data with outliers and all models}  also clearly show that the  STMoE model provides a precise robust fit. 
For the STMoE, even if the fit differs very slightly compared to the case with outliers, the obtained fits for both situations (with and without outliers) are very reasonable.
Moreover, we notice that, as showed in \citet{Song2014}, for this situation with outliers,  the $t$ mixture of regressions fails; The  fit is affected severely by the outliers.
However, for the proposed  STMoE model, the ten high leverage outliers have no significant impact on the fitted experts.  
This is because here the mixing proportions depend on the inputs, which is not the case for the regression mixture model  described in \citet{Song2014}. 
\begin{figure}[H]
   \centering 
   \begin{tabular}{cc}
   \includegraphics[width=6.2cm]{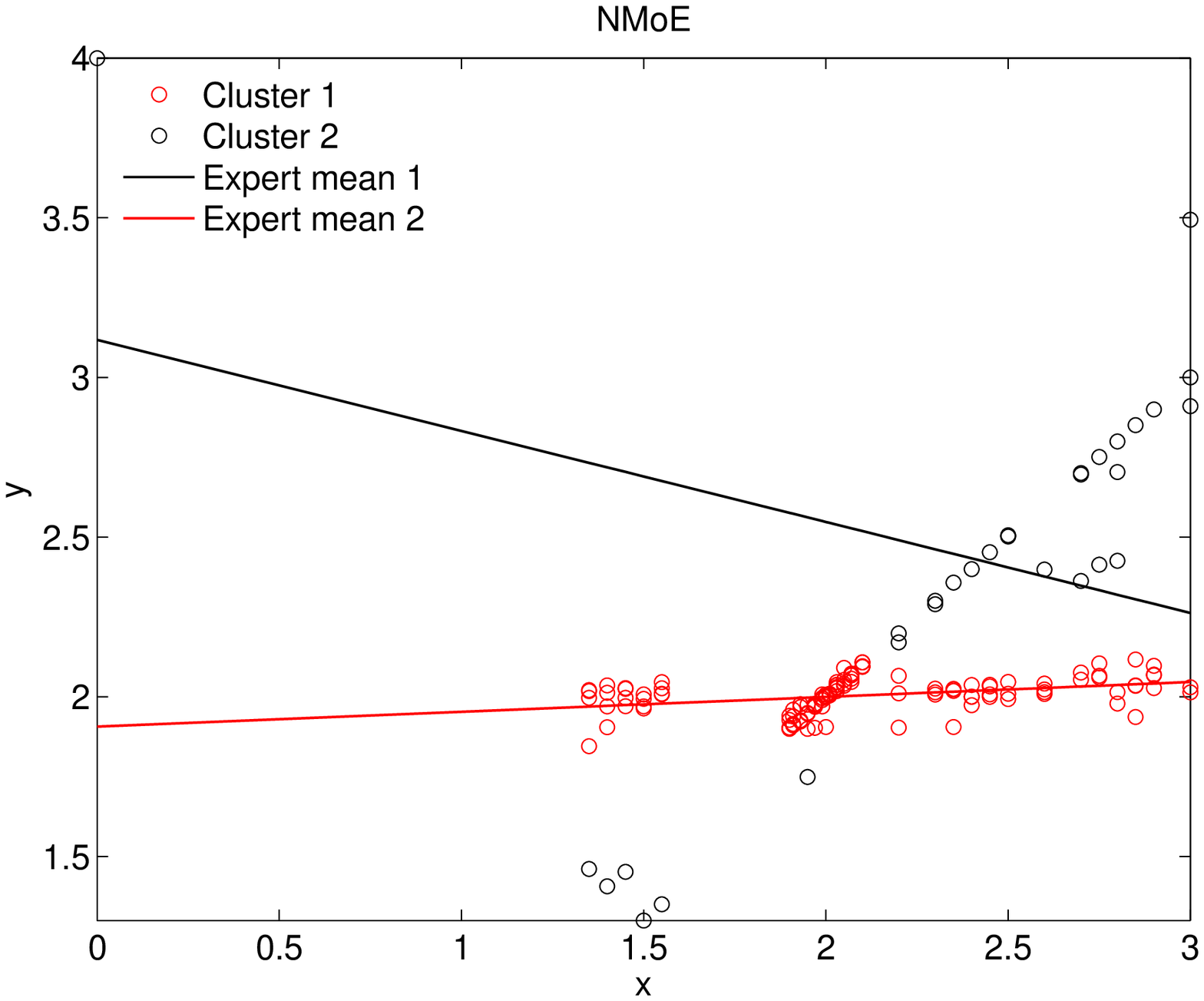} & 
 \includegraphics[width=6.2cm]{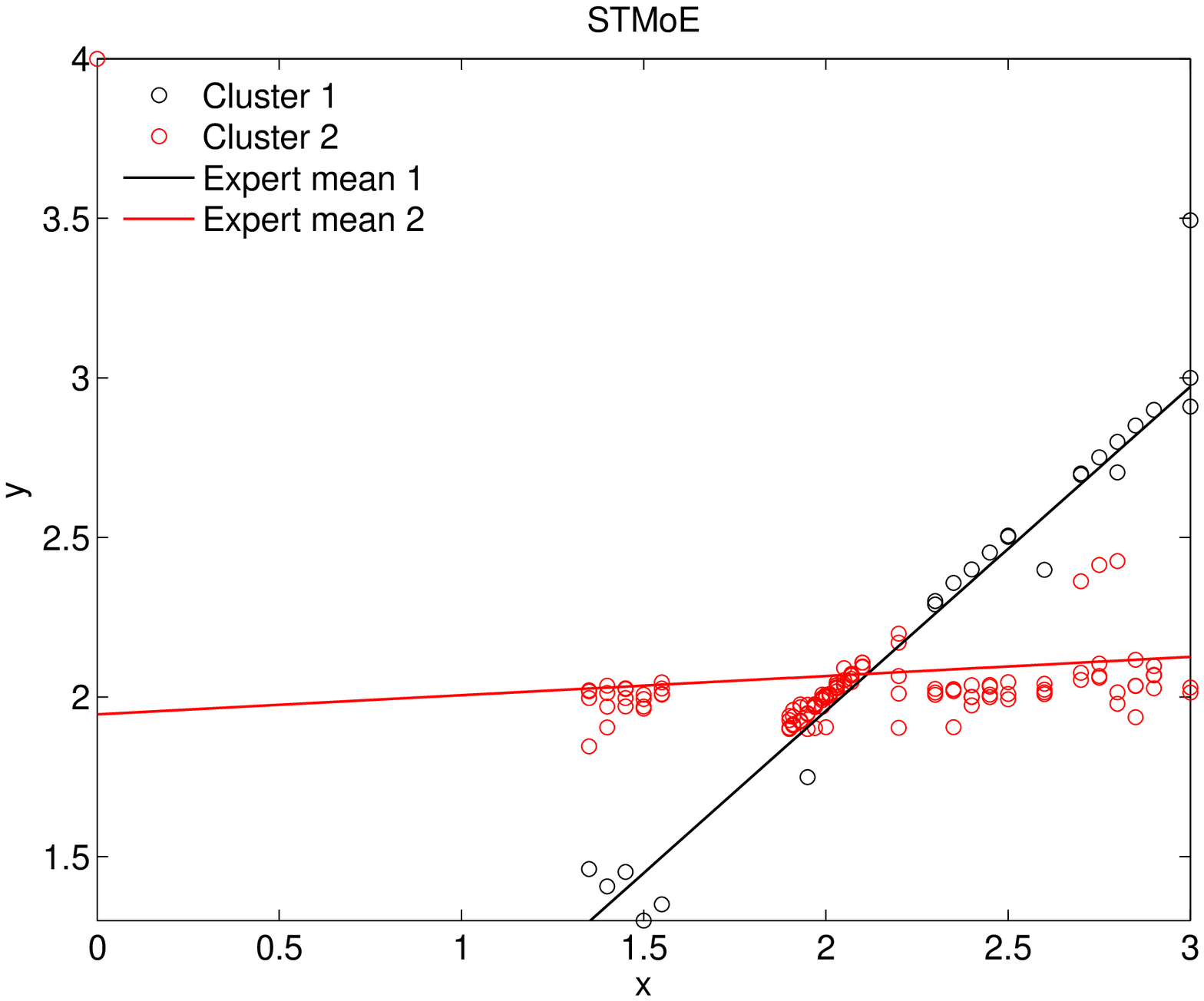}\\
    \end{tabular}
      \caption{\label{fig. Tone data with outliers and all models}Fitting MoLE to the tone data set with ten added outliers $(0,4)$ with  the NMoE model fit (left) and the STMoE model fit (right). The predictor $x$ is the actual tone ratio and the response $y$ is the perceived tone ratio.}
\end{figure}
Figure \ref{fig. Tone data with outliers and loglik} shows the log-likelihood profiles for each of the two  models, which show a similar behavior than the one in the case without outliers.
\begin{figure}[H]
   \centering 
   \begin{tabular}{cc}
   \includegraphics[width=6.2cm]{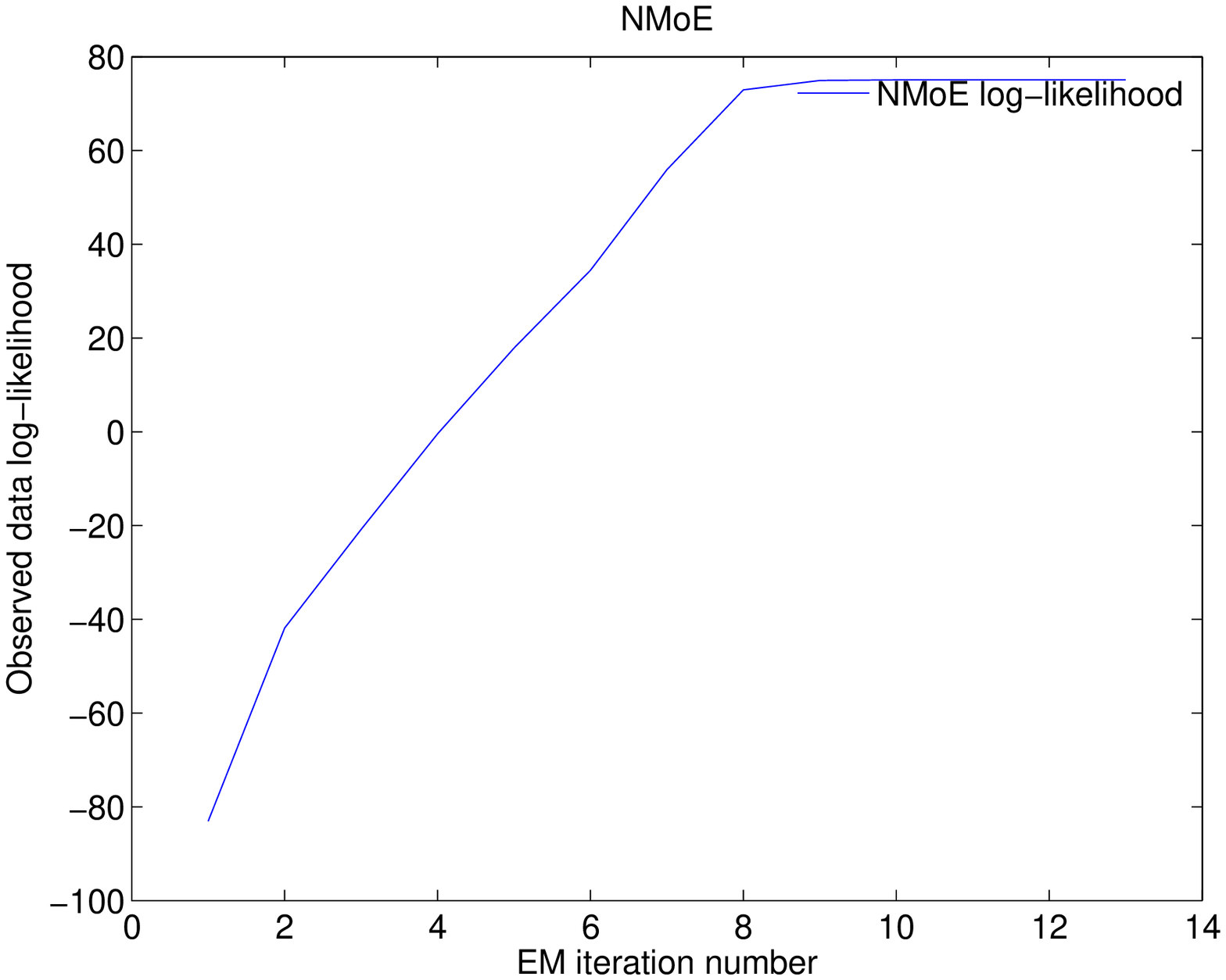} & \includegraphics[width=6.2cm]{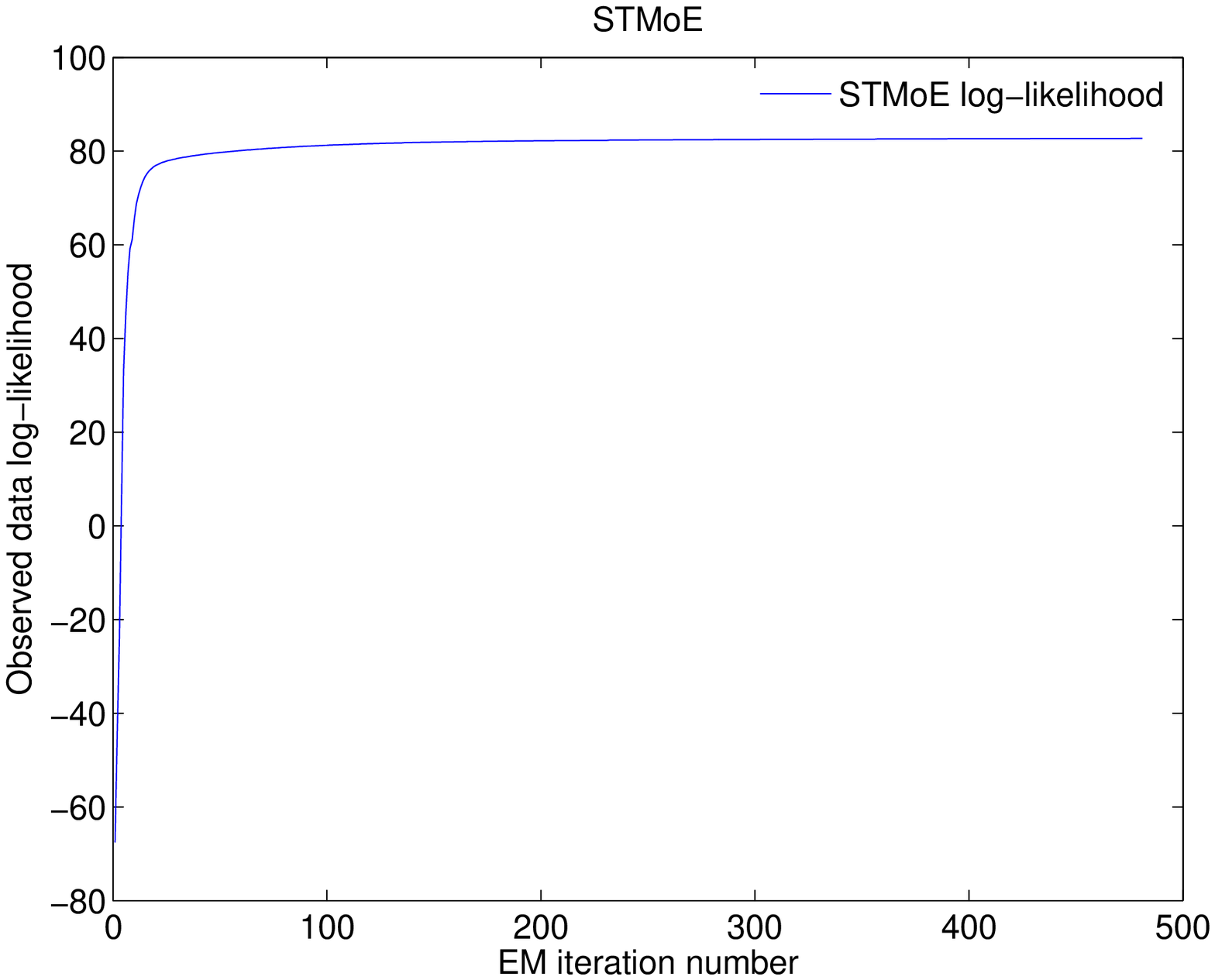}\\
   \end{tabular}
      \caption{\label{fig. Tone data with outliers and loglik}The log-likelihood during the EM iterations when fitting the MoLE models to the tone data set with ten added outliers $(0,4)$ with  the NMoE model (left) and  the STMoE model (right).}
\end{figure}

The values of estimated MoE parameters in this case with outliers are given in Table \ref{tab. Estimated parameters for the tone perception data set with outliers}.
One can  see that the SNMoE model parameters are  identical to those of the NMoE, with a skewness close to zero.
The regression coefficients for the second expert component are very similar for the two models. 
For the STMoE model, it retrieves a skewed normal component  while the second component is approaching a $t$ distribution with a small degrees of freedom. 
{\setlength{\tabcolsep}{3pt
\begin{table}[H]
\centering
{\small \begin{tabular}{l c c  c c c c c c c c c c}
\hline
param. & $\alpha_{10}$ & $\alpha_{11}$ & $\beta_{10}$ & $\beta_{11}$ & $\beta_{20}$ & $\beta_{21}$ & $\sigma_{1}$& $\sigma_{2}$ & $\lambda_{1}$ & $\lambda_{2}$ & $\nu_{1}$ & $\nu_{2}$ \\ 
model		& & & & & & & & & & & & \\
 \hline
 \hline
NMoE  	&  0.811	&  0.150	 & 3.117 & -0.285	&	1.907 & 0.046  & 0.700 & 0.050  & - & -	 & - & - \\
STMoE	&  -3.004 &	0.732  & -0.246  & 1.016 & 1.808	 &	0.060 & 0.212	&	0.088 &	156.240 & 1.757	&	81.355 & 1.630\\
 \hline
\end{tabular}}
\caption{\label{tab. Estimated parameters for the tone perception data set with outliers} Values of the estimated MoE parameters  for the  tone perception data set with added outliers.}
\end{table}
}
		    
\subsubsection{Temperature anomalies data set}
In this experiment, we examine another real-world data set related to climate change analysis. 
The NASA GISS Surface Temperature (GISTEMP) analysis provides a measure of the changing global surface temperature with monthly resolution for the period since 1880, when a reasonably global distribution of meteorological stations was established.
 The GISS analysis is updated monthly, however the data presented here\footnote{Source: \citet{TemperatureAnomalyData}, \url{http://cdiac.ornl.gov/ftp/trends/temp/hansen/gl_land.txt}} are updated annually as issued from the Carbon Dioxide Information Analysis Center (CDIAC), which has served as the primary climate-change data and information analysis center of the U.S. Department of Energy since 1982.
The data consist of $n = 135$ yearly measurements of the global annual temperature anomalies (in degrees C) computed using data from land meteorological stations for the period of $1882-2012$. 
These data have been analyzed earlier by \citet{Hansen1999,Hansen2001} and recently by \citet{Nguyen2016-MoLE} by using the Laplace mixture of linear experts (LMoLE). 

To apply the proposed non-normal mixture of expert model, we consider mixtures of two experts as in \citet{Nguyen2016-MoLE}. 
This number of components is also the one provided by the model selection criteria as shown later in Table \ref{tab. Model selection temperature anomalies data}. 
Indeed, as mentioned by \citet{Nguyen2016-MoLE},  \citet{Hansen2001} found that the data could be segmented into two periods of global warming (before 1940 and after 1965), separated by a transition period where there was a slight global cooling (i.e. 1940 to 1965). Documentation of the basic analysis method is provided by \citet{Hansen1999,Hansen2001}. 
We set the response $y_i (i=1,\ldots,135)$ as the temperature anomalies and the covariates $\bsx_i = \bsr_i = (1,x_i)^T$ where $x_i$ is the year of the $i$th observation.

Figures \ref{fig. temperature anomalies data and models experts}, 
\ref{fig. temperature anomalies data and models means}, and 
\ref{fig. temperature anomalies data and models loglik}
respectively show, for each of the two MoE models, the two fitted linear expert components, 
the corresponding means and confidence regions computed as plus and minus twice the estimated (pointwise) standard deviation as presented in Section \ref{sec: Prediction using the STMoE},
and the log-likelihood profiles.
One can observe that the two models are successfully applied on the data set and provide very similar results.  These results are also similar to those found by \citet{Nguyen2016-MoLE} who used a Laplace mixture of linear experts.
\begin{figure}[H]
   \centering 
   \begin{tabular}{cc}
   \includegraphics[width=6.2cm]{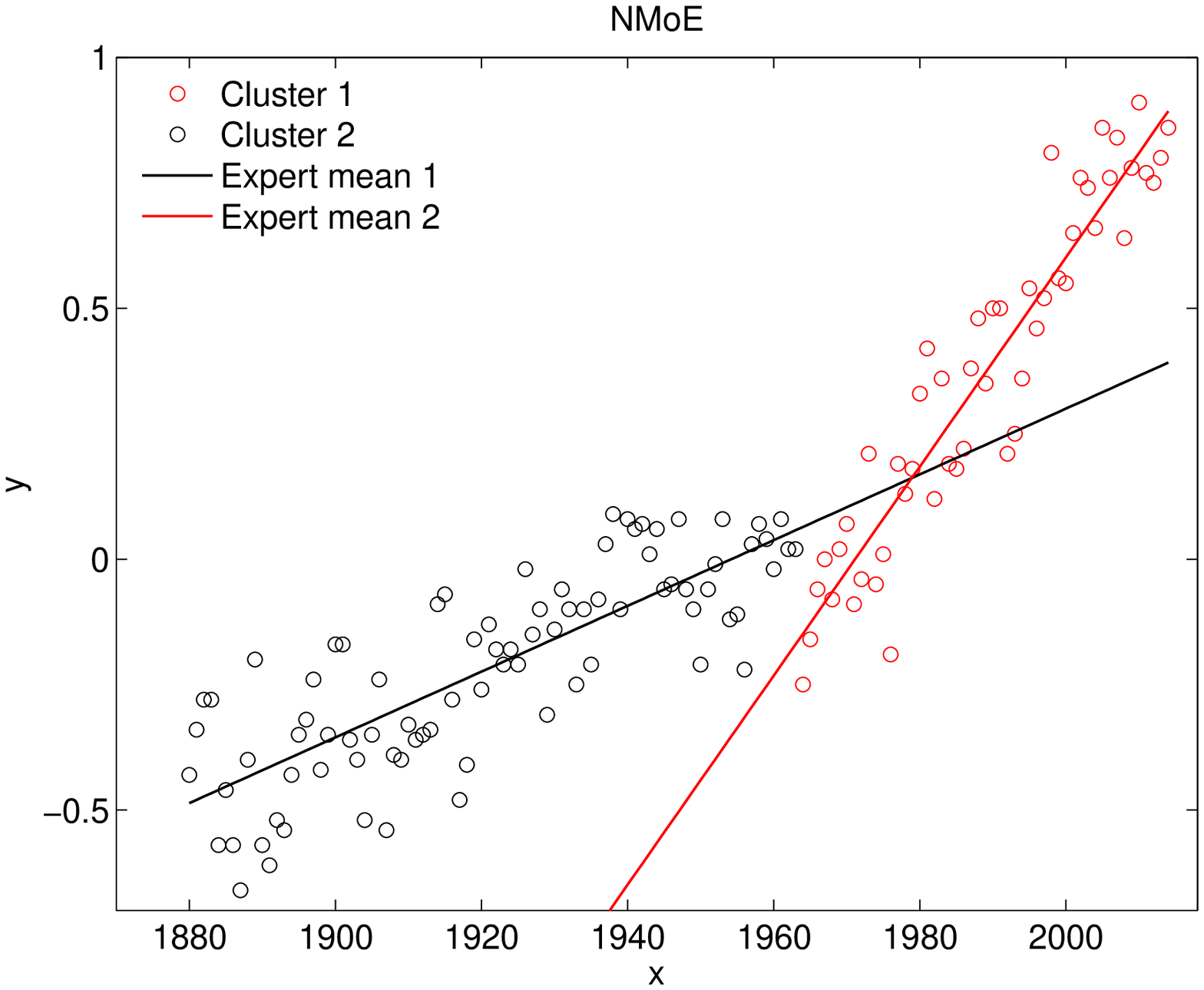} & 
\includegraphics[width=6.2cm]{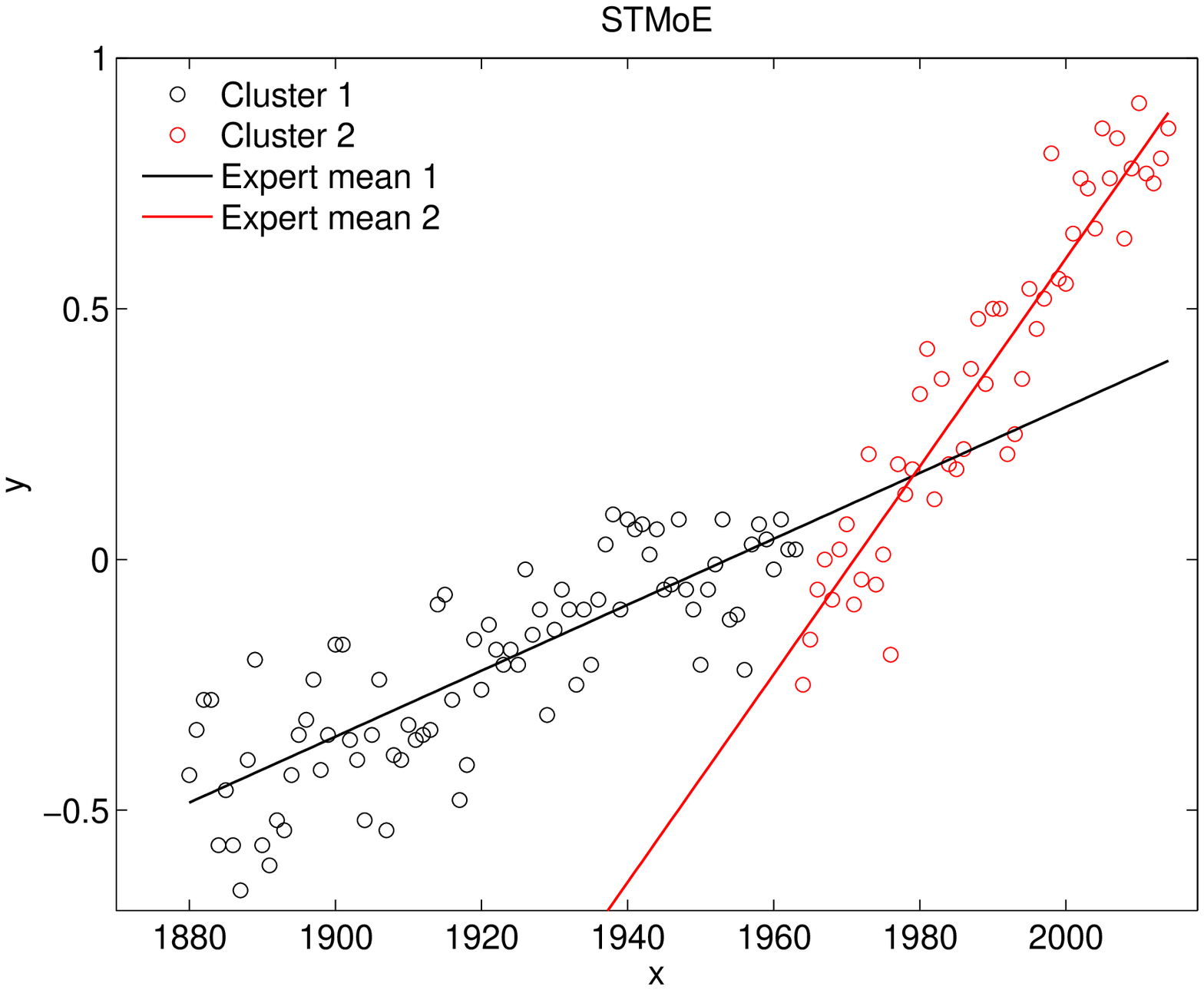}\\    
\end{tabular}
      \caption{\label{fig. temperature anomalies data and models experts}Fitting the MoLE models to the temperature anomalies data set with  the NMoE model (left) and the STMoE model (right). The predictor $x$ is the year and the response $y$ is the temperature anomaly.}
\end{figure}
\begin{figure}[H]
   \centering 
   \begin{tabular}{cc}
   \includegraphics[width=6.2cm]{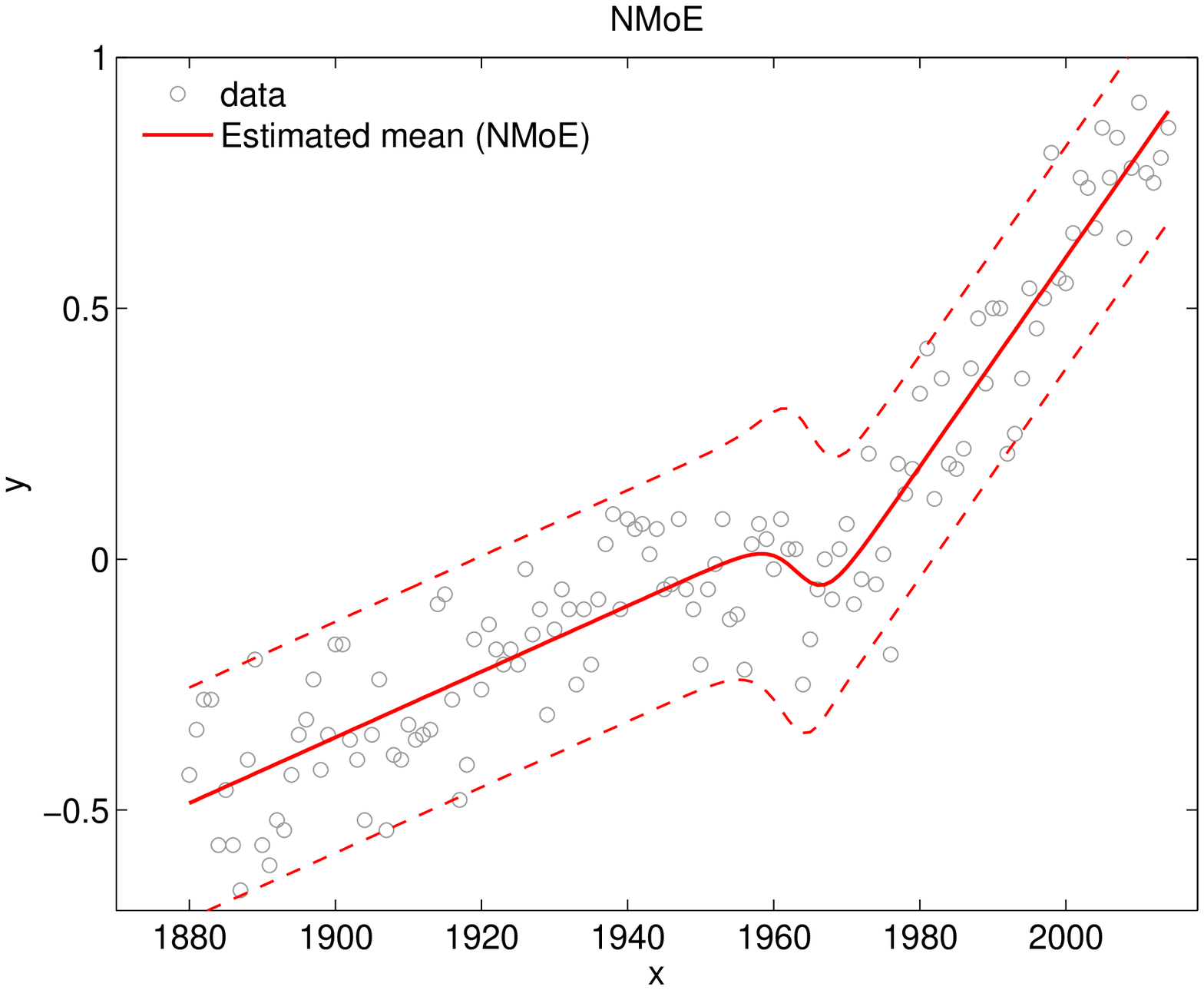} & 
   \includegraphics[width=6.2cm]{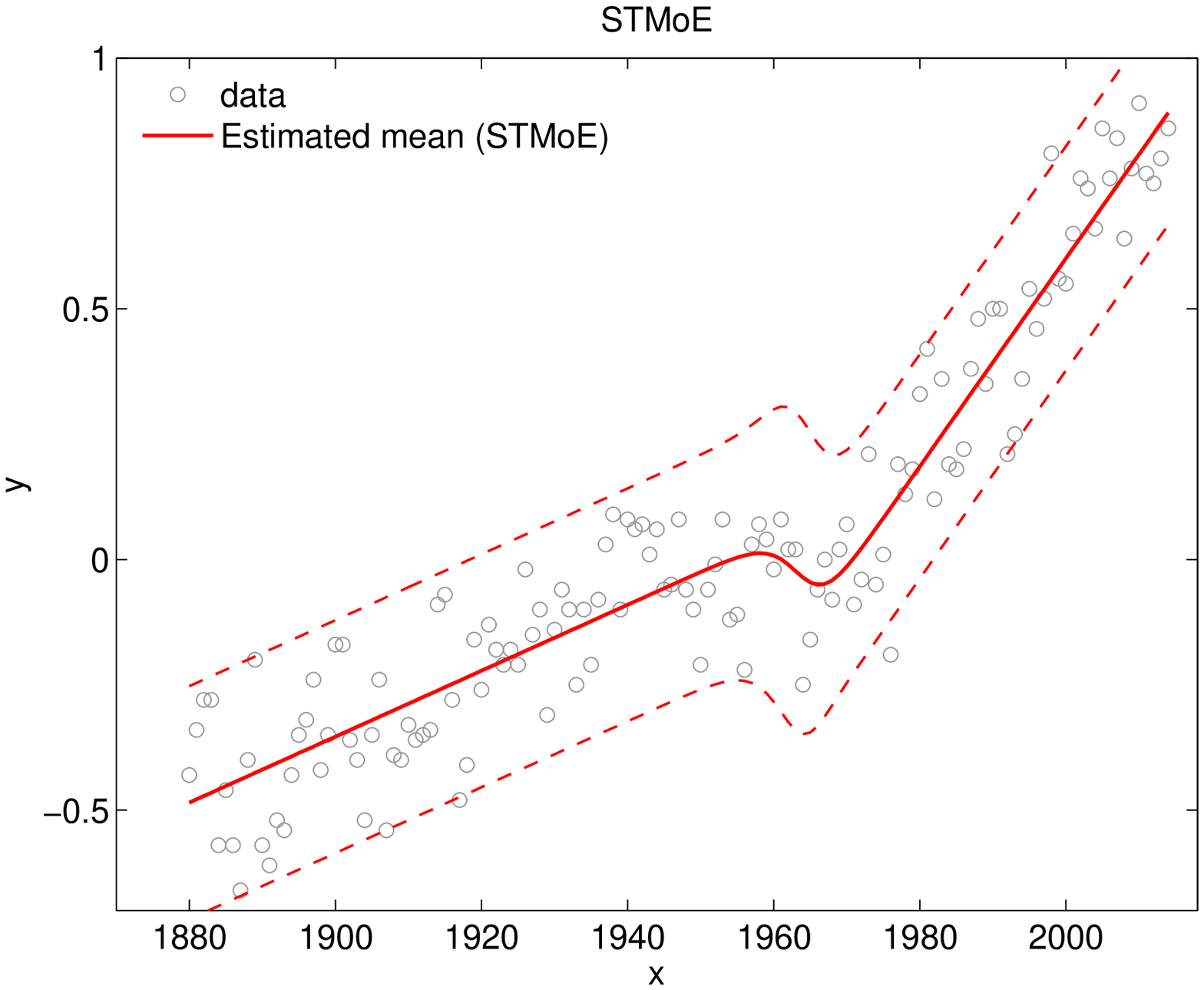} 
   \end{tabular}
      \caption{\label{fig. temperature anomalies data and models means}The fitted MoLE models to the temperature anomalies data set with  the NMoE model fit (left) and the STMoE model fit  (right). The predictor $x$ is the year and the response $y$ is the temperature anomaly.  The shaded region represents plus and minus twice the estimated (pointwise) standard deviation as presented in Section \ref{sec: Prediction using the STMoE}.}
\end{figure}
\begin{figure}[H]
   \centering 
   \begin{tabular}{cc}
   \includegraphics[width=6.2cm]{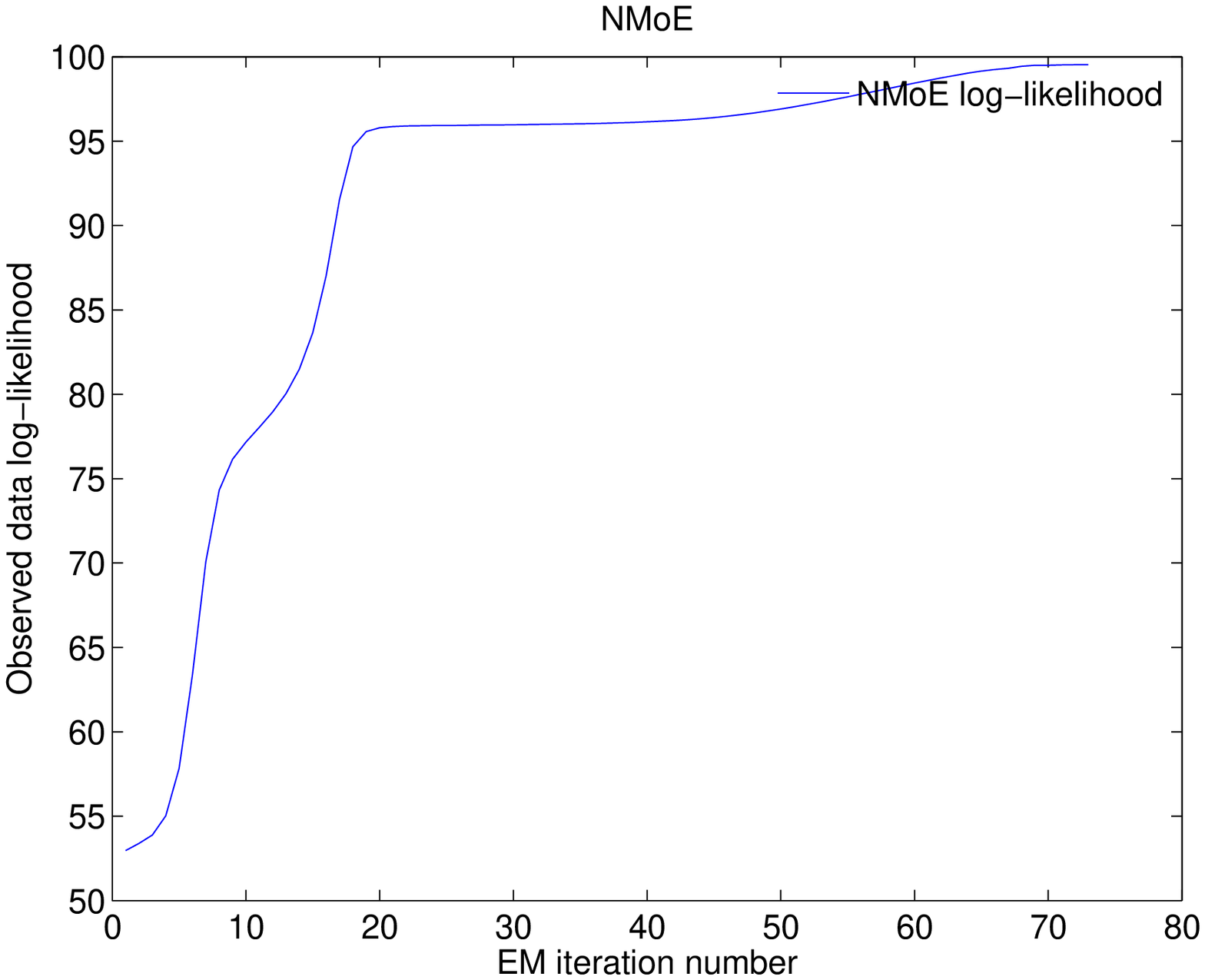} &
     \includegraphics[width=6.2cm]{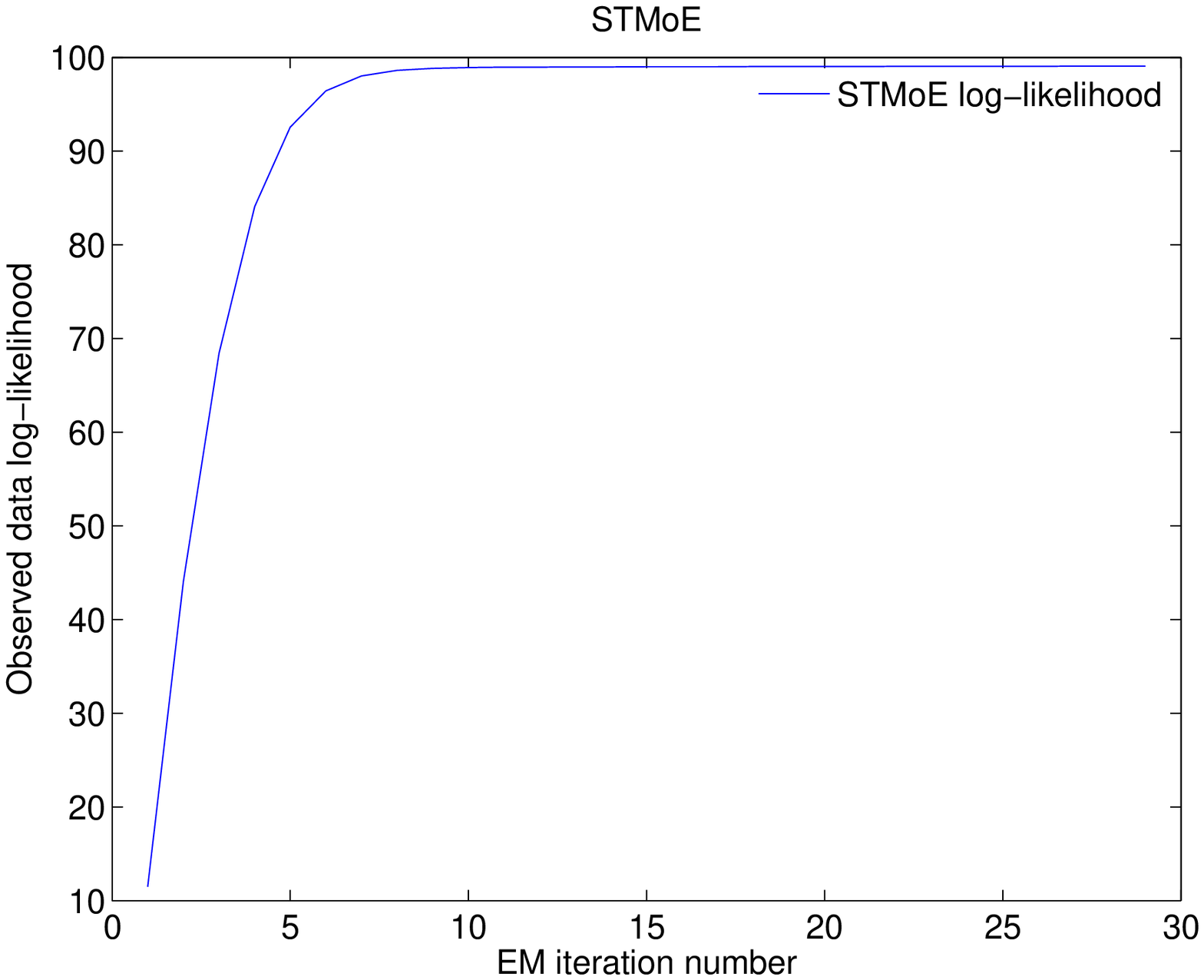}\\
   \end{tabular}
      \caption{\label{fig. temperature anomalies data and models loglik}The log-likelihood during the EM iterations when fitting the MoLE models to the temperature anomalies data set with  the NMoE model (left) and  the STMoE model (right).}
\end{figure}
The values of estimated MoE parameters for the temperature anomalies data set are given in Table \ref{tab. estimated parameters for the temperature anomalies data set}.
One can see that the parameters common for the two models are quasi-identical. 
It can also be seen 
the STMoE model provides a solution with a skewness close to zero. This may support the hypothesis of non-asymmetry for this data set.
Then, the STMoE solution provides a degrees of freedom more than 17, which tends to approach a normal distribution.
On the other hand, the regression coefficients are also similar to those found by \citet{Nguyen2016-MoLE} who used a Laplace mixture of linear experts.
{\setlength{\tabcolsep}{3pt
\begin{table}[H]
\centering
{\small
\begin{tabular}{l c c  c c c c c c c c c c}
\hline
param. & $\alpha_{10}$ & $\alpha_{11}$ & $\beta_{10}$ & $\beta_{11}$ & $\beta_{20}$ & $\beta_{21}$ & $\sigma_{1}$& $\sigma_{2}$ & $\lambda_{1}$ & $\lambda_{2}$ & $\nu_{1}$ & $\nu_{2}$ \\ 
model		& & & & & & & & & & & & \\
 \hline
 \hline
NMoE  	& 946.483 & -0.481  & -12.805 & 0.006 & -41.073 & 0.020 & 0.115 & 0.110 & - & -  & -  & -  \\
STMoE	& 931.966 & -0.474 & -12.848 & 0.006  & -40.876 & 0.020 & 0.113 & 0.105 & 0.024 & -0.015 & 41.048 & 17.589\\
 \hline
\end{tabular}}
\caption{\label{tab. estimated parameters for the temperature anomalies data set}Values of the estimated MoE parameters for the temperature anomalies data set.}
\end{table}
}

We also performed a model selection procedure on the temperature anomalies data set to choose the best number of MoE components from values between 1 and 5. Table  \ref{tab. Model selection temperature anomalies data} gives the obtained values of the used model selection criteria, that is BIC, AIC, and ICL. 
One can see that, except the result provided by the AIC for the NMoE model which provide a high number of components, all the others results provide evidence for two components in the data.  
{\setlength{\tabcolsep}{4pt
\begin{table}[H]
\centering
{\small 
\begin{tabular}{l |ccc | ccc}
\hline
	& \multicolumn{3}{c|}{NMoE}	&  \multicolumn{3}{c}{STMoE}\\
\cline{2-7}
K			 &   BIC 	 & 	AIC 	&    ICL  &   BIC 	 & 	AIC 	&    ICL \\
\hline
\hline
1 & 46.062 &  50.420 &  46.062 &	 40.971 &  48.234 &  40.971 \\
2 & \textbf{79.916} &  91.537 &  \textbf{79.624} & \textbf{69.638} & \textbf{87.069}  &  \textbf{69.341} \\
3 & 71.396 &  90.280 &  58.487 & 54.126 & 81.726  & 30.655 \\
4 & 66.727 &  92.875 &  54.752 & 42.308 &  80.0773 &  20.494 \\
5 & 59.510 &  \textbf{92.920} &  51.242 & 28.037 &  75.974 &  -8.881	 \\ 
\hline 
\end{tabular}
}
\caption{\label{tab. Model selection temperature anomalies data}Choosing the number of expert components $K$ for the temperature anomalies data by using the information criteria BIC, AIC, and ICL. Underlined value indicates the highest value for each criterion.}
\end{table}}

\section{Concluding remarks and future work}
\label{sec: Conclusion}

 In this paper we proposed a new non-normal MoE model, which generalizes the normal MoE model
 and attempts to simultaneously accommodate heavy tailed data with possible outliers and  asymmetric distribution.
The proposed STMoE is based on the flexible skew $t$ distribution which is suggested for possibly non-symmetric, heavy tailed and noisy data.
 We developed a CEM algorithm for model inference and  described the use of the model in non-linear regression and prediction as well as in model-based clustering.
The developed model was successfully applied and validated on simulation studies and two real data sets.
The results obtained on simulated data confirm the good performance of the model in terms of density estimation, non-linear regression function approximation and clustering.
 In addition, the simulation results provide evidence of the robustness of the STMoE model to outliers, compared to the standard alternative NMoE model.
The proposed model was  also successfully applied to two different real data sets, including situations with outliers. 
The model selection using information criteria tends to promote using BIC and ICL against  AIC which may perform poorly in the analyzed data. 
The obtained results support  the potential benefit of the proposed approach for practical applications.
\\
 One interesting future direction is to extend the proposed model to the hierarchical mixture of experts framework \citep{jordanHME}. Another natural  future extension of this work is also to consider the case of   MoE for multiple regression on multivariate responses rather than simple univariate regression. 
In that case, one may consider the multivariate skew-t and the multivariate Normal inverse Gaussian distribution  \citep{OHagan2016} which may be more stable in high-dimensional settings compared to the multivariate skew-t.

\appendix

\section{The skew-normal distribution}
\label{ssec: Skew-normal distribution}

 As introduced by \citep{Azzalini1985,Azzalini1986}, a random variable $Y$ follows a univariate skew-normal distribution with location parameter $\mu \in \R$, scale parameter $\sigma^2\in (0,\infty)$ and skewness parameter $\lambda \in \R$ if it has the density
\begin{eqnarray}
f(y;\mu,\sigma^2,\lambda) &=& \frac{2}{\sigma} \phi(\frac{y-\mu}{\sigma}) 
 \Phi \left(\lambda (\frac{y-\mu}{\sigma})\right)
\label{eq: Skew-normal density}
\end{eqnarray}
where $\phi(.)$ and $\Phi(.)$ denote, respectively, the probability density function (pdf) and the cumulative distribution function (cdf) of the standard normal distribution. It can be seen from (\ref{eq: Skew-normal density}) that when $\lambda = 0$, the skew-normal  reduces to the normal distribution.
As presented by \citet{Azzalini1986, Henze1986}, if 
\begin{equation}
Y = \mu + \delta |U| + \sqrt{1 - \delta^2} E
\label{eq: stochastic representation skew-normal}
\end{equation}
where $\delta= \frac{\lambda}{\sqrt{1 +\lambda^2}}$, $U$ and $E$ are independent random variables following the normal distribution $\text{N}(0,\sigma^2)$, then $Y$ follows the  skew-normal distribution with pdf $\text{SN}(\mu,\sigma^2,\lambda)$ given by  (\ref{eq: Skew-normal density}). In the above,  $|U|$ denotes the magnitude of $U$. 
This stochastic representation of the skew-normal distribution leads to the following hierarchical representation in an incomplete data framework,  as presented in \citet{Lin07univSkewNMixture}: 
\begin{equation}
\begin{tabular}{lll}
$Y|u$ & $\sim$ & $\text{N}\left(\mu + \delta |u|, (1-\delta^2)\sigma^2\right),$\\
$U$ & $\sim$ & $\text{N}(0,\sigma^2)$.
\end{tabular} 
\label{eq: hierarchical representation skew-normal}
\end{equation}

\section{Stochastic representation of the STMoE model}
\label{apx: stochastic representation of the STMoE}
The skew $t$ mixture of experts model is characterized as follows. 
Suppose that conditional on a  categorical variable $Z_i =z_i \in \{1,\ldots,K\}$ representing the hidden label of the component generating the $i$th observation and which,  conditional on some predictor $\bsr_i$,  follows the multinomial distribution (\ref{eq: Multinomial}):
\begin{equation}
Z_i|\bsr_i \sim \text{Mult}\!\left(1;\pi_{1}(\bsr_i;\bsalpha),\ldots, \pi_{K}(\bsr_i;\bsalpha)\right)
\label{eq: Multinomial}
\end{equation}where each of the probabilities $\pi_{z_i}(\bsr_i;\bsalpha) = \Pro(Z_i=z_i|\bsr_i)$ is given by the multinomial logistic function (\ref{eq: multinomial logistic}). 
Now suppose a random variable $Y_i$ having the following representation: 
\begin{equation}
Y_i =  \mu(\bsx_i;\bsbeta_{z_i}) + \sigma_{z_i} \frac{E_i}{\sqrt{W_i}}
\label{eq: stochastic representation skew-t MoE}
\end{equation}where   $E_i$ and $W_i$ are independent univariate random variables with, respectively, a standard skew-normal distribution $E_{i} \sim \text{SN}(\lambda_{z_i})$, and a Gamma distribution $W_i \sim \text{Gamma}(\frac{\nu_{z_i}}{2},\frac{\nu_{z_i}}{2})$, and $\bsx_i$ and $\bsr_i$ are some given covariate variables. Then, the variable $Y_i$ is said to follow the skew $t$ mixture of experts (STMoE) defined by (\ref{eq: skew-t MoE}).

\section*{References}
\bibliographystyle{elsarticle-harv}
\bibliography{STMoE-NeuComp}
\end{document}